\documentclass[10pt]{article}
\usepackage[letterpaper,left=60pt,right=60pt,top=54pt,bottom=54pt]{geometry}
\usepackage{amsmath,amssymb,graphicx,booktabs,natbib,xcolor,hyperref}
\usepackage{microtype}

\let\classnormalsize\normalsize
\renewcommand{\normalsize}{\classnormalsize\fontsize{10.05pt}{12.1pt}\selectfont}
\normalsize
\graphicspath{{assets/}}
\hypersetup{pdftitle={Predicting Pulse Responses in Water Radiolysis with Compact World Models},pdfauthor={Humza Nusrat and Omar Nusrat},colorlinks=true,linkcolor=blue,citecolor=blue,urlcolor=blue}
\title{\vspace{-14pt}Predicting Pulse Responses in Water Radiolysis with Compact World Models}
\author{{\normalsize\textbf{Humza Nusrat}$^{1,2}$\qquad\textbf{Omar Nusrat}$^{3}$}\\
 {\normalsize$^{1}$Department of Radiation Oncology, Henry Ford Health, Detroit, Michigan, USA}\\
 {\normalsize$^{2}$Department of Radiology, Michigan State University, East Lansing, Michigan, USA}\\
 {\normalsize$^{3}$Toronto Metropolitan University, Toronto, Ontario, Canada}\\
 {\normalsize\href{mailto:hnusrat1@hfhs.org}{hnusrat1@hfhs.org}}}
\date{November 2026}
\begin{document}
\raggedbottom
\maketitle
\begin{abstract}
\normalsize
A world model that compresses a particle cloud must retain enough geometry to predict its response to an intervention. We tested which geometry transfers when the energy of a radiation track or the timing of a later molecular pulse changes. We trained at 200\,eV per electron and predicted the change in seven species counts at 20\,ns against repeated Geant4-DNA simulations of water radiolysis. At 400\,eV, a model retaining each species' center and spread (600 bytes per branch, 4\,377 weights) had 57\% lower normalized mean squared error (NMSE) than equivariant pooling with C64 (608 bytes, 64\,517 weights) and 42\% lower error than C32. All five moment fits beat all five C64 fits. Their mean errors were similar at 100\,eV. Giving the energies equal weight, the prespecified NMSE reduction versus C64 was 0.0528 (paired 95\% CI 0.0326--0.0790). Learned moments also reduced error by 66\% on exploratory three-pulse schedules. In post-hoc fits at 400\,eV, narrowing C64 at fixed weight decay lowered NMSE from 0.1966 to 0.1623, while replacing pair-specific encounters with a common scale raised moment-model NMSE from 0.0801 to 0.1344. On supplied 400\,eV clouds, a preselected moment fit scored 0.0672 versus 0.0685 for independent reaction time simulation (IRT) with 64 paths per branch and used one nineteenth of its CPU time. The five-fit mean was 0.0801. Estimated infinite-path IRT squared bias was lower (0.0265 versus 0.0589). For mean passive hydroxyl-radical yield across 64 sources, direct simulation gave lower error at lower cost once source generation was included.

\end{abstract}
\section{Introduction}

A radiation track leaves reactive molecules close enough to meet as they diffuse through water. A second track can change the fate of the first. Its effect depends on which molecules survive until it arrives and which reactants it can reach \citep{kreipl2009interaction}. A particle simulator retains every position. A world model with a fixed memory budget must decide which spatial information to carry forward. An intervention tests that choice directly. Two clouds with similar undisturbed yields can respond differently to the same addition.

Water radiolysis lets us measure that difference. We can restart stochastic chemistry from the same observed cloud, add molecules to one branch and let the other branch evolve undisturbed. Repeated simulations estimate the mean difference in seven species counts and its sampling uncertainty. The reaction network conserves hydrogen and oxygen atoms, including exchange with reservoir water, and electric charge. Changing the electron energy produces clouds with different numbers and arrangements of molecules. Changing when an incoming cloud arrives tests whether the model can carry an earlier intervention into a later one.

We trained two kinds of compact state on 128 source families at 200\,eV per electron. One retains the center and spread of each species and learns corrections to reaction rates. The other learns how to pool positions across species and how the resulting vectors, tensors and scalars evolve. The largest pooled model, C64, uses almost the same branch memory as the moment model. Their different numbers of weights and different encounter rules give two possible reasons for a transfer gap. We also compared these models with independent reaction time simulation, a fast alternative to step-by-step chemistry.

State memory and fitted weights answer different practical questions. Forecasting from many clouds requires a state for each branch, while every branch can share the same fitted weights. We tested two explanations for the transfer gap in exploratory fits that kept branch memory fixed. We varied the width and regularization of the pooled model. In the moment model we replaced species-specific encounter volumes with a common scale while retaining the geometry used to place products.

First, on fresh source families backed by 81\,920 step-by-step paths, learned moments reduced response error at 400\,eV by 57\% relative to C64 and 42\% relative to the smaller C32. Second, the advantage over C64 persisted when a single pulse was split into three arrivals. Replacing pair-specific encounters with a common scale increased the moment model's error on both shifts. For the moment model itself, training only on passive trajectories increased 400\,eV intervention error sixfold at the same simulation budget. Third, tighter integration changed the primary models' standardized response by at most $3.45\times10^{-6}$, and pairing methods within source families separated the transfer effect from variation between clouds. Fourth, on supplied clouds a preselected moment fit used about one nineteenth of the CPU time of 64-path reaction time simulation at similar observed error. When the target was a mean across a finite bank of sources, direct simulation gave a more accurate estimate at lower cost.

\section{Related work}

Diffusion models of water radiolysis approximate how the products of a track spread and react \citep{schwarz1969spur}. Simulating pairs of tracks lets one vary the delay and separation between them \citep{kreipl2009interaction}. Geant4-DNA
follows individual particles \citep{incerti2010project,bernal2015trackstructure,tran2024chemistryreview}
and has been used to study irradiation schedules and oxygen conditions
\citep{le2024pulse,chappuis2024pulse}. MIRaCLE starts from a stochastic
track and evolves a continuum description of the chemistry to predict yields over time
\citep{perkins2026miracle}. Our target is the difference in counts between
two continuations of one observed cloud, only one of which receives a pulse.

Moment closures and correlation models use spatial summaries to approximate
chemical dynamics \citep{ernst2018boltzmann,ernst2019moment,kim2021correlation}.
A physical evolution law can also incorporate a learned correction
\citep{yin2021aphynity}. In our comparison, the moment predictor
retains the position and spread of each species and learns corrections to
reaction rates. The other predictors pool geometry across species and
learn how those pooled summaries evolve.

World-model studies ask which physical distinctions an intervention
requires \citep{thorpe2026viable}, which predictions suffice for a decision
\citep{wei2026distinguish}, and how closely latent dynamics reproduce environmental
transitions \citep{gonzalezlaiz2026limits}. Learned dynamics can predict
reaction--diffusion fields as their inputs change \citep{regazzoni2024ldnets}.
RD-JEPA adapts a pretrained predictor to a new reaction--diffusion equation
using trajectories from the new system \citep{si2026rdjepa}. We keep the reaction network fixed and change either the source or the pulse schedule after fitting. Comparing active and passive training on the same architecture tests whether seeing an intervention during training helps predict its effect. Comparing validation at 200\,eV with transfer to a new energy or a later time tests where that validation ceases to predict accuracy.

\section{Radiation-chemistry prediction task}
\label{sec:environment}

\subsection{Observed sources and supplied additions}
We used Geant4-DNA to simulate electron transport and step-by-step (SBS)
chemistry in pure liquid water \citep{allison2016geant4,geant4v1143,karamitros2014diffusion}.
At 0.02\,ns we recorded the species and position of every molecule in the
cloud produced by two electrons. We tracked seven species,
$e_{\mathrm{aq}}^{-}$, $^{\bullet}\mathrm{OH}$, H, H$_2$, H$_3$O$^{+}$,
OH$^{-}$, and H$_2$O$_2$.

\begin{figure}[!htbp]
\centering
\includegraphics[width=.88\textwidth]{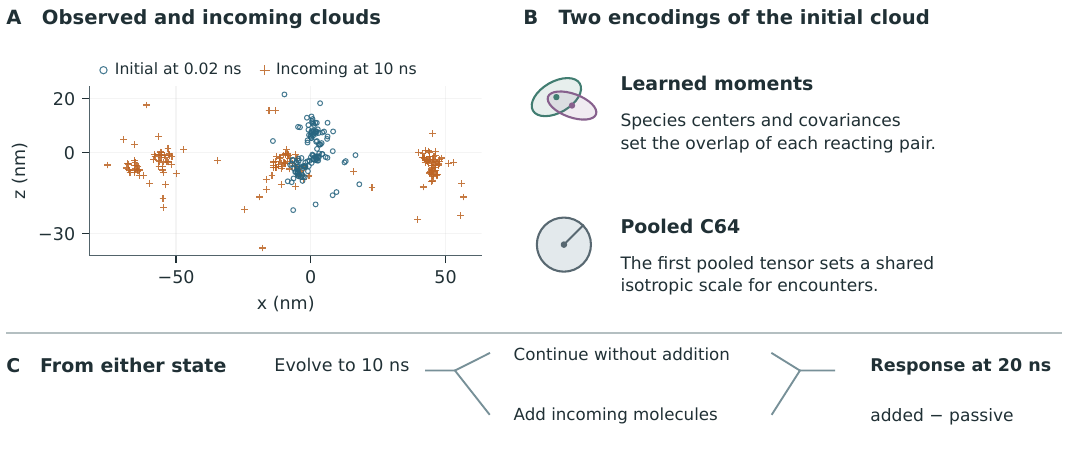}
\caption{Pulse response from a compact state. (A) The first 400\,eV confirmation family, projected about the initial centroid. The three clouds arrive together at 10\,ns at offsets $-50$, 0 and $+50$\,nm. (B) Learned moments retain species geometry. C64 pools geometry across species. (C) Subtracting passive from forced counts at 20\,ns gives the predicted response.}
\label{fig:source-state}
\end{figure}

A source family pairs the initial cloud with three incoming clouds, each
generated independently by one electron. The action sets the arrival time
and placement of the incoming clouds. In the main experiment all three arrive at time $\tau$, with their origins
at $x$ offsets $(-r,0,+r)$. We call this addition a pulse
(Fig.~\ref{fig:source-state}). The predictor encodes the initial cloud once and forecasts two branches
through 20\,ns without further observations. The passive branch evolves
undisturbed. The pulse branch receives the supplied molecules at $\tau$.
Independent SBS continuations estimate the mean counts on each branch.
The difference between these means estimates the pulse response.

The confirmation bank spans 100, 200 and 400\,eV per electron. At those energies, the initial clouds contained medians of 30, 60 and 117.5 molecules. The three incoming clouds together contained 48, 90.5 and 173. Median initial root-mean-square radii about the centroid were 10.1, 10.0 and 11.7\,nm. At zero offset the median pulse radii were 9.0, 9.7 and 11.4\,nm. Separating their three origins by 50\,nm increased the corresponding radii to 41.7--42.8\,nm (Appendix Table~\ref{tab:v13-cloud}).

\subsection{Family separation and source transfer}
All actions and branches from one family stayed in the same split. We
fitted the models on 128 families and selected checkpoints on 32 other
families, all at 200\,eV per primary electron. Six actions per family
combined $\tau=1$ or 10\,ns with $r=0$, 10, or 50\,nm. We averaged
four SBS continuations per branch for each training target and 16 for
each validation target.

To test source transfer, we changed the energy per primary electron and
kept the actions familiar. The development and confirmation banks each contained
64 families at 100\,eV, 32 at 200\,eV, and 64 at 400\,eV. The 200\,eV
families tested new clouds at the training energy. At the other energies,
both the track structure and the number of molecules changed. The five electrons supplied total energy $5E$ at primary energy $E$. Each family supplied four actions ($\tau=1$ or 10\,ns and $r=0$
or 50\,nm), with 64 SBS continuations per branch.  Development results informed the design. For the original confirmation comparison, we fixed the predictors and numerical settings before examining its outcomes. A separate follow-up
kept the energy at 200\,eV and changed the addition schedule on previously
held-out families (Appendix~\ref{app:action-transfer}).

We compared yields from reconstructed clouds with yields from uninterrupted chemistry and checked the native reference against an independent Geant4 application. Both comparisons met the prespecified agreement margin of $\pm0.10$ molecule per 100\,eV (Appendix Table~\ref{tab:native-consistency} in Section~\ref{sec:native-checks}).

\section{Predictive states and fitting}
\label{sec:models}

A cloud state contains seven species counts $n$, a clock $t$ and a ledger $w$ that records exchange with reservoir water. Spatial models also carry geometry about the initial centroid. The stoichiometric matrix $S$ and reservoir coefficients $\nu_w$ convert reaction hazards $a$ to $\dot n=Sa$ and $\dot w=\nu_w^{\mathrm T}a$. For a reaction between $A$ and $B$,
\begin{equation}
 a_r=\kappa_r n_A^+ n_B^+ g_r c_r,\qquad
 c_r=\exp\!\left[6\tanh\!\left(H_r(v)/6\right)\right],
\end{equation}
where $n_s^+=\max(n_s,0)$, $\kappa_r$ is the pair rate constant and $g_r$ is an inverse encounter volume. The learned spatial models use the bounded multiplier $c_r$ shown above. Their rate networks $H$ read log counts, geometry and two clock inputs, $\log(1+t)$ and $\log^2(1+t)/10$, with $t$ in ns. Physical moments fix $c_r=1$.

\textit{Encoding the cloud.}
The moment models store each species' center and raw second moment. C16, C32 and C64 pool positions across species into $(3,1,1)$, $(6,2,2)$ and $(12,4,4)$ vector, tensor and scalar channels, respectively. Their scalar encoder also receives 98 initial species-pair distance summaries at two length scales. All encodings are invariant to particle order \citep{zaheer2017deepsets}. A particle graph simulator retains one node per particle \citep{sanchez2020gns}. Here the coordinates are discarded after encoding, so recurrent memory stays fixed as the cloud grows.

\textit{Advancing the state.}
The moment and pooled models estimate how readily two reactants can meet. For reaction $r$ between species $A$ and $B$, let $d_r$ join their centers, $C_A,C_B$ be their spatial covariances, and $R_r$ be the capture radius. With $M_r=C_A+C_B+R_r^2 I$, their inverse encounter volumes (in $\mathrm{nm}^{-3}$) are
\[
 g_r^{\mathrm{mom}}=\frac{\exp(-d_r^{\mathrm T}M_r^{-1}d_r/2)}
 {(2\pi)^{3/2}\sqrt{\det M_r}},\qquad
 g_r^{\mathrm{pool}}=\big[4\pi(\sigma^2+R_r^2)\big]^{-3/2}.
\]
For learned moments, $v$ includes all nine log overlaps, species covariance traces and squared distances of species centers from the origin. Let $f_s=n_s\mu_s$ and $Q_s=n_sT_s$ denote the extensive first and second moments in units scaled by $L=10\,\mathrm{nm}$. With consumption rate $q_s$, product coefficients $\beta_{sr}$ and diffusion constant $D_s$, their evolution is
\[
 \dot f_s=-q_s\mu_s+\sum_r\beta_{sr}a_r p_r,\qquad
 \dot Q_s=-q_sT_s+\sum_r\beta_{sr}a_r U_r+2n_sD_s I/L^2 .
\]
The product of the two smoothed reactant densities supplies the encounter mean $p_r$ and raw second moment $U_r$. Diffusion increases only the second moment.

The pooled models use one variance, $\sigma^2=L^2\max(\mathrm{tr}\,T_1,0)/3$, from the first uncentered tensor channel. Their rate and flow networks read all channels through rotation-invariant contractions, including every vector pair. The number of pair features therefore grows quadratically with the vector count (Table~\ref{tab:current-state-storage}). The resulting vector and tensor flows rotate with the cloud \citep{villar2021scalars}.

\textit{Adding the pulse.}
Every predictor adds the incoming species counts exactly. The moment
models sum the incoming and resident count-weighted first and second
moments for each species. A pooled model encodes the incoming cloud about
the initial centroid. It merges the old and incoming vectors and
tensors according to their species-weighted populations and applies
a learned correction when merging scalar channels
(Appendix~\ref{app:state-definitions-current}).

\begin{table}[!ht]
\centering
\normalsize
\setlength{\tabcolsep}{3pt}
\caption{Memory per branch and shared model weights in float64. A paired forecast retains two states and one set of weights. B denotes bytes.}
\label{tab:current-state-storage}
\begin{tabular}{@{}lrrrr@{}}
\toprule
Predictor & State (B) & Rate inputs & Parameters & Weights (B)\\
\midrule
Counts-only & 72 & 9 & 3\,273 & 26\,184\\
C16 & 224 & 24 & 20\,501 & 164\,008\\
C32 & 352 & 63 & 29\,683 & 237\,464\\
C64 & 608 & 213 & 64\,517 & 516\,136\\
Physical moments & 600 & --- & 0 & 0\\
Learned moments & 600 & 32 & 4\,377 & 35\,016\\
\bottomrule
\end{tabular}
\end{table}

State memory includes counts, clock, water ledger, geometry and origin. The supplied pulse and temporary solver arrays are additional. We trained learned moments, C32 and C64 on the same families and five fixed seeds. The loss combines standardized errors in both count trajectories with one-half the squared error in their terminal response. Every fit had at most 120 epochs of six updates. Validation selected the checkpoint. We average the five individual-fit losses. Earlier C16 and C32 fits used a first-order update to compare active and passive training at the same path budget (Appendix~\ref{sec:fitting-controls}).

For the 15 fits in the primary study and their forecasts, we used an adaptive
third-order strong-stability-preserving Runge--Kutta solver in float64
\citep{gottlieb1998tvd}. Step doubling controls local error between
additions. Each addition enters as an exact jump. Refining the integration
changed the terminal response by at most $3.45\times10^{-6}$ in standardized
root-mean-square units (Appendix~\ref{sec:numerical-checks}).

\section{Evaluation and computing cost}
\label{sec:evaluation}

\subsection{Response error and practical baselines}
For each source family $h$ and action $a$, we predict the difference in
20\,ns counts between the pulse ($F$) and passive ($P$) branches and
compare it with the same difference in SBS mean counts. Normalized mean squared error (NMSE) averages the
squared discrepancy over the seven species,
\begin{equation}
 \ell_{ha}=\frac17\sum_s\left[
 \frac{\widehat n^F_{has}-\widehat n^P_{has}
       -\bar Y^F_{has}+\bar Y^P_{has}}{b_s}\right]^2.
\end{equation}
The species scale $b_s$ is the root-mean-square passive training count,
with a minimum of one molecule. With this fixed scale, larger absolute count errors at higher energies
contribute more to NMSE.

The first two baselines predict zero response or the incoming species
counts. A third accounts for reactions among the incoming molecules.
On the development bank we simulated the pulse alone with 64 SBS paths
per case, placing its molecules at their supplied positions. Its mean counts estimate the
response when the initial cloud is absent.

On the confirmation bank we also used Geant4's synchronized independent
reaction time (IRT) model \citep{ramosmendez2020irt,tran2021synchronousirt}.
IRT samples pair reaction times without resolving each Brownian step. It shares the molecular inputs and reaction network with SBS. Let $L$ be raw NMSE and $V_S$ the estimated variance of the SBS reference mean. Then $L-V_S$ estimates the error of the forecast actually produced, including any finite-path noise in IRT. Let $V_I$ be the sum of the two single-path IRT branch variances on the same species scales. Then $L-V_S-V_I/64$ estimates the squared discrepancy between the infinite-path IRT and SBS means (Appendix~\ref{sec:native-error}). We also report each method's raw NMSE divided by the zero-response NMSE at that energy. This measures error relative to the size of the observed pulse response.

The prespecified comparison gives the 100 and 400\,eV groups equal weight.
We average losses over five fits and four actions per family for learned
moments and C64. Because the four actions share a source, the paired
bootstrap resamples whole families within each energy, keeping all
actions and both methods together. The interval uses 4000 draws and
conditions on the five fixed fits. The C32 comparison, the 200\,eV control and the remaining
contrasts are secondary and use unadjusted intervals.
Appendix~\ref{sec:reproduction-materials} identifies the family-level
data and replay code.

\subsection{Estimating mean yield across sources}
For the bank mean, we estimate passive hydroxyl-radical (OH) yield at
20\,ns across all sources at one energy, in molecules per 100\,eV. The
model gives a prediction $f_i$ for each of the $N$ sources. On a uniform random subset $S$ of $m$ sources, we also estimate
each source's mean yield $\bar Y_{i,R}$ from $R$ independent SBS paths.
The mean difference on this subset corrects the mean model prediction
over the bank \citep{peherstorfer2016multifidelity},
\begin{equation}
 \widehat\mu=\frac1N\sum_{i=1}^{N} f_i+
              \frac1m\sum_{i\in S}(\bar Y_{i,R}-f_i).
 \label{eq:bank-estimator}
\end{equation}
We compare this estimator with SBS alone under two allocations. One
uses the same number of sampled sources. The other spreads the same
path budget across every source, removing variation due to source
selection (Appendix~\ref{sec:yield-variance}).

\subsection{Measured workloads}
We timed four actions in each of the first eight 400\,eV confirmation families, with both clouds supplied. Three repetitions used a fresh worker and one CPU thread. We timed one fit per learned method (seed 1701, selected in advance), unfitted moments, IRT at 1, 8, 32 and 64 paths per branch, and SBS at 64 paths. For the 400\,eV bank mean we also charged the cost of producing every required source and compared allocations of 64 or 512 SBS paths (Appendix~\ref{sec:yield-variance}).

\section{Results}
\subsection{Pulse response after a change in source energy}
\label{sec:transfer-results}
With equal weight on the two transfer energies, learned moments reduced mean 20\,ns response NMSE by 0.0528 relative to C64 (paired 95\% CI 0.0326--0.0790). The reduction relative to C32 was 0.0288 (secondary comparison, unadjusted 95\% CI 0.0159--0.0453). Both intervals condition on the five fixed fits.

\begin{figure}[!htbp]
\centering
\includegraphics[width=.94\textwidth]{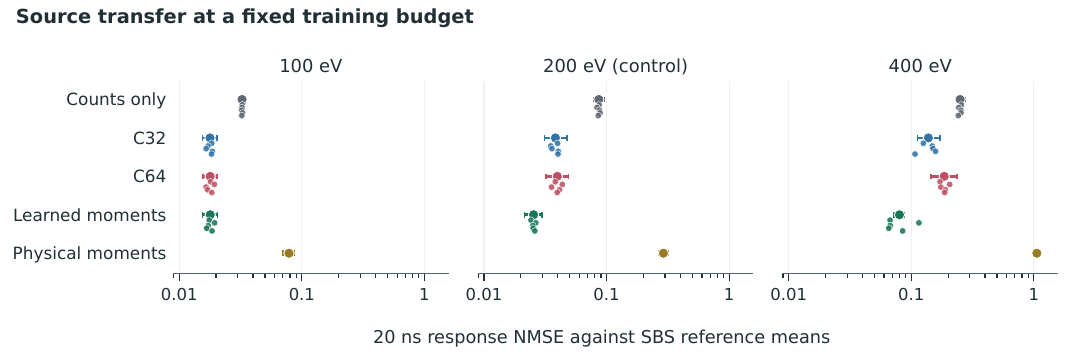}
\caption{Error in the 20\,ns pulse response on fresh families. Large points give the mean NMSE. Small points show the five individual fits at vertically offset positions. C32, C64 and learned moments each had five fits under the same training budget. Counts-only dynamics uses the original fits, and physical moments have no fitted parameters. Bars are 95\% intervals from resampling families with fits held fixed. All panels use the same logarithmic axis and species normalization.}
\label{fig:source-transfer}
\end{figure}

At 400\,eV, learned moments had lower error than C32 and C64 for every species. Mean response NMSE was 0.0801, 0.1379 and 0.1858, respectively (Fig.~\ref{fig:source-transfer}, Table~\ref{tab:fresh-transfer-base-held}). The worst moment fit (0.1155) beat the best C64 fit (0.1718). For OH, the corresponding root mean squared errors were 2.34, 3.32 and 4.40 molecules (Appendix Table~\ref{tab:reported-species-energy400}).

At 100\,eV, averaging over actions concealed two effects of opposite sign. At 1\,ns and zero offset, learned moments scored 0.0141 against C64's 0.0217 (paired difference $-0.00764$, 95\% CI $[-0.01093,-0.00436]$). At 10\,ns and 50\,nm, C64 had lower error, 0.0135 versus 0.0186 (moments minus C64 $+0.00510$, CI $[0.00114,0.00906]$). Across actions, both scored 0.0179. The squared response measured against zero increased nearly thirteenfold from 100 to 400\,eV on the fixed training scales. As a fraction of that reference, moments and C64 scored 2.60\% each at 100\,eV and 0.905\% versus 2.098\% at 400\,eV. The equal-energy difference on this relative scale was $-0.595$ percentage points (95\% CI $[-0.916,-0.324]$). These secondary intervals resample families without adjustment for multiple comparisons (Appendix~\ref{app:mechanism}).

\begin{table}[!ht]
\centering
\normalsize
\setlength{\tabcolsep}{3pt}
\caption{Raw 20\,ns response NMSE by energy per primary electron. Within each bank, the three columns cover 64, 32 and 64 families, respectively, with four actions per family. Each learned-model entry averages the losses of five fits. IRT uses 64 paths per branch. The pulse-alone simulation uses 64 paths per case on the development families.}
\label{tab:fresh-transfer-base-held}
\begin{tabular}{@{}lrrr@{}}
\toprule
Predictor & 100\,eV & 200\,eV & 400\,eV\\
\midrule
\multicolumn{4}{@{}l}{\textit{Fresh confirmation bank}}\\
Zero response & 0.6882 & 2.3957 & 8.8558\\
Added counts only & 0.4017 & 1.5075 & 5.6846\\
Counts-only dynamics & 0.0326 & 0.0865 & 0.2503\\
C32 & 0.0178 & 0.0383 & 0.1379\\
C64 & 0.0179 & 0.0397 & 0.1858\\
Learned moments & 0.0179 & 0.0254 & 0.0801\\
Physical moments & 0.0786 & 0.2909 & 1.0576\\
IRT & 0.0146 & 0.0319 & 0.0685\\
\midrule
\multicolumn{4}{@{}l}{\textit{Development bank}}\\
Pulse alone (SBS) & 0.0198 & 0.0663 & 0.1924\\
First-order-trained moments & 0.0158 & 0.0262 & 0.0818\\
\bottomrule
\end{tabular}
\end{table}

Fourteen of 15 fits reached the 720-update limit. C64's five selected checkpoints fell between epochs 90 and 115. Moment fit 1702 had the lowest validation error at 200\,eV and the highest transfer error at 400\,eV. On the 1\,ns, 50\,nm action, its excess appeared after the exact count jump. Its median H$+$H rate multiplier rose 3.16-fold across the jump, while the other four fits' pooled median fell to 0.19 of its pre-jump value. Across families, the change in cloud spread scarcely predicted excess terminal error (Appendix~\ref{app:mechanism}).

\paragraph{Exploratory tests of the encounter rule and network size.}
At 400\,eV, replacing pair-specific overlap with one centered cloud variance raised moment-model NMSE by 0.0543 (95\% CI $[0.0392,0.0696]$). On three-pulse schedules it raised NMSE by 0.0401 ($[0.0302,0.0511]$). Every original moment fit beat every shared-overlap fit on both tests, with the same species moments, product-placement rule and 4\,377 weights. At full width, adding the specified weight decay raised C64's 400\,eV NMSE from 0.1858 to 0.1966. Seed 1701 accounted for most of this increase. At that same decay, narrowing improved every 400\,eV fit and reduced mean error by 0.0342 (95\% CI $[0.0125,0.0620]$) to 0.1623. Narrowing left three-pulse error close to its full-width value (Table~\ref{tab:ablation-main}, Appendix Table~\ref{tab:v13-mechanism-contrasts}).
\begin{table}[!htbp]
\centering
\setlength{\tabcolsep}{4pt}
\caption{Mean response NMSE over five fits per learned predictor. The two new schedules use 64 held-out 200\,eV families. Each weight occupies eight bytes. Full-width and narrow C64 require 516\,136 and 157\,224 bytes of weights. The moment controls require 35\,016 bytes with clock inputs and 34\,248 without.}
\label{tab:ablation-main}
\begin{tabular}{@{}lrrrrrr@{}}
\toprule
& & \multicolumn{3}{c}{Familiar pulses} & \multicolumn{2}{c}{New schedules}\\
Predictor & Weights & 100 eV & 200 eV & 400 eV & New delay & Three pulses\\
\midrule
Physical moments & 0 & 0.0786 & 0.2909 & 1.0576 & 0.2030 & 0.2699\\
C32 & 29\,683 & 0.0178 & 0.0383 & 0.1379 & 0.0279 & 0.0687\\
C64 & 64\,517 & 0.0179 & 0.0397 & 0.1858 & 0.0288 & 0.0754\\
Learned moments & 4\,377 & 0.0179 & 0.0254 & 0.0801 & 0.0234 & 0.0255\\
\midrule
\multicolumn{7}{@{}l}{\textit{Post-hoc controls}}\\
C64, full width with decay & 64\,517 & 0.0179 & 0.0402 & 0.1966 & 0.0288 & 0.0762\\
C64, narrow with decay & 19\,653 & 0.0194 & 0.0396 & 0.1623 & 0.0293 & 0.0772\\
Moments, shared overlap & 4\,377 & 0.0225 & 0.0400 & 0.1344 & 0.0370 & 0.0656\\
Moments, passive training & 4\,377 & 0.0454 & 0.1393 & 0.4819 & 0.0743 & 0.1173\\
Moments, no clock inputs & 4\,281 & 0.0186 & 0.0295 & 0.0782 & 0.0238 & 0.0249\\
\bottomrule
\end{tabular}
\end{table}

\paragraph{The initial cloud and the training trajectories.}
On the development families, simulating the pulse alone gave a mean NMSE of 0.1061 across the two transfer energies. First-order-trained moment models that retained the initial cloud gave 0.0488 on the same families (paired difference $-0.0573$, descriptive 95\% CI $[-0.0649,-0.0489]$).

\begin{figure}[!htb]
\centering
\includegraphics[width=.94\textwidth]{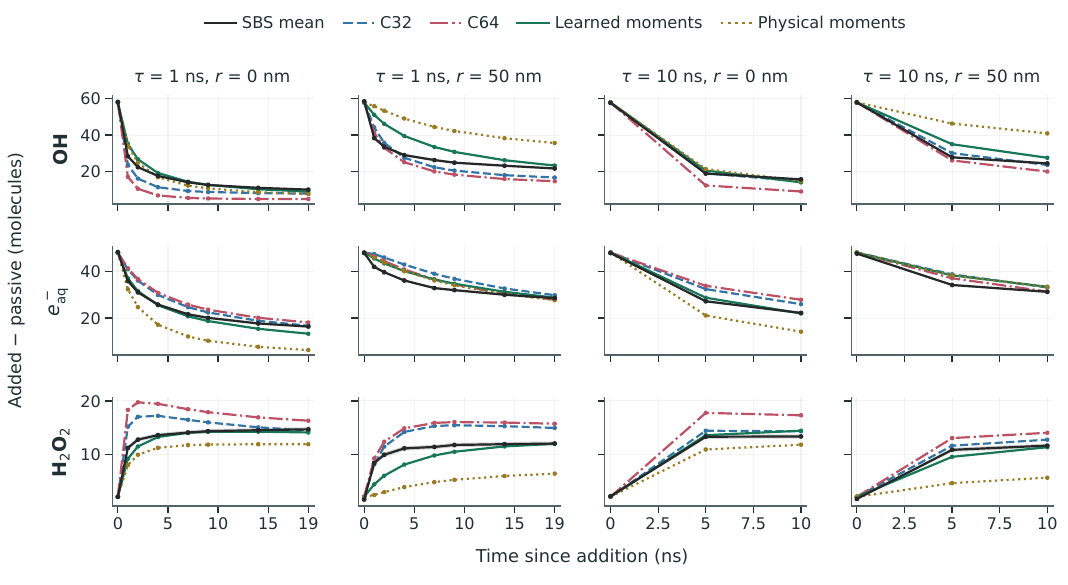}
\caption{Predicted and simulated responses to the four additions in the first 400\,eV confirmation family. Columns vary the arrival time $\tau$ and offset $r$. Rows give the change in counts of three species. Black curves show the SBS mean and shading shows $\pm1$ pointwise Monte Carlo standard error from 64 paths per branch. Colored curves show the preselected fit for each learned model and the unfitted physical moments. Each column starts immediately after the addition and ends at 20\,ns of absolute time. Points mark the available observations.}
\label{fig:representative}
\end{figure}

With eight training paths per case and the same intervention-based validation rule, learned moments trained only on passive trajectories had six times the error of the pulse-trained fits at 400\,eV (0.4819 versus 0.0801, paired difference 0.4018, 95\% CI $[0.3751,0.4280]$). Earlier first-order C32 fits showed the same dependence on intervention data, with passive-to-active error ratios of 5.6 at 100\,eV and 5.4 at 400\,eV (Appendix Table~\ref{tab:reported-context-base}). Figure~\ref{fig:representative} shows the pulse response on one 400\,eV family.

\subsection{Pulse response after a change in arrival time}
\label{sec:action-transfer}
The source-transfer test used familiar pulse schedules. We next changed the schedule at the training energy of 200\,eV. At each offset, the same three incoming clouds arrived together at 3 or 5\,ns or one at a time at 1, 5 and 10\,ns. The models had trained only on single additions at 1 or 10\,ns. This exploratory test used 64 held-out families with independent SBS reference trajectories from an earlier evaluation. We specified the analysis before scoring the current fits (Appendix~\ref{app:action-transfer}).

Among the four original predictors, learned moments had the lowest mean response NMSE under both schedules (Table~\ref{tab:ablation-main}). For a single addition at an unseen time, the paired NMSE difference between learned moments and C64 was $-0.0054$ (descriptive 95\% CI $[-0.0083,-0.0028]$). With three successive additions, learned moments had 66\% lower mean error than C64 (0.0255 versus 0.0754, paired difference $-0.0499$, descriptive 95\% CI $[-0.0570,-0.0429]$).

\subsection{Accuracy and cost of forecasting from a supplied cloud}
\begin{figure}[!htbp]
\centering
\includegraphics[width=.92\textwidth]{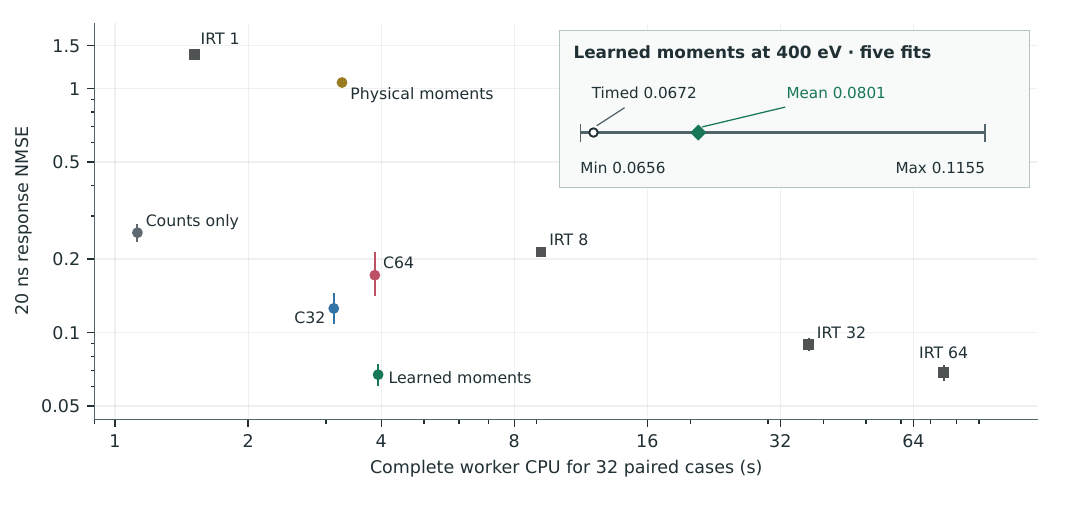}
\caption{Supplied-cloud forecasts at 400\,eV. Vertical bars give 95\% family-bootstrap intervals across 64 families. Horizontal bars give CPU-time quartiles for three repetitions on 32 cases. IRT labels give paths per branch. Errors below 64 paths are projected from the 64-path sample. The inset shows the timed moment fit, the five-fit mean and their range.}
\label{fig:accuracy-cost}
\end{figure}

At 400\,eV the preselected moment fit scored 0.0672 against the finite SBS reference and 64-path IRT scored 0.0685 across 64 families. The five-fit moment mean was 0.0801 (Fig.~\ref{fig:accuracy-cost}). On 32 supplied-cloud cases the fit took 3.94 CPU-seconds and IRT took 74.9. The model's median loaded-solver time was 3.02 seconds, or 94\,ms per paired case. These timings compare our single-threaded Python/PyTorch implementation with the compiled Geant4 engine and include worker startup in both totals.

Removing SBS reference noise put realized forecast MSE at 0.0460 for the timed fit, 0.0589 for the five-fit mean and 0.0473 for 64-path IRT. The timed fit minus IRT was $-0.0013$ (paired 95\% CI $[-0.0099,0.0073]$). For the five-fit mean the difference was $+0.0117$ ($[0.0024,0.0212]$). Removing IRT's own sampling noise gave an estimated infinite-path squared bias of 0.0265. At 200\,eV, learned moments beat the realized 64-path forecast by 0.00645 on average, although the estimated infinite-path IRT error was lower (Appendix~\ref{app:mechanism}).

\subsection{Cost of estimating a mean yield}
For mean passive OH yield, a model correction with coefficient one improved on SBS restricted to the same eight sampled sources. With 64 sources and at least 64 native paths, the budget also allowed a census. At 400\,eV, the native child took 1.06 CPU-seconds to generate all 64 clouds, compared with 1.94 seconds for the loaded moment solver to forecast those clouds. At both path budgets, the census gave lower estimated error and used less total CPU than the model-assisted estimate (Table~\ref{tab:yield-main}).

\begin{table}[!ht]
\centering\normalsize
\caption{Mean passive OH yield at 20\,ns across 64 sources at 400\,eV. The model-assisted estimator uses the preselected moment fit on every source and native paths on eight sources. The census distributes the same paths evenly across all 64 sources. CPU includes source acquisition. MSE is in $(\mathrm{molecules}/100\,\mathrm{eV})^2$.}
\label{tab:yield-main}
\begin{tabular}{@{}rlrr@{}}
\toprule
Native paths & Estimator & Estimated MSE & CPU (s)\\
\midrule
64 & Model-assisted & 0.00292 & 22.1\\
   & Native census  & 0.00206 & 19.3\\
512 & Model-assisted & 0.00112 & 151.2\\
    & Native census  & 0.000258 & 142.2\\
\bottomrule
\end{tabular}
\end{table}
Appendix Fig.~\ref{fig:utility-cost} and Appendix Table~\ref{tab:reported-utility-cost-join} report intervals and results for the other allocations.

\subsection{Forecasts beyond the training horizon}
On four previously examined 400\,eV families, preselected learned moments beat physical moments at 20\,ns (0.0625 versus 0.679 response NMSE), but their ranking reversed at 1\,$\mu$s (0.961 versus 0.388). The squared-log clock input grew from 0.93 to 4.77. Holding only the learned clock inputs at their 20\,ns values lowered the median pulse-branch OH$+$H rate multiplier from 41.29 to 6.78 and improved all four families (mean NMSE 0.511). Refitting without clock inputs also improved all four families. The matched seed scored 0.473 at 1\,$\mu$s, and the five-fit mean was 0.486. On the 64 fresh 400\,eV families at 20\,ns, the clock-free model scored 0.0782 versus 0.0801 for learned moments (Appendix~\ref{app:mechanism}).

\section{Discussion}
The transfer advantage depended in part on how the model computed an encounter. Giving every reactant pair a common cloud scale raised error at 400\,eV and after three pulses, even though the moment state, product geometry and number of weights stayed fixed. In the pooled models, expanding C32 to C64 raised error at 400\,eV despite similar validation scores at epoch 80. At fixed weight decay, narrowing C64 recovered some accuracy at that energy and did little for three successive pulses. At full width, the tested decay gave no consistent improvement across seeds. Under this fitting budget, the pair-specific encounter rule helped on both shifts, whereas the benefit of a narrower pooled network depended on the shift.

The intervention and the forecast horizon each tested something that the training-energy validation score missed. Even with species-specific geometry and an intervention-based validation rule, passive training left the moment model poorly prepared for a pulse. The gap persisted when three clouds arrived in succession. At the longer horizon, both freezing the clock inputs and fitting without them cut the error on the four 400\,eV families. Unfitted physics still had the lowest observed mean. These comparisons support evaluating a world model on the actions, source changes and times that its users will query.

The task also determines which speed comparison matters. On supplied clouds, the prespecified moment fit and 64-path IRT gave similar realized errors at very different CPU costs. IRT retained the lower estimated squared bias at infinite replication. For a passive mean across 64 clouds, acquiring the sources was cheap enough to simulate every one. This study covers low-energy tracks in pure water. A single Gaussian per species preserves the combined first two moments of a separated pulse but cannot retain its distinct spatial modes. Neither a particle graph simulator nor a continuum solver was evaluated as an accuracy upper bound, and the measured CPU costs apply to the implementations we timed.

The supplement contains the family scores, solver checks and figure data. The code, weights and reproduction data are archived in \href{https://github.com/hnusrat1/water-radiolysis-world-models/releases/tag/v1.0.0}{GitHub release v1.0.0} (Appendix~\ref{sec:reproduction-materials}). The change of source energy used fresh confirmation families. The schedule, mechanism and long-horizon experiments were exploratory. We used OpenAI Codex in the research and preparation of the manuscript, including simulation and fitting code, statistical checks, literature searches, reference checks and review of spelling and grammar. The authors are responsible for the results and text.

\newgeometry{margin=48pt}
\appendix
\section{Supporting methods and evidence provenance}
\label{sec:supporting-methods}

\subsection{Study cohorts and inference}
The model receives the initial molecular cloud and each addition. It predicts both branches without observing the subsequent chemistry. The four actions for a given source family reuse the same initial and incoming clouds, so we resampled whole families. For the primary contrast we averaged the five individual-fit losses for each method, averaged the four actions within each family, then weighted 100 and 400\,eV equally. We drew 4000 stratified bootstrap samples of families to form a paired percentile interval. This interval conditions on the fitted models. The ranges across fits describe variation with training seed. The 200\,eV control and the other contrasts are secondary. Table~\ref{tab:support-cohorts} gives the family and path counts.

\begin{table}[!hb]
\centering
\small
\begin{tabular}{lrrr}
Study & Families & Cases & Native paths \\
\hline
Training / validation & 128 / 32 & 768 / 192 & 4 / 16 per arm \\
Development SBS bank & 160 & 640 & 81\,920 \\
Confirmation SBS bank & 160 & 640 & 81\,920 \\
Confirmation IRT & Same 160 & Same 640 & 81\,920 \\
Pulse alone (SBS) & Development 160 & 640 & 40\,960 \\
Long-horizon SBS + IRT & Development 12 & 36 branches & 2304 \\
\end{tabular}
\caption{Source families and native simulations used for fitting, development, confirmation and long-horizon comparisons. The development and confirmation banks each contain 64, 32, and 64 families at 100, 200, and 400\,eV, respectively. Each bank uses 640 physical source events and four actions per family. SBS and IRT response references use 64 paths per arm. The pulse-alone control uses 64 paths per case. Each long-horizon engine uses 32 paths for each of three branches per family. Training and validation use six actions per family. The counts exclude repeated timing runs.}

\label{tab:support-cohorts}
\end{table}

\subsection{Fitting, state size, and numerical qualification}
\label{sec:fitting-controls}
We fit C32, C64, and learned moments with the same training data, loss, optimization budget, float64 solver, and seeds 1701--1705. The training loss averaged the forced and passive trajectory MSEs over all queries with equal branch weights, then added 0.5 times the terminal-response MSE. We scaled each species by its RMS count across passive training cases and times, with a one-molecule floor. We rounded these scales to float32 before conversion. The models, coordinates and times also passed through float32 before float64 integration.

Adam used a learning rate of 0.002, a logical batch size and microbatch size of 128, and a gradient-norm cap of 10. Each epoch comprised six updates. We validated at epoch zero and every five epochs thereafter, up to 120 epochs. The validation score was terminal-response MSE plus 0.1 times the equally weighted branch-trajectory MSE over postinitial queries. We kept a checkpoint only when its score improved strictly and stopped after eight checks without improvement.

For each fit we recorded the selected epoch, CPU and wall time in the fitting loop, and peak resident set size (RSS) of the reporting worker. The loop began after data loading, model initialization and optimizer setup. It included optimization, validation and checkpoint writing. Wall time reflects the execution schedule. After an interruption, the saved counters excluded the interval between worker attempts and discarded work from the partial epoch. We measured inference separately.

\begin{table*}[t]
\centering\small
\setlength{\tabcolsep}{4pt}
\begin{tabular}{lrrrrrr}
\toprule
Model & Seed & Selected epoch & Final epoch & CPU (h) & Wall (h) & Peak RSS (GiB)\\
\midrule
C32 & 1701 & 40 & 80 & 2.36 & 2.42 & 7.85 \\
C32 & 1702 & 110 & 120 & 5.47 & 7.19 & 11.43 \\
C32 & 1703 & 120 & 120 & 4.79 & 5.14 & 11.18 \\
C32 & 1704 & 95 & 120 & 4.24 & 4.40 & 11.45 \\
C32 & 1705 & 100 & 120 & 3.97 & 4.05 & 11.47 \\
\midrule
C64 & 1701 & 90 & 120 & 6.28 & 9.10 & 9.21 \\
C64 & 1702 & 100 & 120 & 7.08 & 12.29 & 12.12 \\
C64 & 1703 & 115 & 120 & 8.65 & 15.88 & 11.93 \\
C64 & 1704 & 105 & 120 & 4.16 & 5.03 & 11.79 \\
C64 & 1705 & 100 & 120 & 9.89 & 12.44 & 8.06 \\
\midrule
Learned moments & 1701 & 115 & 120 & 6.09 & 6.42 & 3.21 \\
Learned moments & 1702 & 115 & 120 & 5.88 & 6.34 & 3.37 \\
Learned moments & 1703 & 120 & 120 & 4.82 & 4.82 & 3.78 \\
Learned moments & 1704 & 115 & 120 & 4.87 & 4.87 & 3.78 \\
Learned moments & 1705 & 115 & 120 & 4.85 & 4.85 & 3.78 \\
\bottomrule
\end{tabular}
\caption{Recorded fitting resources for all controlled fits. 
CPU and wall counters cover optimization, validation, and checkpoint 
writing after setup. Resumed counters exclude intervals between 
worker attempts and discarded work from an interrupted epoch. 
Peak RSS is for the process writing the final 
fit record. Wall times reflect the actual resource schedule; these 
records are separate from the inference benchmarks.}
\label{tab:training-resources}
\end{table*}

C64 doubles C32's vector, tensor and scalar channels while retaining the same pooling and update equations. C64 uses 608 bytes per branch and has 64\,517 trainable parameters. Learned moments use 600 bytes per branch and have 4\,377 trainable parameters.

During differentiation, non-reentrant gradient checkpointing recomputes the rollout. For C64 seed 1705, the continuation after epoch 84 also checkpointed the same vector-field evaluations within each Runge--Kutta update. On fixed private test inputs, the implementations agreed bit for bit on forward states, gradients, optimizer updates and random states. On a separate 128-case training batch, they again produced identical forward states. The nested implementation produced finite gradients on that batch, where we did not perform an optimizer update. We differentiated the accepted Runge--Kutta stages with adaptive step decisions fixed and used a straight-through derivative for the Gaussian positive-semidefinite projection.

The earlier first-order study fit counts-only dynamics, learned moments, and active and passive versions of C16 and C32, each with five seeds (1701--1705). Physical moments had no fitted parameters. To compare training with and without additions, we held the number of native training paths at eight per case. Active training minimized the mean squared standardized count error across both branches at all queries, plus 0.5 times the squared standardized terminal-response error. Passive training predicted one undisturbed trajectory and averaged its losses against two independent passive target means.

Both groups used the same 200\,eV intervention cases for checkpoint selection. We scored terminal-response MSE plus 0.1 times the mean branch-trajectory MSE over postinitial queries. The groups therefore differed in whether post-addition trajectories and response errors informed gradient updates. Both used Adam with a learning rate of 0.002, batches of 128 and at most 120 epochs. We validated at epoch zero and every five epochs, stopping after eight checks without improvement.

During fitting, the original first-order integrator used float32, a maximum step of 0.125\,ns and a fractional reactant-consumption bound of 0.05. On the confirmation bank we evaluated those weights with the primary float64 solver. To assess sensitivity to the update, we also evaluated the earlier and new weights with the original first-order update in float64. The first-order update evolves conditional geometry even for absent species. The extensive Gaussian state assigns zero geometry to a species at zero mass. Learned rates can read those geometric entries, so the comparison changes both the integration rule and a boundary convention. It cannot establish a common continuous closure or the direction of numerical bias.

The base controls use maximum step 0.5\,ns, fractional bound 0.5,
relative tolerance $10^{-6}$, absolute tolerance $10^{-8}$, minimum step
$10^{-12}$\,ns, and 100\,000 attempts per evolution call.
Step doubling accepts the two-half-step state when positivity checks
pass and
\[
 \max_j\frac{|y^{\rm fine}_j-y^{\rm coarse}_j|}
 {7[\mathrm{atol}+\mathrm{rtol}\max(|y_j|,|y^{\rm fine}_j|)]}\le1.
\]
Here $y$ contains counts, spatial memory or extensive Gaussian moments,
and the water ledger. Refinement halves the maximum step and fractional
bound and quarters both tolerances. The strict comparison uses
0.125\,ns, fractional bound 0.1, relative tolerance $10^{-8}$, and
absolute tolerance $10^{-10}$. Both retain the base floor and attempt
limit. Original64 sensitivity uses the original first-order update in
float64 with maximum step 0.125\,ns and fractional bound 0.05.
The strict comparison includes all 15 controlled fits, all five context
fits of active C32, passive C32, and learned moments, the first context
fit of counts, active C16, and passive C16, and physical moments.

Before using confirmation outcomes, we tested all 31 earlier predictors on inputs from 48 development cases (eight families), examined manufactured source and boundary cases, and checked a $10^{-14}$\,ns minimum step for all six moment predictors. An independent audit verified 150 development-bank comparisons. On the confirmation bank we tested all four actions for the first eight families at each energy. Terminal responses and postinitial trajectories had to agree within 0.001 standardized RMS and 0.05 molecule at every coordinate. The absolute conserved-budget drift could reach at most $10^{-8}$ and all counts had to remain at least $-10^{-8}$. The independent audit verified all 240 comparisons and checked the sources, weights, trajectories and water ledger before the confirmation outcomes were released.

\subsection{Native baselines and Monte Carlo correction}
\label{sec:native-error}
The pulse-alone simulation asks how much of the response comes from reactions among the incoming molecules. We simulated all three incoming clouds together with SBS at their specified positions, leaving out the initial cloud. Using the pulse-alone mean as the predicted response accounts for reactions within the pulse and omits its interactions with the initial source. We compared this 40\,960-path control with the 31 earlier predictors on the same 640 development cases.

IRT uses the same confirmation sources and actions as SBS and the compact
predictors. Write $L_{64}$ for its squared response discrepancy against
the 64-path SBS mean, $V_I$ for the estimated single-path IRT response
variance, and $V_S$ for the SBS reference-mean variance. Species scaling
and averaging are applied consistently to these quantities. The estimated
IRT mean-discrepancy MSE is
\[
 \widehat B=L_{64}-V_S-V_I/64.
\]
The expected finite-reference MSE at $R$ IRT paths per arm is
$\widehat B+V_I/R+V_S$, for $R=1,8,32,64$. Thus the $R=64$ profile
recovers the raw IRT comparison. The profiles use bias and variance
estimates from the same 64-path sample. Deterministic
predictor errors subtract only $V_S$. The bias-only model-minus-IRT
mean-discrepancy contrast equals its raw 64-path contrast plus $V_I/64$.
For the expected-MSE profile at $R$, the corresponding adjustment to the
raw 64-path contrast is $V_I/64-V_I/R$. Signed estimates remain signed.
The corrections treat distinct native streams as independent
histories; ordered-batch reproducibility is a separate engineering check.

\subsection{Native consistency and longer horizons}
\label{sec:native-checks}
We ran Geant4 11.4.3 with \texttt{G4EmDNAPhysics\_option2} and
\texttt{G4EmDNAChemistry\_option1} in a 1\,mm liquid-water cube. Primary
electrons stopped in water and deposited their full kinetic energy to
within $10^{-6}$\,eV. The initial electron origins were balanced between
0 and 10\,nm separation in each evaluation bank. After reconstructing the
observed species and positions, native continuations inserted the
incoming clouds at the specified absolute clock time. We recorded counts
at $\{0.02,0.1,0.5,1,2,3,5,8,10,15,20\}$\,ns. At a pulse time the
recorded counts include the added molecules.

To check the native yields, we built an independent application from the Geant4-DNA example \texttt{chem4} using the same release, physics models, chemistry and query grid. We scored deposited energy along physical track steps and counted species at the first watched boundary callback. We compared $G_s(t)=100N_s(t)/E_{\mathrm{dep}}$ in molecules per 100\,eV across eight fully stopped source conditions that varied electron energy (100, 200 and 400\,eV), primary number and initial separation. A whole-event bootstrap with 4000 resamples gave simultaneous 95\% bands for differences in post-source OH, $e_{\mathrm{aq}}^{-}$ and H$_2$O$_2$ yields. To check reconstruction itself, we compared one uninterrupted future with eight reconstructed futures from each of 4096 sources at each of 100 and 400\,eV. Here the simultaneous bands accounted for futures sharing a source.

\begin{table}[t]
\centering
\small
\setlength{\tabcolsep}{4pt}
\begin{tabular}{lrrr}
\toprule
Comparison & Within & Max. point & Max. bound\\
\midrule
Application & 240/240 & 0.0184 & 0.0519\\
Reconstruction & 60/60 & 0.0198 & 0.0626\\
\bottomrule
\end{tabular}
\caption{Agreement of chemical yields between independent Geant4 applications and between reconstructed and uninterrupted chemistry. Each comparison uses its own
family of simultaneous 95\% bootstrap bands for post-source OH,
$e_{\mathrm{aq}}^{-}$, and H$_2$O$_2$ yield differences. The second
column counts bands wholly inside the fixed margin. The last two columns give the
largest absolute point difference and confidence limit, respectively,
in molecules/100\,eV.}
\label{tab:native-consistency}
\end{table}

The application check used 16\,384 events per application in each of eight source conditions, or 262\,144 events. The reconstruction check covered 8192 source events with their uninterrupted futures and 65\,536 reconstructed futures. The two checks thus covered 335\,872 native histories. We compared each set of simultaneous bands with the same margin of $\pm0.10$ molecule per 100\,eV. All 240 application contrasts and all 60 reconstruction contrasts met the criterion (Table~\ref{tab:native-consistency}). Earlier campaigns placed 134/240 and 38/60 bands inside that margin, respectively. The archive records those inconclusive results separately. Both checks concern consistency within Geant4.

In an earlier campaign, the species counter reported one extra OH at a boundary where the active particles conserved charge and reducing equivalents. Both applications now count recognized active particles at the same watched callbacks and retain the species-counter histories as diagnostic checks. They also verify charge, reducing equivalents, species identity, clocks and deposited energy. Repeated runs of the entire ordered batch, including the discrepant event, passed with the revised scorer. We retained 81\,920 production events after independently reconciling their direct-count CSV and NPZ data, identities, clocks, invariants and energies. We used the revised scorer for the remaining 253\,952 native executions and regenerated all reference events. The archive includes the earlier campaign's partial outputs.

To examine a longer horizon, we followed the first four development families at each energy through 1000\,ns. Each family contributed three branches. One evolved without an addition, one received a pulse at 1\,ns and 0\,nm offset, and one received a pulse at 10\,ns and 50\,nm offset. We added 50, 100, 200, 500 and 1000\,ns to the original query grid. Among the ten predictors, each learned predictor used its first fixed fit (seed 1701) and physical moments required no fit. All 60 refinement comparisons met the independent audit's absolute numerical criteria from the main evaluation. We report the terminal error, the response at each query, and mean branch loss across the 15 postinitial queries. This last average weights the forced and passive branches equally and gives equal weight to each query despite their unequal spacing. With four families per energy, the bootstrap intervals are descriptive. We report pointwise native Monte Carlo standard errors separately.

\subsection{Mean-yield utility and measured cost}
\label{sec:yield-variance}
We also estimate passive OH yield, $100N_{\mathrm{OH}}/(2E)$, at 20\,ns over each observed bank,
with $N=64,32,64$ sources at 100, 200, and 400\,eV. For source $i$,
let $\mu_i$ and $\sigma_i^2$ be its conditional mean yield and single-path
yield variance. Let $f_i$ be a fixed yield prediction and $\bar Y_{i,R}$
an independent SBS mean yield on a uniform subset $S$ of $m$ sources.
The estimator in Eq.~\eqref{eq:bank-estimator} is unbiased for the finite-bank SBS mean over subset selection and
native futures, with correction coefficient fixed to one
\citep{peherstorfer2016multifidelity}. Let $S_{\mu-f}^{2}$ be the finite-bank
sample variance of the residual source means and
$\overline{\sigma^2}=N^{-1}\sum_i\sigma_i^2$. The estimator's variance is
\begin{equation}
 \operatorname{Var}(\widehat\mu)=
 \left(\frac1m-\frac1N\right)S_{\mu-f}^{2}
 +\frac{\overline{\sigma^2}}{mR}.
\end{equation}
A model can reduce the first term by making the residual source means more nearly uniform. The within-source term depends on the continuation budget $mR$. To estimate these terms, we used four passive streams (256 paths per source) and subtracted reference Monte Carlo variance from the estimated between-source variance. We retained all fitted seeds, all valid $m\in\{4,8,16,32,64\}$ and $R\in\{1,8,64\}$. Paired empirical-family bootstrap intervals compared each model with native-only sampling and with physical moments. The intervals describe sensitivity to reweighting the observed bank.

At $m=N$, the model predictions cancel in the estimator and the between-source term vanishes. With the same continuation budget, a native census can achieve this cancellation whenever $R'=mR/N\geq1$ is an integer. The census then has the same within-source term as the subset estimator. We measured acquisition and execution costs for each allocation.

After the research workers finished, we timed each workload three times on one CPU thread. For conditional forecasts we timed all four actions in the first eight confirmation families at 400\,eV (32 paired cases) and measured accuracy on all 64 confirmation families at that energy. The learned predictors used fit 1701, and physical moments required no fit. For native simulation, $R$ paths per arm gave $2R$ paths per paired case. Loaded-model CPU covers encoding and rollout. Fresh-worker CPU includes the CPU time of child processes it waited for. We report native-child CPU separately, along with fresh-worker wall time. Worker peak RSS may include the largest descendant-process peak. We report absolute worker, native-child and import-baseline RSS separately because their peaks need not coincide.

For mean yield we timed 64 model inputs, SBS corrections on $m=8$ sources at $R=8,64$, and native censuses with $m=64,R=1,8$. The $R=1$ census and $m=8,R=8$ correction each use 64 continuation paths. The $R=8$ census and $m=8,R=64$ correction each use 512. To measure acquisition as well as continuation, we added the cost of forming and exporting the required $m$ or $N$ initial clouds. Only SBS supplies the correction for the SBS mean-yield target. We report IRT timings separately. We retained all three timing repetitions and checked the raw outputs after timing. Each cost corresponds to the accuracy estimate for that workload.

\section{Chemical network and units}
\label{app:chemical-network-current}

The native simulator and the closures use the same pure-water chemistry. The Geant4 option1 constructor supplies nine reactions (Table~\ref{tab:current-reactions}) and the species diffusion coefficients (Table~\ref{tab:current-diffusion})~\citep{geant4v1143}. We use the following species order throughout the equations.

\[
 (e_{\mathrm{aq}}^{-},{}^{\bullet}\mathrm{OH},\mathrm H,
   \mathrm{H_2},\mathrm{H_3O}^{+},\mathrm{OH}^{-},\mathrm{H_2O_2}).
\]
We express counts in molecules, positions in nm and the absolute clock in ns. The reaction and diffusion coefficients remain fixed.

\begin{table}[t]
\centering
\small
\setlength{\tabcolsep}{3pt}
\caption{Option1 reaction coefficients in
$10^9\,\mathrm{M}^{-1}\mathrm{s}^{-1}$, where
$\mathrm M=\mathrm{mol}\,\mathrm{L}^{-1}$~\citep{geant4v1143}.
Water is an implicit reservoir. The identical-reactant coefficients use
the native macroscopic event-rate convention.}
\label{tab:current-reactions}
\begin{tabular}{lr}
\toprule
Reaction & $k_r/10^9$\\
\midrule
$2e_{\rm aq}^{-}+2\mathrm{H_2O}\to\mathrm{H_2}+2\mathrm{OH}^{-}$ & 6.36\\
$e_{\rm aq}^{-}+{}^{\bullet}\mathrm{OH}\to\mathrm{OH}^{-}$ & 29.5\\
$e_{\rm aq}^{-}+\mathrm H+\mathrm{H_2O}\to\mathrm{H_2}+\mathrm{OH}^{-}$ & 25.0\\
$e_{\rm aq}^{-}+\mathrm{H_3O}^{+}\to\mathrm H+\mathrm{H_2O}$ & 21.1\\
$e_{\rm aq}^{-}+\mathrm{H_2O_2}\to{}^{\bullet}\mathrm{OH}+\mathrm{OH}^{-}$ & 11.0\\
$2{}^{\bullet}\mathrm{OH}\to\mathrm{H_2O_2}$ & 5.50\\
$^{\bullet}\mathrm{OH}+\mathrm H\to\mathrm{H_2O}$ & 15.5\\
$2\mathrm H\to\mathrm{H_2}$ & 5.03\\
$\mathrm{H_3O}^{+}+\mathrm{OH}^{-}\to2\mathrm{H_2O}$ & 113\\
\bottomrule
\end{tabular}
\end{table}

\begin{table}[t]
\centering
\small
\caption{Molecular diffusion coefficients
$D_s$ in $\mathrm{nm}^2\mathrm{ns}^{-1}$~\citep{geant4v1143}.}
\label{tab:current-diffusion}
\begin{tabular}{lr}
\toprule
Species & $D_s$\\
\midrule
$e_{\mathrm{aq}}^{-}$ & 4.9\\
$^{\bullet}\mathrm{OH}$ & 2.2\\
$\mathrm H$ & 7.0\\
$\mathrm{H_2}$ & 4.8\\
$\mathrm{H_3O}^{+}$ & 9.46\\
$\mathrm{OH}^{-}$ & 5.3\\
$\mathrm{H_2O_2}$ & 2.3\\
\bottomrule
\end{tabular}
\end{table}

Let $\alpha_{sr}$ and $\beta_{sr}$ be the explicit reactant and product
stoichiometries, with $S=\beta-\alpha$. Each reaction has two reactant
entries, denoted $A_r,B_r$, including repeated species. Convert the
macroscopic rate to molecule-number units by
\begin{equation}
 \kappa_r=\frac{10^{15}k_r}{N_{\mathrm A}},\qquad
 N_{\mathrm A}=6.02214076\times10^{23}\,\mathrm{mol}^{-1}.
 \label{eq:current-rate-conversion}
\end{equation}
The numerical factor converts litres and seconds to
$\mathrm{nm}^3$ and ns. With counts treated as numerical molecule
numbers, $\kappa_r$ is a volume rate in $\mathrm{nm}^3\mathrm{ns}^{-1}$.
Every closure evaluates nonnegative event hazards of the form
\begin{equation}
 a_r=\kappa_r n_{A_r}^{+}n_{B_r}^{+}g_r c_r,
 \qquad n_s^{+}=\max(n_s,0),
 \label{eq:current-hazard}
\end{equation}
where $g_r$ is an inverse encounter volume and $c_r$ is a dimensionless
positive correction. For identical reactants, the closure uses $n_A^2$
without an additional factor $1/2$. This replaces the unknown factorial
moment by a Poisson closure of the mean count. The native integer
candidate-pair calculation instead combines unordered pairs
$N_A(N_A-1)/2$ with its doubled identical-pair coefficient.

The closure's capture smoothing uses the effective radii defined by the
native reaction table~\citep{geant4v1143}:
\begin{equation}
 R_r=\begin{cases}
 \kappa_r/[4\pi(D_{A_r}+D_{B_r})],&A_r\ne B_r,\\
 \kappa_r/(4\pi D_{A_r}),&A_r=B_r.
 \end{cases}
 \label{eq:current-capture-radius}
\end{equation}
The radii have units of nm and follow the same event-rate convention. The constructors initialized the constants in float32 before converting them to float64. The stored constants retain that initial rounding.

Water participates in several reactions but is implicit in the seven species counts. A signed ledger $w$ starts at zero for the observed cloud and records the net production of reservoir water. In reaction-table order, each event changes the ledger by a component of

\[
 \nu_w=(-2,0,-1,1,0,0,1,0,2)^{\mathrm T}.
\]
The count and ledger equations are
\begin{equation}
 \dot n=S a,\qquad \dot w=\nu_w^{\mathrm T}a.
 \label{eq:current-network-update}
\end{equation}
For rows representing H atoms, O atoms, and charge, define
\[
 F=\begin{pmatrix}
 0&1&1&2&3&1&2\\
 0&1&0&0&1&1&2\\
 -1&0&0&0&1&-1&0
 \end{pmatrix},\qquad f_w=(2,1,0)^{\mathrm T}.
\]
The identity $FS+f_w\nu_w^{\mathrm T}=0$ preserves
$Fn+f_ww$ between additions. An incoming cloud with counts $n^I$ adds
$Fn^I$ to this budget and leaves $w$ unchanged.

\section{Predictive states and closure equations}
\label{app:state-definitions-current}

\subsection{Coordinates and shared rate inputs}
Every branch retains seven counts $n\in\mathbb R^7$, a signed water ledger $w$ and an absolute clock $t$. A predictor that uses geometry also retains spatial memory $z$ and the initial centroid $o\in\mathbb R^3$. We compute this centroid over all observed particles with equal weights. For a particle at $x_i$ in nm, the dimensionless position is $r_i=(x_i-o)/L$, with $L=10$\,nm. Every incoming cloud uses the same origin. The counts-only state omits the origin and spatial memory. Section~\ref{sec:supporting-methods} gives the coordinate and time precision.

Let $n_s^+=\max(n_s,0)$ and $\tau=t/(1\,\mathrm{ns})$. The shared
nine-component count/clock input is
\begin{equation}
 u(n,t)=\big(\log(1+n_s^+)_{s=1}^7,
                 \log(1+\tau),\log^2(1+\tau)/10\big).
 \label{eq:current-count-features}
\end{equation}
The logarithms use numerical molecule counts and the stated time unit.
In the original predictors, each multilayer perceptron has two hidden layers of width 48 with SiLU activation. Appendix~\ref{app:mechanism} gives the subsequent controls. Different heads and fitted models have separate parameters.
Nonlinearities act componentwise unless a row softmax is specified.
Define the bounded positive correction
\[
 C_b(h)=\exp\{b\tanh(h/b)\}.
\]
The counts-only hazard in Eq.~\eqref{eq:current-hazard} uses
$g_r=(1000\,\mathrm{nm}^3)^{-1}C_8(H_r(u))$ and $c_r=1$.
The head therefore learns how the inverse encounter volume changes with the counts and the clock, relative to a reference volume of $1000\,\mathrm{nm}^3$.

\subsection{Gaussian moments and extensive evolution}
Moment closures describe spatial chemical kinetics with a small set of statistics \citep{ernst2019moment}. Our Gaussian state retains the center and spread of each species. The dimensionless mean $\mu_s$ and symmetric raw second moment $T_s$ initially equal the within-species means of $r_i$ and $r_ir_i^{\mathrm T}$, respectively. Both are zero for an absent species. Storing the symmetric entries in the order $(xx,yy,zz,xy,xz,yz)$ gives $7(3+6)=63$ geometric values.

The physical covariance used for encounters is
\begin{equation}
 C_s=\mathcal P_+\!\left[L^2(T_s-\mu_s\mu_s^{\mathrm T})\right],
 \label{eq:current-species-covariance}
\end{equation}
where $\mathcal P_+$ symmetrizes the matrix and replaces negative
eigenvalues by zero. This projection applies only when computing the covariance for encounters and preserves the counts and stored raw moments.

For reaction $r$ with reactants $A,B$, set
\[
 \widetilde C_{rA}=C_A+\tfrac12\rho_r^2I,\quad
 \widetilde C_{rB}=C_B+\tfrac12\rho_r^2I,
\]
where $\rho_r=R_r$ uses Eq.~\eqref{eq:current-capture-radius}; the frozen
capture-scale multiplier is one. Write
$M_r=\widetilde C_{rA}+\widetilde C_{rB}$ and
$d_r=L(\mu_A-\mu_B)$. The overlap is
\begin{equation}
 g_r=\frac{\exp(-d_r^{\mathrm T}M_r^{-1}d_r/2)}
 {(2\pi)^{3/2}\sqrt{\det M_r}}.
 \label{eq:current-gaussian-overlap}
\end{equation}
The matrices and squared distances in this expression use nm units, so
$g_r$ is in $\mathrm{nm}^{-3}$. Physical moments set $c_r=1$ in
Eq.~\eqref{eq:current-hazard}. Learned moments use $c_r=C_6(H_r(v_G))$,
where the 32 inputs $v_G$ concatenate $u$, the nine values
$\log\max\{g_r(1\,\mathrm{nm})^3,10^{-30}\}/20$, seven values
$\log[1+\operatorname{tr}(C_s)/(1\,\mathrm{nm}^2)]$, and seven values
$\log(1+\|\mu_s\|^2)$. Every species' geometry enters this head.

Reaction products use the normalized product of the two smoothed Gaussian
densities. Its dimensionless mean $p_r$ and raw second moment $U_r$ are
\begin{align}
 K_r&=(\widetilde C_{rA}^{-1}+\widetilde C_{rB}^{-1})^{-1},\\
 p_r&=K_r(\widetilde C_{rA}^{-1}\mu_A+
                         \widetilde C_{rB}^{-1}\mu_B),\\
 U_r&=K_r/L^2+p_rp_r^{\mathrm T}.
 \label{eq:current-product-moments}
\end{align}
The integrator evolves extensive moments $f_s=n_s\mu_s$ and $Q_s=n_sT_s$.
Writing $c_s=\sum_r\alpha_{sr}a_r$ for reactant consumption, the equations
are
\begin{align}
 \dot f_s&=-\mu_sc_s+\sum_r\beta_{sr}a_rp_r,\\
 \dot Q_s&=-T_sc_s+\sum_r\beta_{sr}a_rU_r
                   +\frac{2n_sD_s}{L^2}I.
 \label{eq:current-extensive-moments}
\end{align}
Each reaction removes reactants uniformly within their species and places products at the encounter moments. Diffusion contributes to the second moment and leaves the first moment unchanged. Counts and water follow Eq.~\eqref{eq:current-network-update}.

At each integrator stage we recover the conditional geometry by dividing $f_s$ and $Q_s$ by $n_s$ for species with $n_s>0$. At $n_s=0$, both extensive moments are zero. The decoder uses denominator one, giving $\mu_s=T_s=0$. This convention matters because the learned rate head reads every species' geometry. When an absent species appears, its geometry can change corrections for reactions involving other species. An incoming cloud adds its counts and extensive first and second moments exactly. The centroid origin and water ledger stay fixed.

\subsection{Compact pooled states}
C16, C32, and C64 use $(n_v,n_T,n_z)=(3,1,1),(6,2,2),(12,4,4)$,
respectively. Their memory contains $n_v$ vectors $v_j\in\mathbb R^3$,
$n_T$ symmetric matrices $T_j$, and $n_z$ scalars $z_j$. For positive
species weights $a_{js}$ and $b_{js}$, each normalized over species by a
softmax, the initial pooled moments are
\begin{align}
 m_j^v&=\sum_s a_{js}n_s,&
 v_j&=\frac{\sum_i a_{j,s_i}r_i}{\max(m_j^v,\epsilon)},\\
 m_j^T&=\sum_s b_{js}n_s,&
 T_j&=\frac{\sum_i b_{j,s_i}r_ir_i^{\mathrm T}}
                 {\max(m_j^T,\epsilon)}.
 \label{eq:current-compact-pooling}
\end{align}
Here $\epsilon$ is machine epsilon at the working precision. Each matrix starts as a pooled raw second moment. As the predictor evolves, it becomes a learned positive-semidefinite summary. Permuting the input particles leaves the pooled values unchanged \citep{zaheer2017deepsets}.

The initial scalar encoder receives seven $\log(1+n_s)$ values and
98 species-pair kernel values. For $\ell=2,10$\,nm, these are
\begin{equation}
 P_{st}^{(\ell)}=\sum_{i\ne k}
 \mathbf1_{s_i=s}\mathbf1_{s_k=t}
 \exp\!\left[-\frac{\|x_i-x_k\|^2}{2\ell^2}\right],
 \label{eq:current-pair-pooling}
\end{equation}
followed by $\log(1+P_{st}^{(\ell)})$. At each length scale, the sum counts distinct particles in all 49 ordered species pairs and excludes self-pairs. The scalar encoder thus has 105 inputs.

Invariant dynamic features concatenate $u$, scalar memory, and the
contractions
\[
 v_i^{\mathrm T}v_j,\quad \operatorname{tr}T_j,\quad
 \|T_j\|_F^2,\quad v_i^{\mathrm T}T_jv_i.
\]
Each contraction first receives
$\phi(x)=\operatorname{sgn}(x)\log(1+|x|)$. The full feature dimension is
$d=9+n_v^2+2n_T+n_vn_T+n_z$, namely 24, 63, and 213.
These scalar inputs yield invariant hazards and equivariant flows
\citep{villar2021scalars}. The encounter factor is
\begin{align}
 \sigma^2&=L^2\max(\operatorname{tr}T_1,0)/3,\\
 g_r&=[4\pi(\sigma^2+R_r^2)]^{-3/2},\qquad
 c_r=C_6(H_r(v_C)),
 \label{eq:current-compact-hazard}
\end{align}
where $v_C$ contains all invariant features. All reactions share the isotropic scale $\sigma^2$ from the first tensor channel. The capture radius $R_r$ and the learned correction $c_r$ still depend on the reaction.

A flow head maps $v_C$ to a row-stochastic matrix $P$, positive rates $\lambda_j,\eta_j$ and scalar targets $q_j$. A row-wise softmax gives $P$, and $\operatorname{softplus}(h-4)$ gives the rates in $\mathrm{ns}^{-1}$. Between additions, the memory evolves as

\begin{align}
 \dot v_j&=\lambda_j\left(\sum_kP_{jk}v_k-v_j\right),\\
 \dot T_j&=\eta_j\left(\frac1{n_v}\sum_kv_kv_k^{\mathrm T}-T_j\right)
                       +\frac{2\bar D_j}{L^2}I,\\
 \dot z_j&=0.05\,\mathrm{ns}^{-1}(\tanh q_j-z_j),
 \label{eq:current-compact-flow}
\end{align}
with
$\bar D_j=\sum_s b_{js}n_sD_s/\max(\sum_s b_{js}n_s,\epsilon)$.
The hazards update the counts and water ledger. Changes in counts affect memory through the invariant features and the weighted diffusion coefficient.

For an incoming cloud, superscript $I$ denotes its encoding at the same
origin. Vector and tensor jumps use
\begin{align}
 v_j^+&=\frac{m_j^vv_j+m_j^{v,I}v_j^I}
                   {\max(m_j^v+m_j^{v,I},\epsilon)},\\
 T_j^+&=\frac{m_j^TT_j+m_j^{T,I}T_j^I}
                   {\max(m_j^T+m_j^{T,I},\epsilon)}.
 \label{eq:current-compact-jump}
\end{align}
Let $N=\sum_sn_s$, $N_I=\sum_sn_s^I$, and let $J$ read the concatenated
old and incoming invariant features at the current clock. The scalar
jump is
\begin{equation}
 z^+=\frac{Nz+N_Iz^I}{\max(N+N_I,\epsilon)}
       +\frac{N_I\tanh J}{\max(N+N_I,10^{-8})}.
 \label{eq:current-scalar-jump}
\end{equation}
Counts add exactly, with unchanged origin, clock, and water ledger.
An empty addition leaves the entire branch unchanged.

\subsection{Complete retained storage}
A paired forecast needs two branch states but shares a single set of weights and fixed buffers (Table~\ref{tab:current-state-storage}). A geometric predictor stores $7+d_z+3+1+1$ float64 values per branch for species counts, spatial memory, the fixed origin, the water ledger and the clock. C16, C32 and C64 use $d_z=16,32,64$ components of spatial memory, respectively. Both moment models use $d_z=63$, with three coordinates and six independent second-moment entries for each of seven species. Counts-only dynamics require $7+1+1$ values per branch. Each trainable weight occupies eight bytes. Every predictor also has 2120 bytes of fixed buffers, including integer reaction-pair indices. C64 doubles C32's channel counts with the same encoding, features, jumps and flows. Encoded queries and integrator stages occupy additional temporary memory.

\section{Supporting results and complete tables}
\label{sec:supporting-results}
\subsection{Numerical qualification}
\label{sec:numerical-checks}
All 240 prespecified comparisons with tighter solver settings met the numerical criteria before we examined confirmation outcomes. At each of the three energies, we checked the first eight source families and all four actions per family. The refined settings gave 138 comparisons across 46 fixed predictors. The strict settings gave 102 across 34 predictors.

For the 20\,ns pulse-minus-passive response and the postinitial trajectories, the largest standardized RMS changes were $3.45\times10^{-6}$ and $3.17\times10^{-6}$, respectively. Both were below the $10^{-3}$ limit (Table~\ref{tab:numerical-qualification}). The corresponding largest count changes were $1.11\times10^{-4}$ and $1.54\times10^{-4}$ molecules, below the 0.05-molecule limit. We also independently checked conservation in all 516 groups of stored trajectories, each defined by its predictor, energy and solver configuration. These checks included the original64 compatibility runs, which used the same weights with the original first-order update and were excluded from the 240 refinement comparisons. Across the 516 blocks, the maximum conservation residual was $3.77\times10^{-12}$, the minimum count was zero, and the forced and passive branches agreed exactly before the first addition.

\begin{table*}[t]
\centering\small
\setlength{\tabcolsep}{5pt}
\begin{tabular}{lrrr}
\toprule
Metric & Base--refined (138) & Base--strict (102) & Required limit\\
\midrule
Maximum response RMS, standardized & $2.49\times10^{-6}$ & $3.45\times10^{-6}$ & $\leq10^{-3}$\\
Maximum trajectory RMS, standardized & $2.50\times10^{-6}$ & $3.17\times10^{-6}$ & $\leq10^{-3}$\\
Maximum response change (molecules) & $7.31\times10^{-5}$ & $1.11\times10^{-4}$ & $\leq0.05$\\
Maximum trajectory change (molecules) & $1.15\times10^{-4}$ & $1.54\times10^{-4}$ & $\leq0.05$\\
Maximum reported budget residual & $2.39\times10^{-12}$ & $3.87\times10^{-12}$ & $\leq10^{-8}$\\
Minimum reported count & $0$ & $0$ & $\geq-10^{-8}$\\
\bottomrule
\end{tabular}
\caption{Extrema over all 138 base--refined and 102 base--strict comparisons, taken from the independent audit.
Response changes use the 20\,ns forced-minus-passive contrast. Trajectory changes use both arms at the ten stored postinitial times. RMS changes use the species scales fixed at training. The last two rows summarize the values the solver recorded. The text gives the independent conservation check on all 516 stored-trajectory blocks.}

\label{tab:numerical-qualification}
\end{table*}

\subsection{Numerical-convention sensitivity}
\label{sec:convention-sensitivity}
\begin{figure}[t]
\centering
\includegraphics[width=.85\textwidth]{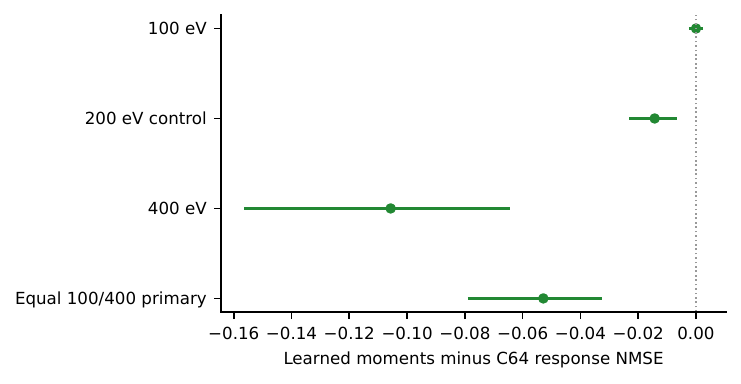}
\caption{Difference in response NMSE between learned moments and C64 under the main solver. Negative values
favor learned moments. The bottom row gives the prespecified comparison with equal weight on 100
and 400\,eV. Markers give the mean paired difference. Bars give 95\% intervals from resampling
source families while keeping each family's results for both models together, conditional on the
five fixed fits per method. Intervals for the individual energies are descriptive and unadjusted.}
\label{fig:primary-support}
\end{figure}

Changing from the fixed base solver to the original first-order update in float64 increased response NMSE at 400\,eV from 0.1858 to 0.3518 for C64 and from 0.1379 to 0.2882 for C32. For learned moments it changed from 0.0801 to 0.0810. Under this first-order update, the learned-moment-minus-C64 difference with equal 100/400\,eV weighting was $-0.1355$ (95\% CI $[-0.1722,-0.1057]$). The change of solver also changes how the update treats absent species (Section~\ref{sec:fitting-controls}). Figure~\ref{fig:primary-support} gives the primary comparison with the fixed base solver.

\subsection{Earlier source and intervention tests}
\label{sec:earlier-fixed-energy}
\label{sec:earlier-actions}
\noindent\textit{Fixed total energy.}
To hold total deposited energy at 400\,eV, an earlier study used four 100\,eV electrons, two 200\,eV electrons, or one 400\,eV electron at a common origin in pure water. Each composition comprised 32 independent source families observed at 0.02\,ns. We followed each source without additions for 64 SBS futures. We scored the original learned-moment and active C32 predictors, with five fixed fits per group (seeds 1701--1705) and qualified float64 integration. This cohort and these fits predate the controlled refits and fresh confirmation bank.

The endpoint here was raw mean squared error in seven-species G-values at 20\,ns, with $G_s=100N_s/400$ in molecules per 100\,eV. Within each composition we averaged the squared errors of individual fits before comparing predictors (Table~\ref{tab:earlier-fixed-energy}). Learned moments had lower point error in all three compositions. Holding deposited energy fixed allowed primary number, charge and species budgets to vary, so this comparison cannot attribute an error difference to track geometry alone.

\begin{table}[t]
\centering\small
\caption{Earlier passive-source G-value MSE differences between the first-order-trained moment and active C32 predictors (moments minus C32), in $(\mathrm{molecules}/100\,\mathrm{eV})^2$.
Losses average seven species, five fixed fits and 32 source families per composition.
The descriptive 95\% intervals use 4000 paired source-family percentile-bootstrap
resamples, conditional on those fixed fits and unadjusted across compositions.}
\label{tab:earlier-fixed-energy}
\begin{tabular}{lrr}
\toprule
Source composition & Difference & Descriptive 95\% CI\\
\midrule
$4\times100$\,eV & $-0.00777$ & $[-0.01107,-0.00419]$\\
$2\times200$\,eV & $-0.01238$ & $[-0.01623,-0.00856]$\\
$1\times400$\,eV & $-0.00938$ & $[-0.01403,-0.00519]$\\
\bottomrule
\end{tabular}
\end{table}

\noindent\textit{Previously unseen additions at 200\,eV.}
A separate test evaluated the original first-order predictors on 64 held-out families from the 200\,eV generator. Each family received all three incoming clouds together at 3 or 5\,ns (384 cases), or one cloud at each of 1, 5 and 10\,ns (192 cases). Both schedules used offsets of 0, 10 and 50\,nm and 64 SBS paths per branch. For the unseen single-addition times, mean terminal-response NMSE was 0.0277 for active C32 and 0.2248 for passive C32. With three additions their errors were 0.0767 and 0.2095, respectively. On this schedule, counts-only dynamics had lower error (0.0602) than active C32 and learned moments had the lowest error of these predictors (0.0253). These descriptive comparisons use the earlier fits at the training energy. Appendix~\ref{app:action-transfer} tests the later fits on the same schedules and families. The main source-transfer comparison uses fresh families and familiar schedules.

\clearpage
\subsection{Native baseline uncertainty}
On the development bank, the first-order-trained moment fits retained their advantage over the pulse-alone simulation after correction for the finite native samples. With equal weighting at 100 and 400\,eV, the learned-minus-pulse NMSE difference was $-0.0522$ (descriptive 95\% CI $[-0.0598,-0.0437]$). For the pulse-alone prediction we subtracted the sampling variances of both its own mean and the SBS reference mean. For a deterministic prediction only the reference mean contributes sampling variance.

On the fresh confirmation bank at 200\,eV, learned moments had lower raw response NMSE than 64-path IRT (paired difference $-0.00645$, 95\% CI $[-0.0119,-0.000702]$). After the separate corrections for sampling variance, the estimated learned-minus-IRT difference in squared discrepancy from the SBS mean was 0.00435 (95\% CI $[-0.000973,0.0100]$). This interval includes zero. The IRT results for fewer paths estimate expected MSE from the same 64-path sample and its estimated variance.

\begin{figure}[!ht]
\centering
\includegraphics[width=\textwidth]{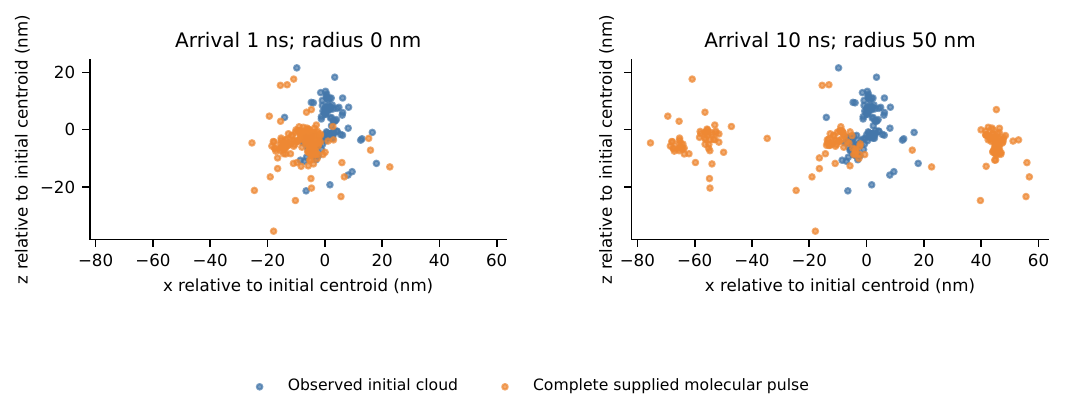}
\caption{Initial cloud and incoming molecules for two actions on source family 60096. The incoming clouds
arrive at 1\,ns with $r=0$\,nm or at 10\,ns with $r=50$\,nm. Both panels project the coordinates
onto the $x,z$ plane about the initial centroid, with the same axis limits and equal scales on
both axes.}
\label{fig:source-clouds}
\end{figure}

\subsection{Complete quantitative records}
The tables give errors by fit, action and species, timings for each repetition, and allocations of sources and paths for estimating mean yield. The research archive includes the same fields and units in machine-readable form. It reports CPU and wall time, peak memory, worker components and matched repetitions for workloads that include source acquisition.

\subsection{Reproduction materials}
\label{sec:reproduction-materials}
The submission supplement includes the 128 family records and paired source-transfer intervals (\path{ICWM_transfer_statistics_v1.zip}). The \path{headline_replay} folder gives individual-fit comparisons, timings and OH-yield allocations. The \path{figure_replay} and \path{training_diagnostic} folders include the observations and code for the forecast, cost and training figures. The frozen schedule plan, numerical checks and family losses are in \path{action_transfer}.

The \path{mechanistic_diagnostics} folder contains geometry, terminal errors, native sampling variances and rate traces. The clock experiment can also be rerun directly from the supplied clouds and weights. The \path{mechanistic_ablations} folder records the exploratory analysis plans, source identities, selected checkpoint hashes, solver checks and per-family losses. Each replay has a short README. The complete code and data release, including the statistical replay from stored trajectories, is available at \href{https://github.com/hnusrat1/water-radiolysis-world-models/releases/tag/v1.0.0}{GitHub release v1.0.0}.

\begin{figure}[!ht]
\centering
\includegraphics[width=\textwidth]{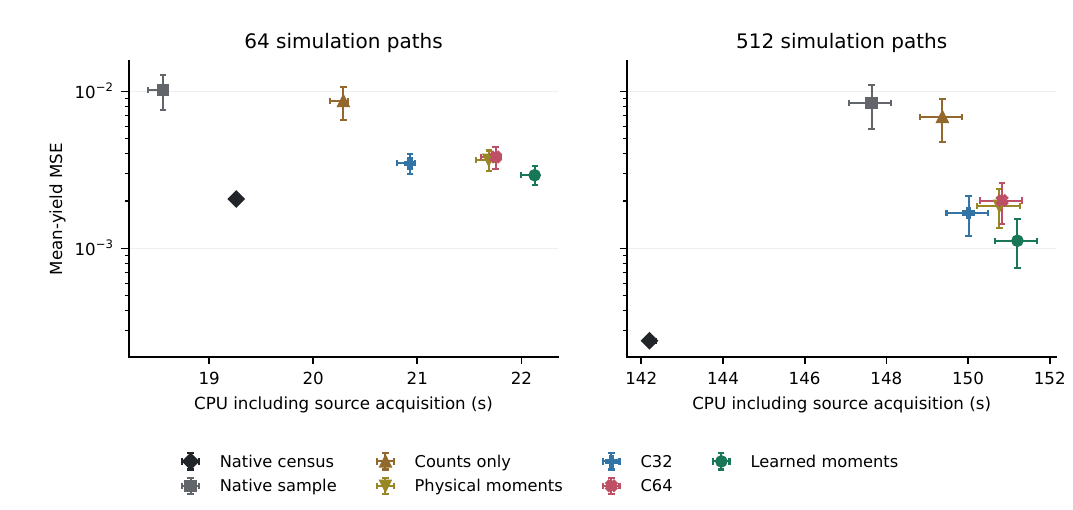}
\caption{Estimating mean passive OH yield across $N=64$ sources at 400\,eV. Each panel fixes the total number of SBS paths. Native sampling and model-assisted estimates use eight sources with 8 or 64 paths each. The census uses all 64 sources with 1 or 8 paths each. Learned models use fit 1701. Vertical bars give 95\% family-bootstrap intervals in $(\mathrm{OH\ molecules}/100\,\mathrm{eV})^2$. Horizontal bars give the quartiles of complete CPU cost over three repetitions, including source acquisition.}
\label{fig:utility-cost}
\end{figure}

\noindent\textbf{Cost of estimating mean yield.} At 400\,eV the 64- and 512-path censuses each sampled the entire bank. Figure~\ref{fig:utility-cost} plots their estimated error against complete CPU time including source acquisition, using the same timing records as the table. Table~\ref{tab:reported-utility-cost-join} gives the mean-yield errors in $(\mathrm{OH\ molecules}/100\,\mathrm{eV})^2$. Table~\ref{tab:reported-accuracy-cost-join} gives the corresponding costs and errors for the separate task of conditional forecasting.

\clearpage
\begin{table}[p]
\centering
\small
\setlength{\tabcolsep}{4pt}
\caption{Response NMSE at 20\,ns for individual fits on the confirmation bank with the main solver. Columns
give energy per primary electron. Seed numbers identify fits. Rows labeled first-order used the
earlier update during training. Physical moments are unfitted.}
\label{tab:reported-fits-base}
\begin{tabular}{lrrrr}
\toprule
Predictor & Seed & 100 eV & 200 eV control & 400 eV \\
\midrule
Counts & 1701 & 0.03271 & 0.08704 & 0.2565 \\
Counts & 1702 & 0.03253 & 0.08373 & 0.2442 \\
Counts & 1703 & 0.03231 & 0.08761 & 0.2543 \\
Counts & 1704 & 0.03287 & 0.08846 & 0.2544 \\
Counts & 1705 & 0.0324 & 0.08561 & 0.2421 \\
First-order C16 active & 1701 & 0.0185 & 0.0415 & 0.2067 \\
First-order C16 active & 1702 & 0.01913 & 0.041 & 0.2566 \\
First-order C16 active & 1703 & 0.01987 & 0.04054 & 0.2248 \\
First-order C16 active & 1704 & 0.01871 & 0.03794 & 0.1684 \\
First-order C16 active & 1705 & 0.01894 & 0.0388 & 0.221 \\
First-order C16 passive & 1701 & 0.1181 & 0.322 & 0.9055 \\
First-order C16 passive & 1702 & 0.1279 & 0.3516 & 0.9994 \\
First-order C16 passive & 1703 & 0.1253 & 0.3426 & 1.031 \\
First-order C16 passive & 1704 & 0.1133 & 0.2795 & 0.7775 \\
First-order C16 passive & 1705 & 0.1245 & 0.3107 & 0.9314 \\
First-order C32 active & 1701 & 0.01976 & 0.04341 & 0.1539 \\
First-order C32 active & 1702 & 0.02185 & 0.04098 & 0.1675 \\
First-order C32 active & 1703 & 0.02297 & 0.04373 & 0.2048 \\
First-order C32 active & 1704 & 0.02229 & 0.04808 & 0.2467 \\
First-order C32 active & 1705 & 0.0212 & 0.0445 & 0.1515 \\
First-order C32 passive & 1701 & 0.1209 & 0.3458 & 1.041 \\
First-order C32 passive & 1702 & 0.1194 & 0.3054 & 0.9265 \\
First-order C32 passive & 1703 & 0.1275 & 0.3388 & 1.102 \\
First-order C32 passive & 1704 & 0.1168 & 0.309 & 0.9814 \\
First-order C32 passive & 1705 & 0.1203 & 0.3327 & 0.9682 \\
First-order moments & 1701 & 0.01771 & 0.02752 & 0.07657 \\
First-order moments & 1702 & 0.01939 & 0.02564 & 0.1262 \\
First-order moments & 1703 & 0.01736 & 0.02493 & 0.06659 \\
First-order moments & 1704 & 0.01678 & 0.02485 & 0.06645 \\
First-order moments & 1705 & 0.0182 & 0.0257 & 0.0807 \\
Physical moments & 0 & 0.0786 & 0.2909 & 1.058 \\
C32 & 1701 & 0.01841 & 0.03986 & 0.1255 \\
C32 & 1702 & 0.01717 & 0.03509 & 0.1484 \\
C32 & 1703 & 0.01659 & 0.03584 & 0.1504 \\
C32 & 1704 & 0.01863 & 0.04044 & 0.1579 \\
C32 & 1705 & 0.01834 & 0.0402 & 0.1073 \\
Learned moments & 1701 & 0.01759 & 0.0242 & 0.0672 \\
Learned moments & 1702 & 0.01943 & 0.0265 & 0.1155 \\
Learned moments & 1703 & 0.01726 & 0.02508 & 0.06736 \\
Learned moments & 1704 & 0.01674 & 0.02521 & 0.06559 \\
Learned moments & 1705 & 0.01853 & 0.02602 & 0.08502 \\
C64 & 1701 & 0.01802 & 0.03826 & 0.1718 \\
C64 & 1702 & 0.01936 & 0.0436 & 0.2058 \\
C64 & 1703 & 0.01658 & 0.03561 & 0.1744 \\
C64 & 1704 & 0.01707 & 0.04138 & 0.1899 \\
C64 & 1705 & 0.01843 & 0.03962 & 0.1871 \\
\bottomrule
\end{tabular}
\end{table}
\clearpage

\begin{table}[p]
\centering
\small
\setlength{\tabcolsep}{4pt}
\caption{Response NMSE at 20\,ns for individual fits on the confirmation bank with the earlier first-order
update in float64. Columns give energy per primary electron. Seed numbers identify fits. Rows
labeled first-order used the earlier update during training. Physical moments are unfitted.}
\label{tab:reported-fits-original64}
\begin{tabular}{lrrrr}
\toprule
Predictor & Seed & 100 eV & 200 eV control & 400 eV \\
\midrule
Counts & 1701 & 0.03271 & 0.0873 & 0.2579 \\
Counts & 1702 & 0.0325 & 0.08383 & 0.2453 \\
Counts & 1703 & 0.03231 & 0.08783 & 0.2558 \\
Counts & 1704 & 0.03288 & 0.08875 & 0.256 \\
Counts & 1705 & 0.03239 & 0.0858 & 0.2432 \\
First-order C16 active & 1701 & 0.01813 & 0.04044 & 0.1554 \\
First-order C16 active & 1702 & 0.01859 & 0.03815 & 0.1909 \\
First-order C16 active & 1703 & 0.01927 & 0.03856 & 0.1646 \\
First-order C16 active & 1704 & 0.01847 & 0.03768 & 0.1403 \\
First-order C16 active & 1705 & 0.01867 & 0.03807 & 0.1674 \\
First-order C16 passive & 1701 & 0.118 & 0.3214 & 0.9044 \\
First-order C16 passive & 1702 & 0.1277 & 0.3507 & 0.9974 \\
First-order C16 passive & 1703 & 0.1251 & 0.3417 & 1.031 \\
First-order C16 passive & 1704 & 0.1132 & 0.279 & 0.7773 \\
First-order C16 passive & 1705 & 0.1243 & 0.3104 & 0.9319 \\
First-order C32 active & 1701 & 0.0185 & 0.03983 & 0.1233 \\
First-order C32 active & 1702 & 0.01745 & 0.03477 & 0.2319 \\
First-order C32 active & 1703 & 0.01814 & 0.03653 & 0.1885 \\
First-order C32 active & 1704 & 0.01856 & 0.03909 & 0.1603 \\
First-order C32 active & 1705 & 0.01911 & 0.03983 & 0.132 \\
First-order C32 passive & 1701 & 0.1208 & 0.3449 & 1.04 \\
First-order C32 passive & 1702 & 0.1193 & 0.3045 & 0.9263 \\
First-order C32 passive & 1703 & 0.1273 & 0.338 & 1.104 \\
First-order C32 passive & 1704 & 0.1167 & 0.3084 & 0.9823 \\
First-order C32 passive & 1705 & 0.1201 & 0.3315 & 0.9657 \\
First-order moments & 1701 & 0.01776 & 0.02788 & 0.07786 \\
First-order moments & 1702 & 0.01946 & 0.0257 & 0.1276 \\
First-order moments & 1703 & 0.01733 & 0.02506 & 0.06728 \\
First-order moments & 1704 & 0.01676 & 0.02503 & 0.06684 \\
First-order moments & 1705 & 0.01818 & 0.026 & 0.08244 \\
Physical moments & 0 & 0.07935 & 0.2937 & 1.066 \\
C32 & 1701 & 0.01869 & 0.04355 & 0.1399 \\
C32 & 1702 & 0.01701 & 0.05103 & 0.437 \\
C32 & 1703 & 0.01703 & 0.05154 & 0.4024 \\
C32 & 1704 & 0.01953 & 0.04745 & 0.2982 \\
C32 & 1705 & 0.01847 & 0.04464 & 0.1634 \\
Learned moments & 1701 & 0.01756 & 0.02432 & 0.06746 \\
Learned moments & 1702 & 0.01948 & 0.02665 & 0.1169 \\
Learned moments & 1703 & 0.01721 & 0.02523 & 0.06809 \\
Learned moments & 1704 & 0.01672 & 0.02539 & 0.0659 \\
Learned moments & 1705 & 0.0185 & 0.02635 & 0.08678 \\
C64 & 1701 & 0.01815 & 0.05053 & 0.3395 \\
C64 & 1702 & 0.01899 & 0.05552 & 0.3839 \\
C64 & 1703 & 0.01709 & 0.05167 & 0.4153 \\
C64 & 1704 & 0.01744 & 0.05205 & 0.2615 \\
C64 & 1705 & 0.01897 & 0.04969 & 0.3587 \\
\bottomrule
\end{tabular}
\end{table}
\clearpage

\begin{table}[p]
\centering
\small
\setlength{\tabcolsep}{4pt}
\caption{Pulse response at 100.0\,eV per primary electron under the main solver. Each of the four actions
specifies an arrival time (ns) and an offset $r$ (nm). For each predictor we report raw NMSE and
NMSE after subtracting the estimated sampling variance of the SBS reference mean. Learned rows
average losses from five fixed fits. Physical moments give one unfitted forecast per case.
Brackets give unadjusted 95\% family-bootstrap intervals.}
\label{tab:reported-actions-energy100}
\begin{tabular}{lrrr}
\toprule
Predictor & Action & Raw NMSE [95\% CI] & Signed corrected [95\% CI] \\
\midrule
Counts & 1, 0 & 0.02501 [0.01974, 0.03134] & 0.01921 [0.01398, 0.02558] \\
Counts & 1, 50 & 0.04877 [0.04463, 0.05304] & 0.04337 [0.03917, 0.04763] \\
Counts & 10, 0 & 0.03119 [0.02692, 0.03596] & 0.02553 [0.02125, 0.03023] \\
Counts & 10, 50 & 0.02528 [0.0215, 0.02939] & 0.0202 [0.01637, 0.02437] \\
C32 & 1, 0 & 0.02152 [0.01697, 0.02664] & 0.01572 [0.01124, 0.02081] \\
C32 & 1, 50 & 0.01672 [0.01297, 0.02138] & 0.01131 [0.007602, 0.01596] \\
C32 & 10, 0 & 0.0193 [0.01623, 0.02306] & 0.01363 [0.01056, 0.01738] \\
C32 & 10, 50 & 0.01379 [0.01095, 0.0172] & 0.008705 [0.005892, 0.01207] \\
C64 & 1, 0 & 0.02171 [0.01716, 0.02675] & 0.01591 [0.01135, 0.02097] \\
C64 & 1, 50 & 0.01641 [0.01255, 0.02121] & 0.01101 [0.00716, 0.01578] \\
C64 & 10, 0 & 0.01998 [0.0168, 0.02397] & 0.01432 [0.01115, 0.0182] \\
C64 & 10, 50 & 0.01347 [0.01055, 0.01688] & 0.008393 [0.005565, 0.01175] \\
Learned moments & 1, 0 & 0.01407 [0.01117, 0.01748] & 0.008272 [0.005366, 0.01175] \\
Learned moments & 1, 50 & 0.0179 [0.01354, 0.02288] & 0.01249 [0.008106, 0.01748] \\
Learned moments & 10, 0 & 0.02111 [0.01725, 0.02602] & 0.01544 [0.01162, 0.0204] \\
Learned moments & 10, 50 & 0.01857 [0.01404, 0.02387] & 0.01349 [0.008956, 0.01881] \\
Physical moments & 1, 0 & 0.05478 [0.04628, 0.0633] & 0.04898 [0.04045, 0.05757] \\
Physical moments & 1, 50 & 0.05957 [0.0524, 0.06727] & 0.05416 [0.0471, 0.06179] \\
Physical moments & 10, 0 & 0.1434 [0.1238, 0.164] & 0.1378 [0.1183, 0.1583] \\
Physical moments & 10, 50 & 0.05662 [0.04909, 0.0648] & 0.05153 [0.04412, 0.05971] \\
\bottomrule
\end{tabular}
\end{table}
\clearpage

\begin{table}[p]
\centering
\small
\setlength{\tabcolsep}{4pt}
\caption{Errors in the 20\,ns response of each species at 100.0\,eV per primary electron under the main
solver. Raw RMSE (molecules) takes the square root after averaging squared errors over fixed fits
and cases for all four actions. Signed MSE (molecules squared) subtracts the estimated sampling
variance of the SBS reference mean and can be negative. Brackets give unadjusted 95\%
family-bootstrap intervals.}
\label{tab:reported-species-energy100}
\begin{tabular}{lrrr}
\toprule
Predictor & Species & Raw RMSE [95\% CI] & Signed MSE [95\% CI] \\
\midrule
Counts & $e_{\rm aq}$ & 1.748 [1.668, 1.83] & 2.976 [2.701, 3.267] \\
Counts & $\mathrm{OH}$ & 2.434 [2.35, 2.513] & 5.788 [5.385, 6.176] \\
Counts & $\mathrm{H}$ & 0.375 [0.331, 0.4215] & 0.09323 [0.06237, 0.1299] \\
Counts & $\mathrm{H}_2$ & 0.3481 [0.3287, 0.3665] & 0.1054 [0.09216, 0.1187] \\
Counts & $\mathrm{H}_3\mathrm{O}^{+}$ & 1.636 [1.564, 1.711] & 2.628 [2.399, 2.879] \\
Counts & $\mathrm{OH}^{-}$ & 0.4039 [0.3614, 0.4465] & 0.1083 [0.07549, 0.1446] \\
Counts & $\mathrm{H}_2\mathrm{O}_2$ & 0.7849 [0.739, 0.8297] & 0.5772 [0.5077, 0.6494] \\
C32 & $e_{\rm aq}$ & 0.8819 [0.7711, 0.9925] & 0.6968 [0.5139, 0.9046] \\
C32 & $\mathrm{OH}$ & 1.261 [1.167, 1.354] & 1.455 [1.223, 1.699] \\
C32 & $\mathrm{H}$ & 0.3368 [0.2978, 0.3765] & 0.06607 [0.04139, 0.09443] \\
C32 & $\mathrm{H}_2$ & 0.1976 [0.1795, 0.2178] & 0.02323 [0.01634, 0.03146] \\
C32 & $\mathrm{H}_3\mathrm{O}^{+}$ & 0.7603 [0.6689, 0.85] & 0.5303 [0.3995, 0.6755] \\
C32 & $\mathrm{OH}^{-}$ & 0.383 [0.3409, 0.424] & 0.09186 [0.06159, 0.1253] \\
C32 & $\mathrm{H}_2\mathrm{O}_2$ & 0.5213 [0.4744, 0.5718] & 0.2329 [0.1864, 0.2882] \\
C64 & $e_{\rm aq}$ & 0.8315 [0.7225, 0.9408] & 0.6105 [0.4415, 0.8032] \\
C64 & $\mathrm{OH}$ & 1.341 [1.237, 1.442] & 1.661 [1.393, 1.942] \\
C64 & $\mathrm{H}$ & 0.3431 [0.3038, 0.384] & 0.07036 [0.04505, 0.09961] \\
C64 & $\mathrm{H}_2$ & 0.1891 [0.1703, 0.2096] & 0.01995 [0.01322, 0.02807] \\
C64 & $\mathrm{H}_3\mathrm{O}^{+}$ & 0.7186 [0.6267, 0.8082] & 0.4686 [0.3448, 0.6055] \\
C64 & $\mathrm{OH}^{-}$ & 0.3808 [0.339, 0.4222] & 0.09018 [0.0604, 0.123] \\
C64 & $\mathrm{H}_2\mathrm{O}_2$ & 0.5362 [0.4832, 0.5913] & 0.2487 [0.1946, 0.3108] \\
Learned moments & $e_{\rm aq}$ & 0.8457 [0.7256, 0.9632] & 0.6344 [0.4457, 0.848] \\
Learned moments & $\mathrm{OH}$ & 1.077 [1.002, 1.155] & 1.024 [0.8667, 1.199] \\
Learned moments & $\mathrm{H}$ & 0.3141 [0.2728, 0.3552] & 0.05125 [0.02735, 0.07821] \\
Learned moments & $\mathrm{H}_2$ & 0.1754 [0.1533, 0.1989] & 0.01493 [0.007782, 0.02359] \\
Learned moments & $\mathrm{H}_3\mathrm{O}^{+}$ & 0.7709 [0.6662, 0.8743] & 0.5467 [0.3959, 0.7162] \\
Learned moments & $\mathrm{OH}^{-}$ & 0.3979 [0.3566, 0.4389] & 0.1035 [0.07273, 0.1379] \\
Learned moments & $\mathrm{H}_2\mathrm{O}_2$ & 0.5423 [0.4959, 0.5888] & 0.2552 [0.2069, 0.3078] \\
Physical moments & $e_{\rm aq}$ & 1.276 [1.199, 1.351] & 1.547 [1.358, 1.742] \\
Physical moments & $\mathrm{OH}$ & 3.209 [3.033, 3.375] & 10.16 [9.066, 11.25] \\
Physical moments & $\mathrm{H}$ & 0.4454 [0.3854, 0.5108] & 0.1509 [0.1024, 0.2125] \\
Physical moments & $\mathrm{H}_2$ & 0.2774 [0.2557, 0.2989] & 0.06113 [0.04975, 0.07308] \\
Physical moments & $\mathrm{H}_3\mathrm{O}^{+}$ & 1.149 [1.052, 1.24] & 1.272 [1.059, 1.49] \\
Physical moments & $\mathrm{OH}^{-}$ & 0.8413 [0.7831, 0.9012] & 0.6529 [0.5597, 0.7572] \\
Physical moments & $\mathrm{H}_2\mathrm{O}_2$ & 1.195 [1.114, 1.278] & 1.389 [1.202, 1.594] \\
\bottomrule
\end{tabular}
\end{table}
\clearpage

\begin{table}[p]
\centering
\small
\setlength{\tabcolsep}{4pt}
\caption{Pulse response at 200.0\,eV per primary electron under the main solver. Each of the four actions
specifies an arrival time (ns) and an offset $r$ (nm). For each predictor we report raw NMSE and
NMSE after subtracting the estimated sampling variance of the SBS reference mean. Learned rows
average losses from five fixed fits. Physical moments give one unfitted forecast per case.
Brackets give unadjusted 95\% family-bootstrap intervals.}
\label{tab:reported-actions-energy200-control}
\begin{tabular}{lrrr}
\toprule
Predictor & Action & Raw NMSE [95\% CI] & Signed corrected [95\% CI] \\
\midrule
Counts & 1, 0 & 0.07703 [0.06335, 0.09098] & 0.06577 [0.05208, 0.07966] \\
Counts & 1, 50 & 0.1248 [0.1072, 0.1436] & 0.1137 [0.09583, 0.1327] \\
Counts & 10, 0 & 0.08438 [0.0708, 0.09789] & 0.07345 [0.05985, 0.08706] \\
Counts & 10, 50 & 0.05975 [0.04548, 0.07636] & 0.04925 [0.03478, 0.06607] \\
C32 & 1, 0 & 0.0471 [0.03478, 0.06192] & 0.03584 [0.02356, 0.05068] \\
C32 & 1, 50 & 0.03651 [0.0275, 0.04616] & 0.02536 [0.01629, 0.03506] \\
C32 & 10, 0 & 0.03699 [0.03042, 0.04405] & 0.02606 [0.01953, 0.03312] \\
C32 & 10, 50 & 0.03254 [0.02093, 0.04843] & 0.02205 [0.01039, 0.03789] \\
C64 & 1, 0 & 0.04706 [0.03459, 0.06237] & 0.0358 [0.02337, 0.05111] \\
C64 & 1, 50 & 0.03871 [0.02941, 0.04937] & 0.02756 [0.01829, 0.03811] \\
C64 & 10, 0 & 0.0358 [0.0279, 0.0444] & 0.02487 [0.01699, 0.03338] \\
C64 & 10, 50 & 0.03719 [0.02541, 0.05265] & 0.02669 [0.01473, 0.0422] \\
Learned moments & 1, 0 & 0.02448 [0.01865, 0.0315] & 0.01322 [0.007321, 0.02023] \\
Learned moments & 1, 50 & 0.02352 [0.01797, 0.02975] & 0.01238 [0.006725, 0.01865] \\
Learned moments & 10, 0 & 0.03178 [0.02429, 0.04031] & 0.02085 [0.01318, 0.02952] \\
Learned moments & 10, 50 & 0.02181 [0.01721, 0.02658] & 0.01132 [0.006764, 0.01601] \\
Physical moments & 1, 0 & 0.1193 [0.09817, 0.1423] & 0.108 [0.08676, 0.1312] \\
Physical moments & 1, 50 & 0.3554 [0.3132, 0.3974] & 0.3442 [0.3022, 0.3863] \\
Physical moments & 10, 0 & 0.4083 [0.3546, 0.4674] & 0.3974 [0.3437, 0.4566] \\
Physical moments & 10, 50 & 0.2806 [0.2506, 0.3127] & 0.2701 [0.2399, 0.3024] \\
\bottomrule
\end{tabular}
\end{table}
\clearpage

\begin{table}[p]
\centering
\small
\setlength{\tabcolsep}{4pt}
\caption{Errors in the 20\,ns response of each species at 200.0\,eV per primary electron under the main
solver. Raw RMSE (molecules) takes the square root after averaging squared errors over fixed fits
and cases for all four actions. Signed MSE (molecules squared) subtracts the estimated sampling
variance of the SBS reference mean and can be negative. Brackets give unadjusted 95\%
family-bootstrap intervals.}
\label{tab:reported-species-energy200-control}
\begin{tabular}{lrrr}
\toprule
Predictor & Species & Raw RMSE [95\% CI] & Signed MSE [95\% CI] \\
\midrule
Counts & $e_{\rm aq}$ & 3.271 [3.111, 3.429] & 10.54 [9.522, 11.6] \\
Counts & $\mathrm{OH}$ & 3.949 [3.801, 4.097] & 15.36 [14.21, 16.55] \\
Counts & $\mathrm{H}$ & 0.634 [0.5366, 0.7324] & 0.2963 [0.1831, 0.43] \\
Counts & $\mathrm{H}_2$ & 0.6744 [0.6236, 0.7264] & 0.4161 [0.3504, 0.489] \\
Counts & $\mathrm{H}_3\mathrm{O}^{+}$ & 3.207 [3.064, 3.347] & 10.18 [9.284, 11.1] \\
Counts & $\mathrm{OH}^{-}$ & 0.5837 [0.5052, 0.663] & 0.2369 [0.1492, 0.337] \\
Counts & $\mathrm{H}_2\mathrm{O}_2$ & 1.159 [1.047, 1.266] & 1.264 [1.016, 1.523] \\
C32 & $e_{\rm aq}$ & 1.394 [1.222, 1.619] & 1.783 [1.332, 2.458] \\
C32 & $\mathrm{OH}$ & 1.401 [1.267, 1.539] & 1.723 [1.369, 2.127] \\
C32 & $\mathrm{H}$ & 0.537 [0.4731, 0.6063] & 0.1826 [0.1191, 0.2611] \\
C32 & $\mathrm{H}_2$ & 0.3201 [0.2788, 0.3694] & 0.06376 [0.03869, 0.09852] \\
C32 & $\mathrm{H}_3\mathrm{O}^{+}$ & 1.221 [1.088, 1.355] & 1.384 [1.078, 1.73] \\
C32 & $\mathrm{OH}^{-}$ & 0.554 [0.4686, 0.6432] & 0.2031 [0.1161, 0.3104] \\
C32 & $\mathrm{H}_2\mathrm{O}_2$ & 0.7873 [0.6761, 0.9082] & 0.5396 [0.3776, 0.7435] \\
C64 & $e_{\rm aq}$ & 1.369 [1.201, 1.56] & 1.712 [1.282, 2.271] \\
C64 & $\mathrm{OH}$ & 1.384 [1.226, 1.556] & 1.675 [1.271, 2.178] \\
C64 & $\mathrm{H}$ & 0.52 [0.4597, 0.5838] & 0.1646 [0.1066, 0.2349] \\
C64 & $\mathrm{H}_2$ & 0.3344 [0.2971, 0.3783] & 0.07311 [0.04945, 0.1048] \\
C64 & $\mathrm{H}_3\mathrm{O}^{+}$ & 1.145 [1.015, 1.274] & 1.205 [0.9252, 1.519] \\
C64 & $\mathrm{OH}^{-}$ & 0.5676 [0.482, 0.6569] & 0.2184 [0.1275, 0.328] \\
C64 & $\mathrm{H}_2\mathrm{O}_2$ & 0.8221 [0.7035, 0.9473] & 0.5956 [0.4153, 0.8153] \\
Learned moments & $e_{\rm aq}$ & 1.167 [0.982, 1.366] & 1.202 [0.8017, 1.702] \\
Learned moments & $\mathrm{OH}$ & 1.18 [1.063, 1.294] & 1.153 [0.8918, 1.436] \\
Learned moments & $\mathrm{H}$ & 0.4662 [0.4046, 0.5292] & 0.1116 [0.05758, 0.1741] \\
Learned moments & $\mathrm{H}_2$ & 0.2568 [0.2302, 0.284] & 0.02725 [0.01387, 0.04232] \\
Learned moments & $\mathrm{H}_3\mathrm{O}^{+}$ & 1.151 [0.9748, 1.329] & 1.218 [0.8431, 1.657] \\
Learned moments & $\mathrm{OH}^{-}$ & 0.4498 [0.3959, 0.5043] & 0.09853 [0.0517, 0.1518] \\
Learned moments & $\mathrm{H}_2\mathrm{O}_2$ & 0.5897 [0.5127, 0.6682] & 0.2675 [0.1829, 0.367] \\
Physical moments & $e_{\rm aq}$ & 2.719 [2.517, 2.931] & 7.234 [6.182, 8.418] \\
Physical moments & $\mathrm{OH}$ & 6.353 [6.047, 6.646] & 40.12 [36.33, 43.93] \\
Physical moments & $\mathrm{H}$ & 0.8812 [0.7421, 1.016] & 0.6708 [0.4449, 0.9231] \\
Physical moments & $\mathrm{H}_2$ & 0.5948 [0.5374, 0.6499] & 0.3151 [0.2499, 0.3838] \\
Physical moments & $\mathrm{H}_3\mathrm{O}^{+}$ & 2.425 [2.267, 2.576] & 5.776 [5.037, 6.523] \\
Physical moments & $\mathrm{OH}^{-}$ & 1.545 [1.435, 1.65] & 2.282 [1.956, 2.618] \\
Physical moments & $\mathrm{H}_2\mathrm{O}_2$ & 2.507 [2.364, 2.645] & 6.207 [5.508, 6.915] \\
\bottomrule
\end{tabular}
\end{table}
\clearpage

\begin{table}[p]
\centering
\small
\setlength{\tabcolsep}{4pt}
\caption{Pulse response at 400.0\,eV per primary electron under the main solver. Each of the four actions
specifies an arrival time (ns) and an offset $r$ (nm). For each predictor we report raw NMSE and
NMSE after subtracting the estimated sampling variance of the SBS reference mean. Learned rows
average losses from five fixed fits. Physical moments give one unfitted forecast per case.
Brackets give unadjusted 95\% family-bootstrap intervals.}
\label{tab:reported-actions-energy400}
\begin{tabular}{lrrr}
\toprule
Predictor & Action & Raw NMSE [95\% CI] & Signed corrected [95\% CI] \\
\midrule
Counts & 1, 0 & 0.2507 [0.2081, 0.2978] & 0.2296 [0.1869, 0.2765] \\
Counts & 1, 50 & 0.3504 [0.3128, 0.3897] & 0.3286 [0.2907, 0.3677] \\
Counts & 10, 0 & 0.251 [0.2208, 0.2844] & 0.2298 [0.1994, 0.2632] \\
Counts & 10, 50 & 0.1491 [0.1268, 0.1735] & 0.1286 [0.1062, 0.1532] \\
C32 & 1, 0 & 0.1283 [0.1095, 0.1497] & 0.1071 [0.0883, 0.1285] \\
C32 & 1, 50 & 0.15 [0.1155, 0.1929] & 0.1282 [0.09396, 0.1712] \\
C32 & 10, 0 & 0.1466 [0.1114, 0.1952] & 0.1253 [0.09025, 0.1739] \\
C32 & 10, 50 & 0.1267 [0.09737, 0.1616] & 0.1061 [0.07679, 0.141] \\
C64 & 1, 0 & 0.1443 [0.118, 0.1728] & 0.1231 [0.09706, 0.1515] \\
C64 & 1, 50 & 0.2072 [0.1567, 0.2691] & 0.1853 [0.1351, 0.247] \\
C64 & 10, 0 & 0.1837 [0.1309, 0.251] & 0.1624 [0.1098, 0.2299] \\
C64 & 10, 50 & 0.2081 [0.1582, 0.2686] & 0.1875 [0.1377, 0.2479] \\
Learned moments & 1, 0 & 0.08647 [0.07567, 0.09762] & 0.06529 [0.0546, 0.07648] \\
Learned moments & 1, 50 & 0.1092 [0.09397, 0.1254] & 0.08742 [0.07194, 0.1038] \\
Learned moments & 10, 0 & 0.04834 [0.04039, 0.0582] & 0.02709 [0.01904, 0.03693] \\
Learned moments & 10, 50 & 0.07647 [0.06448, 0.0892] & 0.05591 [0.0438, 0.0686] \\
Physical moments & 1, 0 & 0.4599 [0.4278, 0.4931] & 0.4387 [0.4066, 0.4719] \\
Physical moments & 1, 50 & 1.63 [1.517, 1.742] & 1.608 [1.495, 1.721] \\
Physical moments & 10, 0 & 0.8543 [0.7957, 0.913] & 0.8331 [0.7744, 0.8916] \\
Physical moments & 10, 50 & 1.286 [1.213, 1.36] & 1.265 [1.192, 1.34] \\
\bottomrule
\end{tabular}
\end{table}
\clearpage

\begin{table}[p]
\centering
\small
\setlength{\tabcolsep}{4pt}
\caption{Errors in the 20\,ns response of each species at 400.0\,eV per primary electron under the main
solver. Raw RMSE (molecules) takes the square root after averaging squared errors over fixed fits
and cases for all four actions. Signed MSE (molecules squared) subtracts the estimated sampling
variance of the SBS reference mean and can be negative. Brackets give unadjusted 95\%
family-bootstrap intervals.}
\label{tab:reported-species-energy400}
\begin{tabular}{lrrr}
\toprule
Predictor & Species & Raw RMSE [95\% CI] & Signed MSE [95\% CI] \\
\midrule
Counts & $e_{\rm aq}$ & 5.995 [5.786, 6.21] & 35.62 [33.16, 38.25] \\
Counts & $\mathrm{OH}$ & 6.269 [6.079, 6.464] & 38.88 [36.54, 41.36] \\
Counts & $\mathrm{H}$ & 0.8252 [0.7307, 0.9216] & 0.4566 [0.3095, 0.6246] \\
Counts & $\mathrm{H}_2$ & 1.218 [1.148, 1.285] & 1.392 [1.226, 1.558] \\
Counts & $\mathrm{H}_3\mathrm{O}^{+}$ & 6.226 [6.055, 6.403] & 38.54 [36.45, 40.78] \\
Counts & $\mathrm{OH}^{-}$ & 0.9883 [0.8855, 1.091] & 0.7882 [0.5947, 1.006] \\
Counts & $\mathrm{H}_2\mathrm{O}_2$ & 1.871 [1.683, 2.07] & 3.345 [2.675, 4.129] \\
C32 & $e_{\rm aq}$ & 3.147 [2.767, 3.53] & 9.587 [7.336, 12.14] \\
C32 & $\mathrm{OH}$ & 3.315 [2.962, 3.673] & 10.57 [8.356, 13.05] \\
C32 & $\mathrm{H}$ & 0.9488 [0.8349, 1.061] & 0.6759 [0.4736, 0.9009] \\
C32 & $\mathrm{H}_2$ & 0.6903 [0.617, 0.7669] & 0.3841 [0.2881, 0.4946] \\
C32 & $\mathrm{H}_3\mathrm{O}^{+}$ & 2.791 [2.514, 3.055] & 7.57 [6.107, 9.114] \\
C32 & $\mathrm{OH}^{-}$ & 0.8997 [0.787, 1.03] & 0.6209 [0.4328, 0.8715] \\
C32 & $\mathrm{H}_2\mathrm{O}_2$ & 1.847 [1.596, 2.136] & 3.258 [2.394, 4.407] \\
C64 & $e_{\rm aq}$ & 3.229 [2.815, 3.634] & 10.11 [7.614, 12.9] \\
C64 & $\mathrm{OH}$ & 4.4 [4.03, 4.759] & 18.94 [15.82, 22.21] \\
C64 & $\mathrm{H}$ & 1.055 [0.9289, 1.177] & 0.8876 [0.639, 1.163] \\
C64 & $\mathrm{H}_2$ & 0.6891 [0.6177, 0.7633] & 0.3825 [0.2886, 0.4901] \\
C64 & $\mathrm{H}_3\mathrm{O}^{+}$ & 2.888 [2.586, 3.188] & 8.121 [6.473, 9.949] \\
C64 & $\mathrm{OH}^{-}$ & 1.104 [0.9189, 1.3] & 1.03 [0.6553, 1.5] \\
C64 & $\mathrm{H}_2\mathrm{O}_2$ & 2.083 [1.801, 2.401] & 4.183 [3.088, 5.607] \\
Learned moments & $e_{\rm aq}$ & 1.831 [1.69, 1.977] & 3.038 [2.542, 3.596] \\
Learned moments & $\mathrm{OH}$ & 2.337 [2.125, 2.563] & 5.037 [4.086, 6.152] \\
Learned moments & $\mathrm{H}$ & 0.7801 [0.7207, 0.8382] & 0.3842 [0.2978, 0.4762] \\
Learned moments & $\mathrm{H}_2$ & 0.6579 [0.608, 0.7051] & 0.3404 [0.2768, 0.4055] \\
Learned moments & $\mathrm{H}_3\mathrm{O}^{+}$ & 1.71 [1.591, 1.828] & 2.707 [2.315, 3.124] \\
Learned moments & $\mathrm{OH}^{-}$ & 0.6834 [0.6308, 0.7404] & 0.2784 [0.208, 0.3603] \\
Learned moments & $\mathrm{H}_2\mathrm{O}_2$ & 1.31 [1.186, 1.432] & 1.562 [1.25, 1.896] \\
Physical moments & $e_{\rm aq}$ & 7.048 [6.854, 7.246] & 49.37 [46.67, 52.19] \\
Physical moments & $\mathrm{OH}$ & 11.29 [10.91, 11.68] & 127.1 [118.7, 135.9] \\
Physical moments & $\mathrm{H}$ & 1.66 [1.538, 1.772] & 2.53 [2.144, 2.909] \\
Physical moments & $\mathrm{H}_2$ & 1.277 [1.208, 1.346] & 1.539 [1.366, 1.719] \\
Physical moments & $\mathrm{H}_3\mathrm{O}^{+}$ & 5.989 [5.828, 6.149] & 35.65 [33.75, 37.59] \\
Physical moments & $\mathrm{OH}^{-}$ & 2.737 [2.661, 2.808] & 7.304 [6.896, 7.694] \\
Physical moments & $\mathrm{H}_2\mathrm{O}_2$ & 5.438 [5.217, 5.647] & 29.42 [27.06, 31.73] \\
\bottomrule
\end{tabular}
\end{table}
\clearpage

\begin{table}[p]
\centering
\small
\setlength{\tabcolsep}{4pt}
\caption{Terminal-response NMSE over all four actions on the confirmation bank with the main solver. Signed
estimates subtract the sampling variance of the SBS reference mean. Brackets give unadjusted 95\%
family-bootstrap intervals. Counts-only dynamics and rows labeled first-order used the earlier
update during training.}
\label{tab:reported-context-base}
\begin{tabular}{lrrr}
\toprule
eV & Predictor / control & Raw NMSE [95\% CI] & Signed corrected [95\% CI] \\
\midrule
100.0 & Counts & 0.03256 [0.02987, 0.03543] & 0.02708 [0.0244, 0.02991] \\
100.0 & First-order C16 active & 0.01903 [0.01662, 0.02173] & 0.01354 [0.01115, 0.01618] \\
100.0 & First-order C16 passive & 0.1218 [0.1113, 0.1331] & 0.1163 [0.1059, 0.1275] \\
100.0 & First-order C32 active & 0.02161 [0.01857, 0.025] & 0.01613 [0.01311, 0.01949] \\
100.0 & First-order C32 passive & 0.121 [0.1102, 0.1326] & 0.1155 [0.1048, 0.127] \\
100.0 & First-order moments & 0.01789 [0.01544, 0.02054] & 0.0124 [0.01, 0.01502] \\
100.0 & Physical moments & 0.0786 [0.07041, 0.08749] & 0.07312 [0.06511, 0.08192] \\
100.0 & C32 & 0.01783 [0.01556, 0.02036] & 0.01234 [0.01008, 0.01482] \\
100.0 & Learned moments & 0.01791 [0.01545, 0.02055] & 0.01242 [0.009989, 0.01502] \\
100.0 & C64 & 0.01789 [0.01561, 0.02038] & 0.01241 [0.01015, 0.01484] \\
100.0 & Zero response & 0.6882 [0.6419, 0.7365] & 0.6827 [0.6366, 0.7309] \\
100.0 & Added counts only & 0.4017 [0.372, 0.4322] & 0.3962 [0.3665, 0.4266] \\
200.0 & Counts & 0.08649 [0.07811, 0.09569] & 0.07553 [0.06702, 0.08486] \\
200.0 & First-order C16 active & 0.03996 [0.03172, 0.05077] & 0.029 [0.02071, 0.03986] \\
200.0 & First-order C16 passive & 0.3213 [0.292, 0.3542] & 0.3103 [0.2812, 0.3432] \\
200.0 & First-order C32 active & 0.04414 [0.0359, 0.05542] & 0.03318 [0.02487, 0.04444] \\
200.0 & First-order C32 passive & 0.3263 [0.2941, 0.3615] & 0.3154 [0.2831, 0.3503] \\
200.0 & First-order moments & 0.02573 [0.02159, 0.03029] & 0.01477 [0.01057, 0.01941] \\
200.0 & Physical moments & 0.2909 [0.2646, 0.3183] & 0.2799 [0.2536, 0.3074] \\
200.0 & C32 & 0.03828 [0.03105, 0.0474] & 0.02733 [0.02004, 0.03654] \\
200.0 & Learned moments & 0.0254 [0.02134, 0.02979] & 0.01444 [0.01031, 0.01889] \\
200.0 & C64 & 0.03969 [0.03211, 0.04906] & 0.02873 [0.0211, 0.03814] \\
200.0 & Zero response & 2.396 [2.192, 2.612] & 2.385 [2.181, 2.6] \\
200.0 & Added counts only & 1.508 [1.384, 1.631] & 1.497 [1.374, 1.62] \\
400.0 & Counts & 0.2503 [0.229, 0.2738] & 0.2291 [0.2079, 0.2526] \\
400.0 & First-order C16 active & 0.2155 [0.1796, 0.2568] & 0.1943 [0.1585, 0.2354] \\
400.0 & First-order C16 passive & 0.929 [0.8677, 0.9905] & 0.9078 [0.8467, 0.9692] \\
400.0 & First-order C32 active & 0.1849 [0.1548, 0.2201] & 0.1637 [0.1336, 0.1988] \\
400.0 & First-order C32 passive & 1.004 [0.9377, 1.069] & 0.9826 [0.9166, 1.048] \\
400.0 & First-order moments & 0.08331 [0.07534, 0.09175] & 0.06211 [0.05419, 0.07048] \\
400.0 & Physical moments & 1.058 [1.013, 1.102] & 1.036 [0.9917, 1.08] \\
400.0 & C32 & 0.1379 [0.1124, 0.1716] & 0.1167 [0.09123, 0.1505] \\
400.0 & Learned moments & 0.08013 [0.07255, 0.08819] & 0.05893 [0.05139, 0.06709] \\
400.0 & C64 & 0.1858 [0.1452, 0.2367] & 0.1646 [0.1241, 0.215] \\
400.0 & Zero response & 8.856 [8.455, 9.273] & 8.835 [8.434, 9.252] \\
400.0 & Added counts only & 5.685 [5.432, 5.941] & 5.663 [5.411, 5.92] \\
\bottomrule
\end{tabular}
\end{table}
\clearpage

\begin{table}[p]
\centering
\small
\setlength{\tabcolsep}{4pt}
\caption{Terminal-response NMSE over all four actions on the confirmation bank with the earlier first-order
update in float64. Signed estimates subtract the sampling variance of the SBS reference mean.
Brackets give unadjusted 95\% family-bootstrap intervals. Counts-only dynamics and rows labeled
first-order used the earlier update during training.}
\label{tab:reported-context-original64}
\begin{tabular}{lrrr}
\toprule
eV & Predictor / control & Raw NMSE [95\% CI] & Signed corrected [95\% CI] \\
\midrule
100.0 & Counts & 0.03256 [0.02988, 0.03544] & 0.02707 [0.0244, 0.0299] \\
100.0 & First-order C16 active & 0.01863 [0.01623, 0.02133] & 0.01314 [0.01075, 0.0158] \\
100.0 & First-order C16 passive & 0.1217 [0.1112, 0.133] & 0.1162 [0.1058, 0.1274] \\
100.0 & First-order C32 active & 0.01835 [0.01584, 0.0211] & 0.01286 [0.0104, 0.01558] \\
100.0 & First-order C32 passive & 0.1208 [0.11, 0.1324] & 0.1153 [0.1046, 0.1268] \\
100.0 & First-order moments & 0.0179 [0.01543, 0.02056] & 0.01241 [0.01001, 0.01503] \\
100.0 & Physical moments & 0.07935 [0.07111, 0.08831] & 0.07386 [0.06578, 0.0827] \\
100.0 & C32 & 0.01815 [0.01597, 0.02058] & 0.01266 [0.01046, 0.0151] \\
100.0 & Learned moments & 0.01789 [0.01542, 0.02055] & 0.01241 [0.009961, 0.01503] \\
100.0 & C64 & 0.01813 [0.01594, 0.02062] & 0.01264 [0.01044, 0.01509] \\
100.0 & Zero response & 0.6882 [0.6419, 0.7365] & 0.6827 [0.6366, 0.7309] \\
100.0 & Added counts only & 0.4017 [0.372, 0.4322] & 0.3962 [0.3665, 0.4266] \\
200.0 & Counts & 0.0867 [0.07829, 0.096] & 0.07574 [0.06723, 0.08507] \\
200.0 & First-order C16 active & 0.03858 [0.0308, 0.04829] & 0.02762 [0.01974, 0.03743] \\
200.0 & First-order C16 passive & 0.3206 [0.2914, 0.3536] & 0.3097 [0.2806, 0.3426] \\
200.0 & First-order C32 active & 0.03801 [0.03072, 0.04745] & 0.02705 [0.01969, 0.03643] \\
200.0 & First-order C32 passive & 0.3255 [0.2933, 0.3605] & 0.3145 [0.2823, 0.3494] \\
200.0 & First-order moments & 0.02594 [0.02178, 0.0306] & 0.01498 [0.01072, 0.01973] \\
200.0 & Physical moments & 0.2937 [0.2671, 0.3213] & 0.2827 [0.2562, 0.3103] \\
200.0 & C32 & 0.04764 [0.0394, 0.05718] & 0.03668 [0.02833, 0.04633] \\
200.0 & Learned moments & 0.02559 [0.02151, 0.03005] & 0.01463 [0.01048, 0.01914] \\
200.0 & C64 & 0.05189 [0.04367, 0.06172] & 0.04093 [0.03254, 0.05077] \\
200.0 & Zero response & 2.396 [2.192, 2.612] & 2.385 [2.181, 2.6] \\
200.0 & Added counts only & 1.508 [1.384, 1.631] & 1.497 [1.374, 1.62] \\
400.0 & Counts & 0.2516 [0.2301, 0.2754] & 0.2304 [0.209, 0.2542] \\
400.0 & First-order C16 active & 0.1637 [0.1299, 0.2035] & 0.1425 [0.1089, 0.1821] \\
400.0 & First-order C16 passive & 0.9283 [0.8673, 0.9899] & 0.9071 [0.846, 0.9685] \\
400.0 & First-order C32 active & 0.1672 [0.1435, 0.1973] & 0.146 [0.1224, 0.176] \\
400.0 & First-order C32 passive & 1.004 [0.9377, 1.069] & 0.9825 [0.9165, 1.047] \\
400.0 & First-order moments & 0.0844 [0.07641, 0.09285] & 0.0632 [0.05522, 0.07165] \\
400.0 & Physical moments & 1.066 [1.021, 1.111] & 1.045 [1, 1.089] \\
400.0 & C32 & 0.2882 [0.2525, 0.3321] & 0.267 [0.2312, 0.3109] \\
400.0 & Learned moments & 0.08102 [0.07328, 0.0892] & 0.05982 [0.05217, 0.06801] \\
400.0 & C64 & 0.3518 [0.2914, 0.4224] & 0.3306 [0.2702, 0.4011] \\
400.0 & Zero response & 8.856 [8.455, 9.273] & 8.835 [8.434, 9.252] \\
400.0 & Added counts only & 5.685 [5.432, 5.941] & 5.663 [5.411, 5.92] \\
\bottomrule
\end{tabular}
\end{table}
\clearpage

\begin{table}[p]
\centering
\small
\setlength{\tabcolsep}{4pt}
\caption{Raw and corrected response NMSE across all four actions on the confirmation bank. The learned
models give deterministic forecasts once fitted. For these models and unfitted physical moments,
the correction subtracts the estimated sampling variance of the SBS reference mean. For IRT it
also subtracts the sampling variance of the IRT mean from 64 paths per branch. The signed
estimates measure squared discrepancy from the underlying SBS mean. Intervals are unadjusted 95\%
family-bootstrap intervals.}
\label{tab:reported-irt-context}
\begin{tabular}{lrrr}
\toprule
eV & Predictor & Raw NMSE [95\% CI] & Signed corrected [95\% CI] \\
\midrule
100.0 & IRT, R=64 & 0.01457 [0.01276, 0.01647] & 0.003367 [0.001647, 0.005216] \\
100.0 & Counts & 0.03256 [0.02995, 0.03541] & 0.02708 [0.02449, 0.02991] \\
100.0 & C32 & 0.01783 [0.01551, 0.02032] & 0.01234 [0.01007, 0.0148] \\
100.0 & C64 & 0.01789 [0.0156, 0.02032] & 0.01241 [0.01018, 0.01482] \\
100.0 & Learned moments & 0.01791 [0.01545, 0.02047] & 0.01242 [0.01001, 0.01496] \\
100.0 & Physical moments & 0.0786 [0.07088, 0.08692] & 0.07312 [0.06545, 0.08135] \\
200.0 & IRT, R=64 & 0.03186 [0.02786, 0.0359] & 0.01009 [0.006219, 0.01408] \\
200.0 & Counts & 0.08649 [0.07809, 0.09537] & 0.07553 [0.06709, 0.08452] \\
200.0 & C32 & 0.03828 [0.03068, 0.04757] & 0.02733 [0.01969, 0.03663] \\
200.0 & C64 & 0.03969 [0.03166, 0.04924] & 0.02873 [0.02076, 0.03825] \\
200.0 & Learned moments & 0.0254 [0.02134, 0.02976] & 0.01444 [0.01027, 0.01887] \\
200.0 & Physical moments & 0.2909 [0.2645, 0.3177] & 0.2799 [0.2535, 0.307] \\
400.0 & IRT, R=64 & 0.06846 [0.06315, 0.07387] & 0.02651 [0.02132, 0.03186] \\
400.0 & Counts & 0.2503 [0.2282, 0.2743] & 0.2291 [0.207, 0.2531] \\
400.0 & C32 & 0.1379 [0.1129, 0.1721] & 0.1167 [0.09196, 0.151] \\
400.0 & C64 & 0.1858 [0.1466, 0.2377] & 0.1646 [0.1253, 0.2165] \\
400.0 & Learned moments & 0.08013 [0.07217, 0.08832] & 0.05893 [0.051, 0.06709] \\
400.0 & Physical moments & 1.058 [1.014, 1.103] & 1.036 [0.9926, 1.082] \\
\bottomrule
\end{tabular}
\end{table}
\clearpage

\begin{table}[p]
\centering
\small
\setlength{\tabcolsep}{4pt}
\caption{Expected IRT response error as the number of paths per branch changes. Every row uses estimates
from the same 64-path simulations. Finite-reference MSE measures error against a sampled SBS mean.
Signed latent MSE removes the estimated sampling variance of that SBS mean and retains the IRT
variance for the indicated path count. Both columns average standardized errors over seven species
and all four actions, with unadjusted 95\% family-bootstrap intervals.}
\label{tab:reported-irt-sampling-profile}
\begin{tabular}{lrrr}
\toprule
eV & Paths per arm & Finite-reference MSE [95\% CI] & Signed latent MSE [95\% CI] \\
\midrule
100.0 & 1 & 0.3746 [0.3614, 0.3874] & 0.3692 [0.3561, 0.3817] \\
100.0 & 8 & 0.05458 [0.05195, 0.05711] & 0.04909 [0.04661, 0.05148] \\
100.0 & 32 & 0.02029 [0.01841, 0.02218] & 0.0148 [0.01297, 0.01668] \\
100.0 & 64 & 0.01457 [0.01276, 0.01647] & 0.009083 [0.007318, 0.01096] \\
200.0 & 1 & 0.7127 [0.6898, 0.736] & 0.7017 [0.6793, 0.7247] \\
200.0 & 8 & 0.1075 [0.1026, 0.1126] & 0.09655 [0.09185, 0.1013] \\
200.0 & 32 & 0.04266 [0.03858, 0.04676] & 0.0317 [0.02776, 0.03569] \\
200.0 & 64 & 0.03186 [0.02786, 0.0359] & 0.0209 [0.01698, 0.02483] \\
400.0 & 1 & 1.376 [1.354, 1.398] & 1.354 [1.332, 1.376] \\
400.0 & 8 & 0.2137 [0.2077, 0.2199] & 0.1925 [0.1866, 0.1985] \\
400.0 & 32 & 0.0892 [0.08376, 0.09462] & 0.068 [0.06269, 0.07337] \\
400.0 & 64 & 0.06846 [0.06315, 0.07387] & 0.04726 [0.04201, 0.05262] \\
\bottomrule
\end{tabular}
\end{table}
\clearpage

\begin{table}[p]
\centering
\small
\setlength{\tabcolsep}{4pt}
\caption{Estimated MSE of the mean passive OH yield at 20\,ns across the 100.0\,eV bank. Each allocation
uses $R$ SBS continuations on each of $m=8$ sampled source families. Learned rows average the MSE
of the individual fits. MSE is in $(\mathrm{OH\ molecules}/100\,\mathrm{eV})^2$. Brackets give
unadjusted 95\% family-bootstrap intervals.}
\label{tab:reported-utility-energy100}
\begin{tabular}{lrr}
\toprule
Predictor / control & R & Finite-bank MSE [95\% CI] \\
\midrule
Native only & 8 & 0.02608 [0.01987, 0.0326] \\
Native only & 64 & 0.01725 [0.01111, 0.02377] \\
Counts & 8 & 0.0212 [0.01712, 0.0259] \\
Counts & 64 & 0.01237 [0.00834, 0.01713] \\
First-order C16 active & 8 & 0.01716 [0.01489, 0.01955] \\
First-order C16 active & 64 & 0.008334 [0.005942, 0.01088] \\
First-order C16 passive & 8 & 0.02401 [0.01832, 0.03074] \\
First-order C16 passive & 64 & 0.01518 [0.009616, 0.02184] \\
First-order C32 active & 8 & 0.016 [0.01392, 0.01823] \\
First-order C32 active & 64 & 0.007176 [0.005103, 0.009382] \\
First-order C32 passive & 8 & 0.02557 [0.01921, 0.03296] \\
First-order C32 passive & 64 & 0.01674 [0.01051, 0.02409] \\
First-order moments & 8 & 0.01283 [0.01159, 0.01417] \\
First-order moments & 64 & 0.004003 [0.002872, 0.005277] \\
Physical moments & 8 & 0.01396 [0.01252, 0.01535] \\
Physical moments & 64 & 0.005134 [0.003782, 0.006416] \\
C32 & 8 & 0.01597 [0.01392, 0.0182] \\
C32 & 64 & 0.00714 [0.005081, 0.00938] \\
Learned moments & 8 & 0.01286 [0.01162, 0.01421] \\
Learned moments & 64 & 0.004031 [0.002906, 0.005312] \\
C64 & 8 & 0.01549 [0.01351, 0.01765] \\
C64 & 64 & 0.006657 [0.004663, 0.008892] \\
\bottomrule
\end{tabular}
\end{table}
\clearpage

\begin{table}[p]
\centering
\small
\setlength{\tabcolsep}{4pt}
\caption{Estimated MSE of the mean passive OH yield at 20\,ns across the 200.0\,eV bank. Each allocation
uses $R$ SBS continuations on each of $m=8$ sampled source families. Learned rows average the MSE
of the individual fits. MSE is in $(\mathrm{OH\ molecules}/100\,\mathrm{eV})^2$. Brackets give
unadjusted 95\% family-bootstrap intervals.}
\label{tab:reported-utility-energy200-control}
\begin{tabular}{lrr}
\toprule
Predictor / control & R & Finite-bank MSE [95\% CI] \\
\midrule
Native only & 8 & 0.01199 [0.009228, 0.01466] \\
Native only & 64 & 0.008013 [0.005256, 0.01064] \\
Counts & 8 & 0.009377 [0.007508, 0.01113] \\
Counts & 64 & 0.005401 [0.003562, 0.007135] \\
First-order C16 active & 8 & 0.007136 [0.005814, 0.008539] \\
First-order C16 active & 64 & 0.003159 [0.001882, 0.004516] \\
First-order C16 passive & 8 & 0.01101 [0.008276, 0.01378] \\
First-order C16 passive & 64 & 0.007037 [0.004263, 0.0098] \\
First-order C32 active & 8 & 0.006781 [0.005824, 0.007795] \\
First-order C32 active & 64 & 0.002805 [0.001855, 0.003804] \\
First-order C32 passive & 8 & 0.01174 [0.008639, 0.01495] \\
First-order C32 passive & 64 & 0.007768 [0.004646, 0.01096] \\
First-order moments & 8 & 0.005599 [0.004961, 0.006201] \\
First-order moments & 64 & 0.001623 [0.001096, 0.002139] \\
Physical moments & 8 & 0.006044 [0.005157, 0.007083] \\
Physical moments & 64 & 0.002068 [0.001268, 0.002993] \\
C32 & 8 & 0.006851 [0.005864, 0.007898] \\
C32 & 64 & 0.002875 [0.001918, 0.003882] \\
Learned moments & 8 & 0.005601 [0.004982, 0.006186] \\
Learned moments & 64 & 0.001625 [0.001105, 0.002128] \\
C64 & 8 & 0.006727 [0.005667, 0.007869] \\
C64 & 64 & 0.00275 [0.00176, 0.003843] \\
\bottomrule
\end{tabular}
\end{table}
\clearpage

\begin{table}[p]
\centering
\small
\setlength{\tabcolsep}{4pt}
\caption{Estimated MSE of the mean passive OH yield at 20\,ns across the 400.0\,eV bank. Each allocation
uses $R$ SBS continuations on each of $m=8$ sampled source families. Learned rows average the MSE
of the individual fits. MSE is in $(\mathrm{OH\ molecules}/100\,\mathrm{eV})^2$. Brackets give
unadjusted 95\% family-bootstrap intervals.}
\label{tab:reported-utility-energy400}
\begin{tabular}{lrr}
\toprule
Predictor / control & R & Finite-bank MSE [95\% CI] \\
\midrule
Native only & 8 & 0.01022 [0.007567, 0.01279] \\
Native only & 64 & 0.008414 [0.005767, 0.011] \\
Counts & 8 & 0.008583 [0.006507, 0.01058] \\
Counts & 64 & 0.006779 [0.004706, 0.008756] \\
First-order C16 active & 8 & 0.003679 [0.00319, 0.004135] \\
First-order C16 active & 64 & 0.001875 [0.001402, 0.002318] \\
First-order C16 passive & 8 & 0.00435 [0.003651, 0.005021] \\
First-order C16 passive & 64 & 0.002546 [0.001864, 0.003213] \\
First-order C32 active & 8 & 0.00338 [0.002996, 0.003733] \\
First-order C32 active & 64 & 0.001575 [0.001203, 0.001925] \\
First-order C32 passive & 8 & 0.004479 [0.003755, 0.005167] \\
First-order C32 passive & 64 & 0.002675 [0.001967, 0.003346] \\
First-order moments & 8 & 0.002863 [0.0025, 0.003289] \\
First-order moments & 64 & 0.001058 [0.0007143, 0.001469] \\
Physical moments & 8 & 0.003667 [0.003129, 0.004201] \\
Physical moments & 64 & 0.001863 [0.001343, 0.002402] \\
C32 & 8 & 0.003529 [0.003102, 0.00393] \\
C32 & 64 & 0.001725 [0.001301, 0.002122] \\
Learned moments & 8 & 0.002876 [0.002516, 0.003292] \\
Learned moments & 64 & 0.001072 [0.0007314, 0.001473] \\
C64 & 8 & 0.003742 [0.003208, 0.004265] \\
C64 & 64 & 0.001938 [0.001406, 0.002447] \\
\bottomrule
\end{tabular}
\end{table}
\clearpage

\begin{table}[p]
\centering
\small
\setlength{\tabcolsep}{4pt}
\caption{Exploratory forecasts against SBS on four previously examined source families per energy, with two
actions per family and one preselected fit per learned predictor. The 20 and 1000\,ns columns give
raw response NMSE. The final column averages branch-trajectory NMSE over 15 times after the
initial observation. The descriptive intervals resample four families, giving 35 possible sets of
family counts.}
\label{tab:reported-long-SBS}
\begin{tabular}{lrrrr}
\toprule
eV & Predictor & 20 ns [95\% CI] & 1000 ns [95\% CI] & Query average [95\% CI] \\
\midrule
100.0 & Counts & 0.04306 [0.02175, 0.06547] & 0.268 [0.208, 0.3301] & 0.0918 [0.08007, 0.1052] \\
100.0 & C32 & 0.03382 [0.01598, 0.05441] & 0.2697 [0.193, 0.3541] & 0.08653 [0.07484, 0.09822] \\
100.0 & C64 & 0.03703 [0.01993, 0.04976] & 0.3092 [0.2383, 0.3919] & 0.1078 [0.08464, 0.1245] \\
100.0 & Learned moments & 0.01808 [0.01272, 0.02244] & 0.08775 [0.05786, 0.1298] & 0.01898 [0.01261, 0.02874] \\
100.0 & Physical moments & 0.0388 [0.01535, 0.06225] & 0.08006 [0.03929, 0.1344] & 0.02978 [0.02226, 0.04189] \\
200.0 & Counts & 0.06017 [0.0331, 0.08986] & 0.9563 [0.8637, 1.033] & 0.2499 [0.2233, 0.2836] \\
200.0 & C32 & 0.04228 [0.02124, 0.06671] & 0.8329 [0.6591, 0.9524] & 0.179 [0.1545, 0.2153] \\
200.0 & C64 & 0.04289 [0.01996, 0.07665] & 1.045 [0.863, 1.203] & 0.2698 [0.2209, 0.3187] \\
200.0 & Learned moments & 0.0308 [0.01344, 0.04413] & 0.2632 [0.2043, 0.322] & 0.04926 [0.04342, 0.05547] \\
200.0 & Physical moments & 0.2274 [0.207, 0.2553] & 0.1451 [0.07663, 0.2133] & 0.1073 [0.06923, 0.1586] \\
400.0 & Counts & 0.1784 [0.1222, 0.2346] & 2.985 [2.723, 3.269] & 0.8335 [0.767, 0.9019] \\
400.0 & C32 & 0.1047 [0.04316, 0.1662] & 1.522 [1.404, 1.619] & 0.4978 [0.4266, 0.5731] \\
400.0 & C64 & 0.1058 [0.05148, 0.1602] & 1.616 [0.3647, 2.867] & 0.7044 [0.4405, 0.9721] \\
400.0 & Learned moments & 0.06248 [0.02251, 0.09491] & 0.9609 [0.8038, 1.112] & 0.1824 [0.1487, 0.2271] \\
400.0 & Physical moments & 0.6793 [0.5409, 0.8176] & 0.3882 [0.324, 0.4524] & 0.3271 [0.3077, 0.3436] \\
\bottomrule
\end{tabular}
\end{table}
\clearpage

\begin{table}[p]
\centering
\small
\setlength{\tabcolsep}{4pt}
\caption{Exploratory forecasts against synchronized IRT on four previously examined source
families per energy, with two actions per family and one preselected fit per learned predictor.
The 20 and 1000\,ns columns give raw response NMSE. The final column averages branch-trajectory
NMSE over 15 times after the initial observation. The descriptive intervals resample four
families, giving 35 possible sets of family counts.}
\label{tab:reported-long-IRT-syn}
\begin{tabular}{lrrrr}
\toprule
eV & Predictor & 20 ns [95\% CI] & 1000 ns [95\% CI] & Query average [95\% CI] \\
\midrule
100.0 & Counts & 0.02633 [0.0182, 0.03446] & 0.2021 [0.149, 0.2551] & 0.07811 [0.06784, 0.08837] \\
100.0 & C32 & 0.02309 [0.015, 0.0331] & 0.2069 [0.1388, 0.275] & 0.07662 [0.0662, 0.09148] \\
100.0 & C64 & 0.0226 [0.01412, 0.03272] & 0.2328 [0.1707, 0.2948] & 0.09323 [0.07557, 0.1067] \\
100.0 & Learned moments & 0.02371 [0.01368, 0.04075] & 0.05264 [0.03732, 0.06795] & 0.02433 [0.01363, 0.03922] \\
100.0 & Physical moments & 0.03047 [0.01343, 0.05385] & 0.02878 [0.01745, 0.03851] & 0.03283 [0.02178, 0.05248] \\
200.0 & Counts & 0.1018 [0.0638, 0.1557] & 0.6794 [0.6279, 0.7537] & 0.228 [0.2104, 0.2457] \\
200.0 & C32 & 0.06715 [0.03542, 0.0986] & 0.5515 [0.4595, 0.6749] & 0.1595 [0.1379, 0.1776] \\
200.0 & C64 & 0.05008 [0.03037, 0.07197] & 0.766 [0.6765, 0.8544] & 0.2399 [0.2103, 0.2694] \\
200.0 & Learned moments & 0.0579 [0.02825, 0.08755] & 0.229 [0.2209, 0.2381] & 0.04874 [0.04483, 0.05416] \\
200.0 & Physical moments & 0.2364 [0.1452, 0.3382] & 0.2226 [0.1161, 0.2964] & 0.08695 [0.06988, 0.104] \\
400.0 & Counts & 0.2079 [0.1227, 0.2987] & 2.468 [2.04, 2.895] & 0.7295 [0.6786, 0.7804] \\
400.0 & C32 & 0.106 [0.0549, 0.1932] & 1.135 [0.9796, 1.29] & 0.405 [0.3629, 0.4743] \\
400.0 & C64 & 0.09982 [0.03664, 0.2029] & 1.264 [0.3667, 2.161] & 0.5819 [0.3611, 0.807] \\
400.0 & Learned moments & 0.05282 [0.02927, 0.07553] & 0.6659 [0.578, 0.7538] & 0.154 [0.1348, 0.171] \\
400.0 & Physical moments & 0.6771 [0.6091, 0.7596] & 0.329 [0.2708, 0.4113] & 0.3457 [0.3243, 0.3671] \\
\bottomrule
\end{tabular}
\end{table}
\clearpage

\begin{table}[p]
\centering
\small
\setlength{\tabcolsep}{4pt}
\caption{Repetition 0 of 34 timing workloads (33 original workloads and the census with one path per
source). All three repetitions used identical inputs. CPU and wall time are seconds for a fresh
worker. The CPU total includes child processes that the worker waited for. Peak resident memory
(RSS, MiB) can reflect the largest child process. Internal CPU time measures the component named
in the last column and is included in the total.}
\label{tab:reported-cost-repeat0}
\begin{tabular}{lrrrrl}
\toprule
Workload & CPU s & Wall s & RSS MiB & Internal CPU s & Boundary \\
\midrule
IRT sample ($m=8$, $R=64$) & 4.661 & 4.667 & 74.61 & 4.113 & Native child \\
IRT sample ($m=8$, $R=8$) & 0.9813 & 0.9849 & 57.83 & 0.7804 & Native child \\
IRT census ($m=64$, $R=8$) & 4.646 & 4.651 & 73.94 & 4.109 & Native child \\
SBS sample ($m=8$, $R=64$) & 146 & 146 & 74.66 & 145.4 & Native child \\
SBS sample ($m=8$, $R=8$) & 17.72 & 17.81 & 59.52 & 17.52 & Native child \\
SBS census ($m=64$, $R=8$) & 141 & 141 & 73.92 & 140.4 & Native child \\
32 pairs, First-order C16 (active) & 6.684 & 6.69 & 417.4 & 5.741 & Loaded solver \\
32 pairs, First-order C16 (passive) & 1.591 & 1.594 & 418 & 0.6588 & Loaded solver \\
32 pairs, First-order C32 (active) & 3.188 & 3.194 & 418.1 & 2.243 & Loaded solver \\
32 pairs, First-order C32 (passive) & 1.664 & 1.672 & 417.7 & 0.7061 & Loaded solver \\
32 pairs, Counts only & 1.141 & 1.146 & 299.8 & 0.2034 & Loaded solver \\
32 pairs, First-order moments & 3.919 & 3.924 & 303.4 & 2.961 & Loaded solver \\
32 pairs, Physical moments & 3.262 & 3.269 & 303 & 2.326 & Loaded solver \\
32 pairs, C32 & 3.125 & 3.13 & 417.8 & 2.186 & Loaded solver \\
32 pairs, C64 & 3.888 & 3.895 & 418.1 & 2.931 & Loaded solver \\
32 pairs, Learned moments & 3.93 & 3.937 & 302.5 & 2.989 & Loaded solver \\
Paired IRT ($R=1$) & 1.531 & 1.535 & 58.98 & 1.297 & Native child \\
Paired IRT ($R=32$) & 36.67 & 36.68 & 135.7 & 33.95 & Native child \\
Paired IRT ($R=64$) & 74.29 & 74.31 & 130 & 68.88 & Native child \\
Paired IRT ($R=8$) & 9.137 & 9.143 & 99.94 & 8.345 & Native child \\
Paired SBS ($R=64$) & 3217 & 3217 & 117.1 & 3212 & Native child \\
64 passive, First-order C16 (active) & 2.843 & 2.849 & 394.6 & 1.889 & Loaded solver \\
64 passive, First-order C16 (passive) & 1.252 & 1.258 & 393.8 & 0.2902 & Loaded solver \\
64 passive, First-order C32 (active) & 1.679 & 1.685 & 394.2 & 0.7304 & Loaded solver \\
64 passive, First-order C32 (passive) & 1.272 & 1.278 & 394 & 0.326 & Loaded solver \\
64 passive, Counts only & 1.074 & 1.078 & 301.5 & 0.1035 & Loaded solver \\
64 passive, First-order moments & 2.893 & 2.899 & 303.3 & 1.931 & Loaded solver \\
64 passive, Physical moments & 2.485 & 2.491 & 303.8 & 1.529 & Loaded solver \\
64 passive, C32 & 1.727 & 1.732 & 394.1 & 0.7732 & Loaded solver \\
64 passive, C64 & 2.524 & 2.53 & 394.8 & 1.575 & Loaded solver \\
64 passive, Learned moments & 2.904 & 2.911 & 304 & 1.947 & Loaded solver \\
Source generation (64 clouds) & 1.229 & 1.234 & 59.45 & 1.059 & Native child \\
Source generation (8 clouds) & 0.551 & 0.5555 & 59.22 & 0.4003 & Native child \\
SBS census ($m=64$, $R=1$) & 17.93 & 17.94 & 59.5 & 17.73 & Native child \\
\bottomrule
\end{tabular}
\end{table}
\clearpage

\begin{table}[p]
\centering
\small
\setlength{\tabcolsep}{4pt}
\caption{Repetition 1 of 34 timing workloads (33 original workloads and the census with one path per
source). All three repetitions used identical inputs. CPU and wall time are seconds for a fresh
worker. The CPU total includes child processes that the worker waited for. Peak resident memory
(RSS, MiB) can reflect the largest child process. Internal CPU time measures the component named
in the last column and is included in the total.}
\label{tab:reported-cost-repeat1}
\begin{tabular}{lrrrrl}
\toprule
Workload & CPU s & Wall s & RSS MiB & Internal CPU s & Boundary \\
\midrule
IRT sample ($m=8$, $R=64$) & 4.655 & 4.658 & 74.69 & 4.13 & Native child \\
IRT sample ($m=8$, $R=8$) & 0.9699 & 0.9728 & 57.84 & 0.7923 & Native child \\
IRT census ($m=64$, $R=8$) & 4.62 & 4.623 & 73.94 & 4.104 & Native child \\
SBS sample ($m=8$, $R=64$) & 147.1 & 147.1 & 74.55 & 146.6 & Native child \\
SBS sample ($m=8$, $R=8$) & 18.03 & 18.03 & 59.45 & 17.85 & Native child \\
SBS census ($m=64$, $R=8$) & 140.8 & 140.9 & 73.92 & 140.3 & Native child \\
32 pairs, First-order C16 (active) & 6.669 & 6.674 & 417.2 & 5.744 & Loaded solver \\
32 pairs, First-order C16 (passive) & 1.585 & 1.587 & 417.5 & 0.6666 & Loaded solver \\
32 pairs, First-order C32 (active) & 3.166 & 3.168 & 417 & 2.242 & Loaded solver \\
32 pairs, First-order C32 (passive) & 1.623 & 1.626 & 417.7 & 0.7011 & Loaded solver \\
32 pairs, Counts only & 1.123 & 1.126 & 300.6 & 0.2039 & Loaded solver \\
32 pairs, First-order moments & 3.904 & 3.907 & 301.8 & 2.981 & Loaded solver \\
32 pairs, Physical moments & 3.245 & 3.249 & 303 & 2.333 & Loaded solver \\
32 pairs, C32 & 3.128 & 3.132 & 417.6 & 2.206 & Loaded solver \\
32 pairs, C64 & 3.87 & 3.874 & 418.2 & 2.951 & Loaded solver \\
32 pairs, Learned moments & 3.937 & 3.94 & 301.9 & 3.016 & Loaded solver \\
Paired IRT ($R=1$) & 1.504 & 1.507 & 58.95 & 1.299 & Native child \\
Paired IRT ($R=32$) & 37.07 & 37.09 & 135.6 & 34.33 & Native child \\
Paired IRT ($R=64$) & 74.92 & 74.94 & 130 & 69.5 & Native child \\
Paired IRT ($R=8$) & 9.195 & 9.2 & 99.89 & 8.428 & Native child \\
Paired SBS ($R=64$) & 3230 & 3231 & 117.3 & 3225 & Native child \\
64 passive, First-order C16 (active) & 2.815 & 2.817 & 394 & 1.873 & Loaded solver \\
64 passive, First-order C16 (passive) & 1.224 & 1.227 & 394.4 & 0.2898 & Loaded solver \\
64 passive, First-order C32 (active) & 1.678 & 1.682 & 394.8 & 0.7464 & Loaded solver \\
64 passive, First-order C32 (passive) & 1.266 & 1.27 & 393.9 & 0.3314 & Loaded solver \\
64 passive, Counts only & 1.055 & 1.059 & 302.2 & 0.1034 & Loaded solver \\
64 passive, First-order moments & 2.864 & 2.868 & 303 & 1.935 & Loaded solver \\
64 passive, Physical moments & 2.451 & 2.455 & 303.1 & 1.525 & Loaded solver \\
64 passive, C32 & 1.696 & 1.698 & 394.3 & 0.7591 & Loaded solver \\
64 passive, C64 & 2.517 & 2.521 & 394.2 & 1.584 & Loaded solver \\
64 passive, Learned moments & 2.892 & 2.894 & 302.8 & 1.939 & Loaded solver \\
Source generation (64 clouds) & 1.21 & 1.213 & 59.41 & 1.065 & Native child \\
Source generation (8 clouds) & 0.53 & 0.5325 & 59.2 & 0.4076 & Native child \\
SBS census ($m=64$, $R=1$) & 18.05 & 18.05 & 59.58 & 17.87 & Native child \\
\bottomrule
\end{tabular}
\end{table}
\clearpage

\begin{table}[p]
\centering
\small
\setlength{\tabcolsep}{4pt}
\caption{Repetition 2 of 34 timing workloads (33 original workloads and the census with one path per
source). All three repetitions used identical inputs. CPU and wall time are seconds for a fresh
worker. The CPU total includes child processes that the worker waited for. Peak resident memory
(RSS, MiB) can reflect the largest child process. Internal CPU time measures the component named
in the last column and is included in the total.}
\label{tab:reported-cost-repeat2}
\begin{tabular}{lrrrrl}
\toprule
Workload & CPU s & Wall s & RSS MiB & Internal CPU s & Boundary \\
\midrule
IRT sample ($m=8$, $R=64$) & 4.668 & 4.671 & 74.62 & 4.147 & Native child \\
IRT sample ($m=8$, $R=8$) & 0.9843 & 0.9866 & 57.88 & 0.8063 & Native child \\
IRT census ($m=64$, $R=8$) & 4.641 & 4.644 & 73.94 & 4.12 & Native child \\
SBS sample ($m=8$, $R=64$) & 148.1 & 148.1 & 74.61 & 147.5 & Native child \\
SBS sample ($m=8$, $R=8$) & 18.11 & 18.12 & 59.39 & 17.93 & Native child \\
SBS census ($m=64$, $R=8$) & 141.3 & 141.3 & 73.86 & 140.8 & Native child \\
32 pairs, First-order C16 (active) & 6.663 & 6.665 & 417.3 & 5.744 & Loaded solver \\
32 pairs, First-order C16 (passive) & 1.582 & 1.586 & 417.3 & 0.6578 & Loaded solver \\
32 pairs, First-order C32 (active) & 3.148 & 3.15 & 418 & 2.23 & Loaded solver \\
32 pairs, First-order C32 (passive) & 1.634 & 1.637 & 417.5 & 0.7149 & Loaded solver \\
32 pairs, Counts only & 1.124 & 1.128 & 299.8 & 0.2044 & Loaded solver \\
32 pairs, First-order moments & 3.915 & 3.919 & 302.2 & 2.998 & Loaded solver \\
32 pairs, Physical moments & 3.29 & 3.294 & 302.8 & 2.373 & Loaded solver \\
32 pairs, C32 & 3.111 & 3.115 & 418.1 & 2.197 & Loaded solver \\
32 pairs, C64 & 3.865 & 3.869 & 418.1 & 2.948 & Loaded solver \\
32 pairs, Learned moments & 3.948 & 3.952 & 301.8 & 3.024 & Loaded solver \\
Paired IRT ($R=1$) & 1.512 & 1.515 & 59.14 & 1.307 & Native child \\
Paired IRT ($R=32$) & 37.32 & 37.34 & 135.6 & 34.57 & Native child \\
Paired IRT ($R=64$) & 74.94 & 74.96 & 130 & 69.52 & Native child \\
Paired IRT ($R=8$) & 9.213 & 9.218 & 99.7 & 8.444 & Native child \\
Paired SBS ($R=64$) & 3234 & 3234 & 117 & 3229 & Native child \\
64 passive, First-order C16 (active) & 2.826 & 2.829 & 393.8 & 1.876 & Loaded solver \\
64 passive, First-order C16 (passive) & 1.217 & 1.221 & 394.3 & 0.2865 & Loaded solver \\
64 passive, First-order C32 (active) & 1.672 & 1.676 & 395 & 0.7413 & Loaded solver \\
64 passive, First-order C32 (passive) & 1.259 & 1.262 & 394.7 & 0.3311 & Loaded solver \\
64 passive, Counts only & 1.065 & 1.067 & 301.7 & 0.1036 & Loaded solver \\
64 passive, First-order moments & 2.871 & 2.875 & 303.6 & 1.943 & Loaded solver \\
64 passive, Physical moments & 2.495 & 2.499 & 302.6 & 1.556 & Loaded solver \\
64 passive, C32 & 1.707 & 1.711 & 394.1 & 0.7698 & Loaded solver \\
64 passive, C64 & 2.525 & 2.529 & 394.1 & 1.6 & Loaded solver \\
64 passive, Learned moments & 2.883 & 2.886 & 302.9 & 1.945 & Loaded solver \\
Source generation (64 clouds) & 1.208 & 1.211 & 59.33 & 1.064 & Native child \\
Source generation (8 clouds) & 0.5261 & 0.5287 & 59.19 & 0.4039 & Native child \\
SBS census ($m=64$, $R=1$) & 18.12 & 18.12 & 59.53 & 17.94 & Native child \\
\bottomrule
\end{tabular}
\end{table}
\clearpage

\begin{table}[p]
\centering
\small
\setlength{\tabcolsep}{4pt}
\caption{CPU time in seconds for each model workload after loading the solver. Columns 0, 1, and 2 identify
the three repetitions. Solver time includes the time to encode the cloud, which the Encode columns
show separately. The full cost records also give the costs of loading inputs, restoring the model,
and writing output.}
\label{tab:reported-cost-loaded-components}
\begin{tabular}{lrrrrrr}
\toprule
Workload & Solver 0 & Solver 1 & Solver 2 & Encode 0 & Encode 1 & Encode 2 \\
\midrule
32 pairs, First-order C16 (active) & 5.741 & 5.744 & 5.744 & 0.01511 & 0.01524 & 0.01541 \\
32 pairs, First-order C16 (passive) & 0.6588 & 0.6666 & 0.6578 & 0.01498 & 0.01514 & 0.0153 \\
32 pairs, First-order C32 (active) & 2.243 & 2.242 & 2.23 & 0.0156 & 0.01518 & 0.01512 \\
32 pairs, First-order C32 (passive) & 0.7061 & 0.7011 & 0.7149 & 0.01512 & 0.01556 & 0.01523 \\
32 pairs, Counts only & 0.2034 & 0.2039 & 0.2044 & 0.000307 & 0.000321 & 0.00031 \\
32 pairs, First-order moments & 2.961 & 2.981 & 2.998 & 0.000776 & 0.000766 & 0.00078 \\
32 pairs, Physical moments & 2.326 & 2.333 & 2.373 & 0.000769 & 0.000777 & 0.000787 \\
32 pairs, C32 & 2.186 & 2.206 & 2.197 & 0.0152 & 0.01512 & 0.0153 \\
32 pairs, C64 & 2.931 & 2.951 & 2.948 & 0.01519 & 0.01548 & 0.01539 \\
32 pairs, Learned moments & 2.989 & 3.016 & 3.024 & 0.000768 & 0.000771 & 0.000793 \\
64 passive, First-order C16 (active) & 1.889 & 1.873 & 1.876 & 0.01876 & 0.01904 & 0.01924 \\
64 passive, First-order C16 (passive) & 0.2902 & 0.2898 & 0.2865 & 0.01841 & 0.01895 & 0.01835 \\
64 passive, First-order C32 (active) & 0.7304 & 0.7464 & 0.7413 & 0.01895 & 0.01861 & 0.01867 \\
64 passive, First-order C32 (passive) & 0.326 & 0.3314 & 0.3311 & 0.01849 & 0.01884 & 0.01864 \\
64 passive, Counts only & 0.1035 & 0.1034 & 0.1036 & 0.000288 & 0.00029 & 0.000296 \\
64 passive, First-order moments & 1.931 & 1.935 & 1.943 & 0.000707 & 0.00071 & 0.000703 \\
64 passive, Physical moments & 1.529 & 1.525 & 1.556 & 0.000727 & 0.000722 & 0.0007 \\
64 passive, C32 & 0.7732 & 0.7591 & 0.7698 & 0.01885 & 0.01891 & 0.01849 \\
64 passive, C64 & 1.575 & 1.584 & 1.6 & 0.01911 & 0.01859 & 0.01847 \\
64 passive, Learned moments & 1.947 & 1.939 & 1.945 & 0.000687 & 0.000703 & 0.000706 \\
\bottomrule
\end{tabular}
\end{table}
\clearpage

\begin{table}[p]
\centering
\small
\setlength{\tabcolsep}{4pt}
\caption{CPU time and elapsed wall time in seconds for each Geant4 child process. Columns 0, 1, and 2
identify the three repetitions. The parent Python process adds overhead to the complete timings.
We timed source generation separately. Bank-mean estimates include the cost of acquiring 8 or 64
sources, as required, from the same repetition.}
\label{tab:reported-cost-native-components}
\begin{tabular}{lrrrrrr}
\toprule
Workload & CPU 0 & CPU 1 & CPU 2 & Wall 0 & Wall 1 & Wall 2 \\
\midrule
IRT sample ($m=8$, $R=64$) & 4.113 & 4.13 & 4.147 & 4.117 & 4.134 & 4.151 \\
IRT sample ($m=8$, $R=8$) & 0.7804 & 0.7923 & 0.8063 & 0.7819 & 0.794 & 0.8079 \\
IRT census ($m=64$, $R=8$) & 4.109 & 4.104 & 4.12 & 4.113 & 4.108 & 4.124 \\
SBS sample ($m=8$, $R=64$) & 145.4 & 146.6 & 147.5 & 145.4 & 146.6 & 147.6 \\
SBS sample ($m=8$, $R=8$) & 17.52 & 17.85 & 17.93 & 17.6 & 17.85 & 17.94 \\
SBS census ($m=64$, $R=8$) & 140.4 & 140.3 & 140.8 & 140.4 & 140.3 & 140.8 \\
Paired IRT ($R=1$) & 1.297 & 1.299 & 1.307 & 1.299 & 1.3 & 1.308 \\
Paired IRT ($R=32$) & 33.95 & 34.33 & 34.57 & 33.96 & 34.35 & 34.58 \\
Paired IRT ($R=64$) & 68.88 & 69.5 & 69.52 & 68.92 & 69.54 & 69.55 \\
Paired IRT ($R=8$) & 8.345 & 8.428 & 8.444 & 8.349 & 8.432 & 8.447 \\
Paired SBS ($R=64$) & 3212 & 3225 & 3229 & 3212 & 3225 & 3229 \\
Source generation (64 clouds) & 1.059 & 1.065 & 1.064 & 1.063 & 1.068 & 1.067 \\
Source generation (8 clouds) & 0.4003 & 0.4076 & 0.4039 & 0.4017 & 0.4093 & 0.4055 \\
SBS census ($m=64$, $R=1$) & 17.73 & 17.87 & 17.94 & 17.73 & 17.87 & 17.94 \\
\bottomrule
\end{tabular}
\end{table}
\clearpage

\begin{table}[p]
\centering
\small
\setlength{\tabcolsep}{4pt}
\caption{Conditional forecasts at 400\,eV. Raw response NMSE compares predictions with SBS on 64
confirmation families (256 cases). CPU times cover the first eight families (32 cases) and give
the median and quartiles of three runs in fresh processes. Learned predictors use the preselected
fit 1701. Physical moments are unfitted. IRT error at $R=1,8,32$ paths per branch is estimated
from the $R=64$ simulations.}
\label{tab:reported-accuracy-cost-join}
\begin{tabular}{lrrr}
\toprule
Predictor / native workload & Accuracy [95\% CI] & CPU median s & CPU Q1--Q3 s \\
\midrule
Counts & 0.2565 [0.2346, 0.2795] & 1.124 & 1.123--1.132 \\
Physical moments & 1.058 [1.014, 1.103] & 3.262 & 3.254--3.276 \\
Paired IRT ($R=1$) & 1.376 [1.354, 1.398] & 1.512 & 1.508--1.522 \\
Paired IRT ($R=32$) & 0.0892 [0.08376, 0.09462] & 37.07 & 36.87--37.2 \\
Paired IRT ($R=64$) & 0.06846 [0.06315, 0.07387] & 74.92 & 74.6--74.93 \\
Paired IRT ($R=8$) & 0.2137 [0.2077, 0.2199] & 9.195 & 9.166--9.204 \\
Paired SBS ($R=64$) & Reference simulator & 3230 & 3224--3232 \\
C32 & 0.1255 [0.1088, 0.1446] & 3.125 & 3.118--3.126 \\
C64 & 0.1718 [0.1411, 0.2129] & 3.87 & 3.868--3.879 \\
Learned moments & 0.0672 [0.0602, 0.07442] & 3.937 & 3.934--3.943 \\
\bottomrule
\end{tabular}
\end{table}
\clearpage

\begin{table}[p]
\centering
\small
\setlength{\tabcolsep}{4pt}
\caption{Error and CPU cost of estimating the mean passive OH yield at 20\,ns across the 400\,eV bank. In
the row labels, $m$ counts sources sampled by SBS and $R$ counts paths per sampled source. Learned
models use the preselected first fit. CPU time sums the costs of acquiring sources, generating
continuations, and making any model predictions within the same repetition. MSE is in
$(\mathrm{molecules}/100\,\mathrm{eV})^2$. Table~\ref{tab:yield-main} compares the two allocations
at 64 and 512 paths.}
\label{tab:reported-utility-cost-join}
\begin{tabular}{lrrr}
\toprule
Allocation / predictor & MSE [95\% CI] & Charged CPU median s & CPU Q1--Q3 s \\
\midrule
Counts only ($m=8$, $R=64$) & 0.006886 [0.004767, 0.008915] & 149.4 & 148.8--149.9 \\
Counts only ($m=8$, $R=8$) & 0.008691 [0.006556, 0.01072] & 20.29 & 20.16--20.34 \\
Physical moments ($m=8$, $R=64$) & 0.001863 [0.001343, 0.002402] & 150.8 & 150.2--151.3 \\
Physical moments ($m=8$, $R=8$) & 0.003667 [0.003129, 0.004201] & 21.69 & 21.56--21.75 \\
Native census ($m=64$, $R=1$) & 0.002062 [0.001979, 0.002145] & 19.26 & 19.21--19.29 \\
Native census ($m=64$, $R=8$) & 0.0002577 [0.0002474, 0.0002681] & 142.2 & 142.1--142.4 \\
Native sample ($m=8$, $R=64$) & 0.008414 [0.005767, 0.011] & 147.6 & 147.1--148.1 \\
Native sample ($m=8$, $R=8$) & 0.01022 [0.007567, 0.01279] & 18.56 & 18.42--18.6 \\
C32 ($m=8$, $R=64$) & 0.001676 [0.001193, 0.00216] & 150 & 149.5--150.5 \\
C32 ($m=8$, $R=8$) & 0.003481 [0.002988, 0.003966] & 20.93 & 20.81--20.98 \\
C64 ($m=8$, $R=64$) & 0.002016 [0.001428, 0.002606] & 150.8 & 150.3--151.3 \\
C64 ($m=8$, $R=8$) & 0.003821 [0.003222, 0.004417] & 21.75 & 21.61--21.8 \\
Learned moments ($m=8$, $R=64$) & 0.001118 [0.0007537, 0.001548] & 151.2 & 150.7--151.7 \\
Learned moments ($m=8$, $R=8$) & 0.002922 [0.00254, 0.003367] & 22.13 & 21.99--22.17 \\
\bottomrule
\end{tabular}
\end{table}
\clearpage

\begin{table}[p]
\centering
\small
\setlength{\tabcolsep}{4pt}
\caption{Numerical checks on the confirmation and long-horizon forecasts. All 240 confirmation checks
passed before we examined prediction errors. All 60 long-horizon checks passed at the prespecified
numerical settings. The audit records give each threshold and comparison.}
\label{tab:reported-numerical-qualifications}
\begin{tabular}{lrr}
\toprule
Qualification & Comparisons & Passed \\
\midrule
Independent confirmation checks & 240 & 240 \\
Long-horizon checks & 60 & 60 \\
\bottomrule
\end{tabular}
\end{table}
\clearpage

\clearpage
\section{Training and validation trajectories}
\label{app:training-diagnostic}

\begin{figure}[!ht]
\centering
\includegraphics[width=0.96\linewidth]{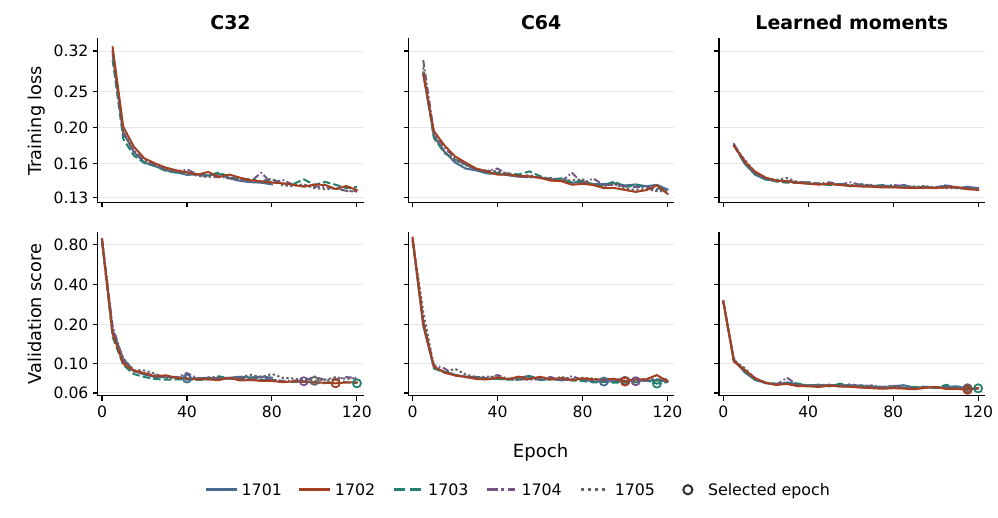}
\caption{Training loss and validation score for the 15 fits at 200\,eV, shown on separate logarithmic scales. Lines join the values recorded every five epochs for each seed. Training values average the six batch losses from the recorded epoch. Open circles mark the selected checkpoints. C32 seed 1701 stopped at epoch 80. All other runs reached epoch 120.}
\label{fig:training-diagnostic}
\end{figure}

Training and checkpoint selection used different losses. During training we averaged the squared standardized count errors across both branch trajectories and added one-half the terminal-response loss. Each target used four native paths per arm. For validation every five epochs, we used all 32 validation families and 16 paths per arm. The validation score gave full weight to the terminal-response loss and one-tenth weight to the trajectory loss across postinitial queries. Figure~\ref{fig:training-diagnostic} compares the five fixed fits per method within each loss.

All 15 fits reached epoch 80. Between epochs 5 and 80, median training loss fell from 0.321 to 0.143 for C32, from 0.282 to 0.143 for C64, and from 0.179 to 0.139 for learned moments. At epoch 80, the last epoch common to all fits, their median validation scores were 0.07598, 0.07594 and 0.06707, respectively. The two pooled states thus had similar validation scores at that epoch. Their selected checkpoints differed in transfer error at 400\,eV (Table~\ref{tab:fresh-transfer-base-held}).

Among the five moment fits, seed 1702 had the lowest selected validation score (0.06342) and the highest held-out response error at all three energies. It selected epoch 115, as did three of the other four fits. At 400\,eV seed 1702 had the highest NMSE on each of the four actions. The largest contributor to its excess over the other fits was the 1\,ns addition at $r=50$\,nm. Its NMSE on that action was 0.1832, compared with a mean of 0.0907 for the other four fits. With equal action weights, this one action accounted for 52\% of the excess.

\clearpage
\section{Changing the addition schedule with frozen predictors}
\label{app:action-transfer}

The source-transfer experiment changed the initial and incoming clouds while keeping addition times and offsets familiar. Here we tested new addition schedules at the training energy of 200\,eV. This comparison is exploratory. The 64 archived families were distinct from the training, validation and fresh source-transfer families, but earlier predictors had been tested on them (Appendix~\ref{sec:earlier-actions}). Before scoring the current predictors, we fixed the plan, model weights, source inputs, native reference, error measure and family resampling procedure.

Within each family we kept the initial cloud, the three incoming clouds and their total molecule counts fixed. At each offset ($r=0$, 10 or 50\,nm), we also held particle positions fixed while changing when the incoming clouds arrived. When they arrived together at 3 or 5\,ns, the two times and three offsets gave six cases per family (384 total). When they arrived one at a time at 1, 5 and 10\,ns, the three offsets gave three cases per family (192 total). The fitted models had seen only single additions at 1 or 10\,ns during optimization. The archived reference used 64 independent SBS continuations per branch and case, or 73\,728 paths across the two schedule groups.

We evaluated the five fixed fits of each learned predictor (C32, C64 and learned moments) and the unfitted physical moments. For each case we scored the 20\,ns pulse-minus-passive response against the native mean using the seven species scales fixed at training. Within each schedule group, we averaged losses over fits and actions for each family, then over families. We paired the methods within families and drew 4000 whole-family bootstrap samples. The descriptive 95\% intervals condition on the fitted models and are unadjusted across the four contrasts (Table~\ref{tab:action-intervals}).

We repeated every action on the first eight families for each predictor with the refined solver. Across the two schedule groups, the largest standardized RMS change in terminal response was $1.09\times10^{-6}$. The largest absolute changes were $2.90\times10^{-5}$ molecules for a terminal response and $4.02\times10^{-5}$ molecules for a branch trajectory. As a check on familiar actions, we repeated the 128 confirmation cases at 200\,eV for the preselected fit of each learned method and for physical moments. Their counts and water ledgers matched the archived trajectories at every query. An independent calculation reconstructed the reference means from all 73\,728 archived paths and reproduced the family-level losses and paired intervals.

\begin{table}[!htbp]
\centering\small
\caption{Raw 20\,ns response NMSE for new pulse schedules at 200\,eV. Both columns use the same 64 families. At each offset, each schedule uses the same initial and incoming molecules in the same positions. Each learned-model entry averages losses across the five frozen fits.}
\label{tab:action-transfer}
\begin{tabular}{@{}lrr@{}}
\toprule
Predictor & One addition at 3 or 5\,ns & Three additions at 1, 5 and 10\,ns\\
\midrule
C32 & 0.0279 & 0.0687\\
C64 & 0.0288 & 0.0754\\
Learned moments & 0.0234 & 0.0255\\
Physical moments & 0.2030 & 0.2699\\
\bottomrule
\end{tabular}
\end{table}

\begin{table}[!htbp]
\centering\small
\caption{Paired differences in 20\,ns response NMSE for the schedule tests at 200\,eV (learned moments minus comparator). Negative differences favor moments. Each interval resamples the same 64 families and conditions on the five fitted models per method.}
\label{tab:action-intervals}
\begin{tabular}{@{}llrr@{}}
\toprule
Schedule & Comparator & Difference & Descriptive 95\% interval\\
\midrule
One addition at an unseen delay & C64 & $-0.0054$ & $-0.0083, -0.0028$\\
One addition at an unseen delay & C32 & $-0.0046$ & $-0.0073, -0.0019$\\
Three successive additions & C64 & $-0.0499$ & $-0.0570, -0.0429$\\
Three successive additions & C32 & $-0.0432$ & $-0.0488, -0.0379$\\
\bottomrule
\end{tabular}
\end{table}

\clearpage
\section{Post-hoc mechanistic ablations and diagnostics}
\label{app:mechanism}
The following exploratory analyses reuse the archived clouds, stochastic reference paths, species scales and fixed fits. Whole-family bootstrap intervals are descriptive and unadjusted.

\begingroup
\makeatletter
\renewenvironment{table}[1][]{\par\addvspace{10pt}\noindent\begin{minipage}{\linewidth}\def\@captype{table}}{\end{minipage}\par\addvspace{10pt}}
\renewenvironment{figure}[1][]{\par\addvspace{10pt}\noindent\begin{minipage}{\linewidth}\def\@captype{figure}}{\end{minipage}\par\addvspace{10pt}}
\makeatother

\subsection{What each new fit changes}
\label{sec:v13-ablation-design}
We designed these controls after seeing the original comparisons. For each control, we fixed the architecture, optimizer, validation rule and scoring code before evaluating its selected checkpoints on the archived transfer trajectories. The molecular inputs and reference histories had already been collected. All fits used the same 128 training and 32 validation families at 200\,eV and the original species scales. Seeds 1701--1705 defined the five fits of each method. Each had up to 120 epochs, with six batches of 128 cases per epoch and a learning rate of 0.002. The original objective and intervention-aware checkpoint selection applied except for the passive training loss described below.

\paragraph{Network capacity at fixed pooled state.}
Full-width C64 with weight decay retains the original 48-unit hidden layers. The narrow variant reduces all hidden layers to 16 units while retaining the same 213 rate inputs, recurrent vectors and tensors, encoder, jump rule and shared isotropic encounter scale. Both use AdamW with weight decay $10^{-3}$. At the stated learning rate, the multiplicative contribution of this decay alone over 720 updates is $(1-2\times10^{-6})^{720}=0.9986$. The two new C64 predictors thus vary width at a common regularization setting. Comparing full-width AdamW to the original Adam fit tests the change of optimizer configuration at full width.

\paragraph{Encounter geometry at fixed moment state.}
For the shared-overlap control, let $N=\sum_s n_s^+$ and $p_s=n_s^+/\max(N,\epsilon)$. We use a count-weighted variance about the combined particle centroid,
\[
 \sigma^2_{\mathrm{tot}}=\frac{L^2}{3}\max\!\left(0,\sum_s p_s\operatorname{tr} T_s-
                   \left\|\sum_s p_s\mu_s\right\|^2\right).
\]
We replace each pair overlap by $[4\pi(\sigma^2_{\mathrm{tot}}+R_r^2)]^{-3/2}$ in the reaction hazard and in the corresponding log-overlap input to the rate head. Species-specific moments still determine product placement and diffusion, and the solver still evolves the extensive moments. Centering the pooled variance removes dependence on displacement of the whole cloud from the original coordinate origin. The contrast tests the encounter rule and its inputs to the learned correction together.

\paragraph{Training targets and absolute time.}
Passive moment fits used two independent sets of four passive paths per training case. We averaged trajectory loss against the two passive means, spending the same eight-path budget as the active fits. Validation still used interventions. A second group retained both active training targets and removed the two clock inputs from the rate head. The rate network then had 30 inputs and 4\,281 weights. The physical clock still governed diffusion, integration and the arrival of every pulse.

For each new checkpoint we compared the base integrator with a refined integrator on all four actions of the first eight confirmation families at each energy. We repeated the comparison across all nine schedules on the first eight action-transfer families. The tests used both branch trajectories and the terminal response, with limits of $10^{-3}$ standardized RMS and 0.05 molecule in absolute counts. We required conservation residuals at most $10^{-8}$, counts of at least $-10^{-8}$ and identical branches before addition. We formed paired differences within families and retained all actions and five fits when resampling. The associated 95\% intervals describe uncertainty across these families, conditional on the fitted weights and without a multiplicity adjustment.

\subsection{Cloud size, action-specific transfer and finite-path IRT error}

\label{app:paired-diagnostics}

Each confirmation family supplied one initial cloud and three incoming clouds. Table~\ref{tab:v13-cloud} describes the geometry at 0.02\,ns. For the incoming pulse, the radius includes the specified displacement of each cloud.

\begin{table}[htbp]
\centering
\caption{Cloud size in the confirmation families. Entries give the median and interquartile range. Counts include all seven tracked species. The RMS radius is measured about the particle centroid of the initial cloud or the combined pulse.}
\label{tab:v13-cloud}
\begin{tabular}{@{}lrrr@{}}
\toprule
Quantity & 100 eV & 200 eV & 400 eV \\
\midrule
Initial count & 30 [28, 32] & 60 [56, 63.25] & 117.5 [111, 122] \\
Pulse count & 48 [44, 50] & 90.5 [86, 93] & 173 [168, 178.25] \\
Initial RMS radius (nm) & 10.1 [9.1, 11.0] & 10.0 [9.6, 10.8] & 11.7 [10.3, 13.2] \\
Pulse RMS radius (nm), $r=0$ & 9.0 [8.3, 10.2] & 9.7 [9.1, 10.2] & 11.4 [10.9, 12.1] \\
Pulse RMS radius (nm), $r=50$ & 41.7 [40.6, 42.8] & 42.2 [40.9, 43.1] & 42.8 [40.7, 43.8] \\
\bottomrule
\end{tabular}
\end{table}

For model $m$ at energy $E$, relative error is $L_{m,E}/L_{0,E}$, where $L_{0,E}$ is the raw NMSE of predicting zero pulse response on the same families. The denominator contains the squared SBS mean response about zero and its sampling error. Table~\ref{tab:v13-relative} uses a separate denominator for each bank and energy. We held all fits fixed, sampled whole families within each energy 4,000 times, and used the same draws for numerator and denominator. All intervals in this subsection are descriptive, unadjusted percentile intervals.

\begin{table}[htbp]
\centering
\caption{Raw response NMSE as a fraction of the zero-response error. For learned methods each entry averages the losses of five fitted predictors. The development and confirmation families are disjoint. The supplement gives intervals and individual-fit estimates.}
\label{tab:v13-relative}
\begin{tabular}{@{}lrrr@{}}
\toprule
Predictor & 100 eV & 200 eV & 400 eV \\
\midrule
\multicolumn{4}{@{}l}{\textit{Confirmation}}\\
Zero response & 1.0000 & 1.0000 & 1.0000 \\
Added counts only & 0.5837 & 0.6293 & 0.6419 \\
Counts-only dynamics & 0.0473 & 0.0361 & 0.0283 \\
C32 & 0.0259 & 0.0160 & 0.0156 \\
C64 & 0.0260 & 0.0166 & 0.0210 \\
Learned moments & 0.0260 & 0.0106 & 0.0090 \\
Physical moments & 0.1142 & 0.1214 & 0.1194 \\
IRT (64 paths per branch) & 0.0212 & 0.0133 & 0.0077 \\
C16, pulse-trained, first-order training & 0.0277 & 0.0167 & 0.0243 \\
C16, passive-trained, first-order training & 0.1770 & 0.1341 & 0.1049 \\
C32, pulse-trained, first-order training & 0.0314 & 0.0184 & 0.0209 \\
C32, passive-trained, first-order training & 0.1758 & 0.1362 & 0.1134 \\
Learned moments, first-order training & 0.0260 & 0.0107 & 0.0094 \\
\midrule
\multicolumn{4}{@{}l}{\textit{Development}}\\
Zero response & 1.0000 & 1.0000 & 1.0000 \\
Added counts only & 0.5756 & 0.6778 & 0.6810 \\
Counts-only dynamics & 0.0573 & 0.0340 & 0.0275 \\
Physical moments & 0.1165 & 0.1225 & 0.1166 \\
IRT (64 paths per branch) & 0.0189 & 0.0139 & 0.0083 \\
C16, pulse-trained, first-order training & 0.0347 & 0.0165 & 0.0213 \\
C16, passive-trained, first-order training & 0.1771 & 0.1516 & 0.1056 \\
C32, pulse-trained, first-order training & 0.0339 & 0.0201 & 0.0183 \\
C32, passive-trained, first-order training & 0.1763 & 0.1547 & 0.1139 \\
Learned moments, first-order training & 0.0253 & 0.0104 & 0.0093 \\
Pulse alone (SBS, 64 paths) & 0.0316 & 0.0264 & 0.0218 \\
\bottomrule
\end{tabular}
\end{table}

The mean of the 100 and 400 eV relative differences (learned moments minus C64) was $-0.00595$ (95\% interval $-0.00916$ to $-0.00324$). Table~\ref{tab:v13-actions} gives the four contrasts in the original NMSE units. Subtracting the SBS reference variance from each loss leaves its paired difference unchanged.

\begin{table}[htbp]
\centering
\caption{Confirmation NMSE by pulse time and offset. The last column gives learned moments minus C64 with a paired 95\% interval.}
\label{tab:v13-actions}
\begin{tabular}{@{}rrrrrr@{}}
\toprule
eV & ns & nm & Moments & C64 & Paired difference \\
\midrule
100 & 1 & 0 & 0.0141 & 0.0217 & $-0.0076\;[-0.0109,-0.0044]$ \\
100 & 1 & 50 & 0.0179 & 0.0164 & $0.0015\;[-0.0034,0.0063]$ \\
100 & 10 & 0 & 0.0211 & 0.0200 & $0.0011\;[-0.0032,0.0056]$ \\
100 & 10 & 50 & 0.0186 & 0.0135 & $0.0051\;[0.0011,0.0091]$ \\
400 & 1 & 0 & 0.0865 & 0.1443 & $-0.0578\;[-0.0849,-0.0324]$ \\
400 & 1 & 50 & 0.1092 & 0.2072 & $-0.0979\;[-0.1660,-0.0451]$ \\
400 & 10 & 0 & 0.0483 & 0.1837 & $-0.1354\;[-0.2030,-0.0856]$ \\
400 & 10 & 50 & 0.0765 & 0.2081 & $-0.1316\;[-0.1928,-0.0802]$ \\
\bottomrule
\end{tabular}
\end{table}

\begin{table}[htbp]
\centering
\caption{Learned moments versus IRT on confirmation families. Both methods subtract the SBS reference variance in the realized forecast comparison. The infinite-path IRT estimate also subtracts the IRT sampling variance. The last column gives the paired moment-model minus IRT difference and 95\% interval.}
\label{tab:v13-realized}
\begin{tabular}{@{}rllrrr@{}}
\toprule
eV & Moment fit & IRT target & Moments & IRT & Paired difference \\
\midrule
200 & Seed 1701 & 64-path forecast & 0.0132 & 0.0209 & $-0.0077\;[-0.0130,-0.0020]$ \\
200 & Seed 1701 & Infinite-path bias & 0.0132 & 0.0101 & $0.0032\;[-0.0021,0.0087]$ \\
200 & Five-fit mean & 64-path forecast & 0.0144 & 0.0209 & $-0.0065\;[-0.0119,-0.0007]$ \\
200 & Five-fit mean & Infinite-path bias & 0.0144 & 0.0101 & $0.0044\;[-0.0010,0.0100]$ \\
400 & Seed 1701 & 64-path forecast & 0.0460 & 0.0473 & $-0.0013\;[-0.0099,0.0073]$ \\
400 & Seed 1701 & Infinite-path bias & 0.0460 & 0.0265 & $0.0195\;[0.0109,0.0279]$ \\
400 & Five-fit mean & 64-path forecast & 0.0589 & 0.0473 & $0.0117\;[0.0024,0.0212]$ \\
400 & Five-fit mean & Infinite-path bias & 0.0589 & 0.0265 & $0.0324\;[0.0232,0.0420]$ \\
\bottomrule
\end{tabular}
\end{table}

\subsection{Extrapolation of the learned clock}
\label{app:clock-diagnostic}

The rate correction reads the absolute time in two forms, $\log(1+t)$ and $\log^2(1+t)/10$, with $t$ in ns. The second feature grows from 0.93 at 20 ns to 4.77 at $1\,\mu$s. To isolate the effect of this input, we reran fit 1701 on the four 400 eV long-horizon families with $\min(t,20)$ as the time supplied to the rate head. Both branches continued to diffuse and react for the full integration time. We used both archived actions per family and the original SBS reference means from 32 histories per branch.

Holding the clock at 20 ns reduced the $1\,\mu$s response NMSE in all four families (Fig.~\ref{fig:clock-horizon}). The mean fell from 0.961 to 0.511. The paired change was $-0.450$ (descriptive 95\% interval $[-0.580,-0.321]$, enumerating the 256 ordered whole-family bootstrap samples). Unfitted moments reached 0.388 on the same cases. The median multiplier for $\mathrm{OH}+\mathrm{H}$ on the pulse branch rose from 2.17 at 20 ns to 41.29 at $1\,\mu$s. With the clock held at 20 ns it reached 6.78. The remaining change reflects the evolving input state.

Halving the maximum step and positivity fraction and dividing both error tolerances by four changed the clamped trajectories by at most $2.81\times10^{-5}$ molecules. The largest seven-species standardized RMS difference at any case and time was $3.31\times10^{-6}$ for branch counts and $1.92\times10^{-6}$ for the response. Both integrations preserved nonnegative counts, with exact branch agreement before the pulse and a maximum conservation residual below $2.11\times10^{-12}$.

\begin{figure}[htbp]
\centering
\includegraphics[width=0.97\linewidth]{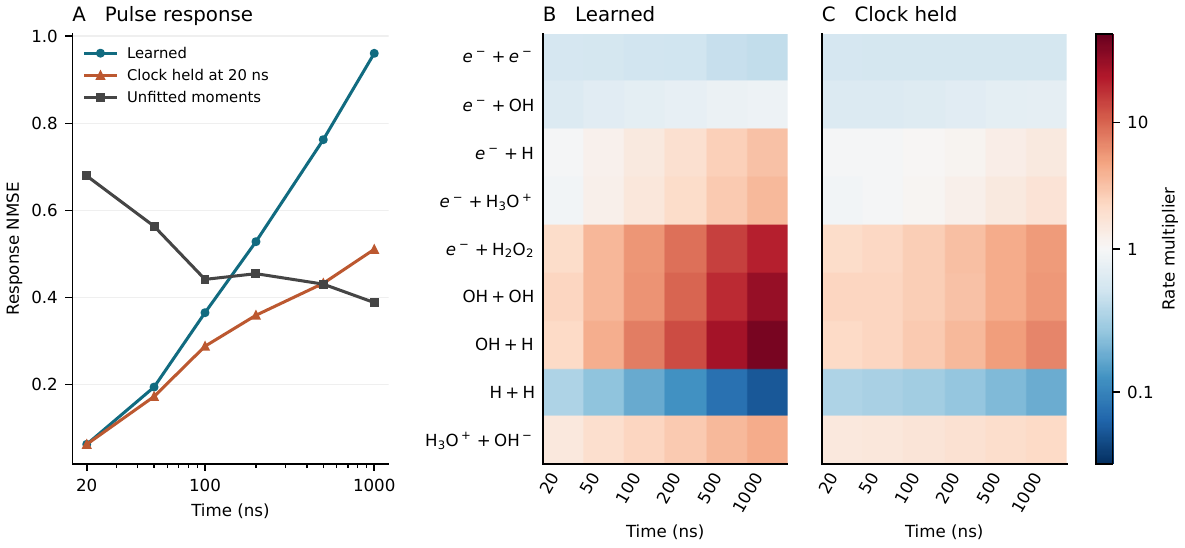}
\caption{Holding the learned clock at its training endpoint. Panel A gives the seven-species raw response NMSE for both actions on four 400 eV families. Panels B and C give the median multiplier on the pulse branch across the eight cases. Each cell refers to one reaction at one query time.}
\label{fig:clock-horizon}
\end{figure}

\subsection{Rate corrections after a separated pulse}
\label{app:seed-diagnostic}

We compared the five moment fits on the 64 fresh 400 eV families given three incoming clouds at 1 ns with 50 nm spacing. At injection, every fit added the known species counts and produced the same immediate response. The excess NMSE of fit 1702 relative to the other four reached 0.066 at 2 ns, peaked at 0.170 at the 8 ns query, and fell to 0.092 at 20 ns (Fig.~\ref{fig:seed-jump}).

The $\mathrm{H}+\mathrm{H}$ correction distinguished the fits at injection. Within each family we divided the multiplier in fit 1702 by the geometric mean in the other four fits. The median ratio changed from 0.82 immediately before injection to 14.86 immediately after. The mean species covariance trace in fit 1702 rose by a median of $810\ \mathrm{nm}^2$ across families, counting species present on both sides of the jump. Larger trace increases had little association with larger excess error at 20 ns (Pearson $r=-0.009$). The multiplier identifies where the fits diverged. Several rates and state features changed together, so this comparison alone cannot assign the later error to one reaction.

At 20 ns the mean signed response errors in molecules for fit 1702 and the other four fits were $+3.30$ and $+0.99$ for OH, $-0.66$ and $-1.27$ for the hydrated electron, $+0.49$ and $-0.74$ for hydronium, and $-1.43$ and $-0.75$ for hydrogen peroxide. After division by the training species scales, hydroxide accounted for 74\% of the total excess squared error. The supplementary traces give the distributions for all seven species and nine reactions.

\begin{figure}[htbp]
\centering
\includegraphics[width=0.97\linewidth]{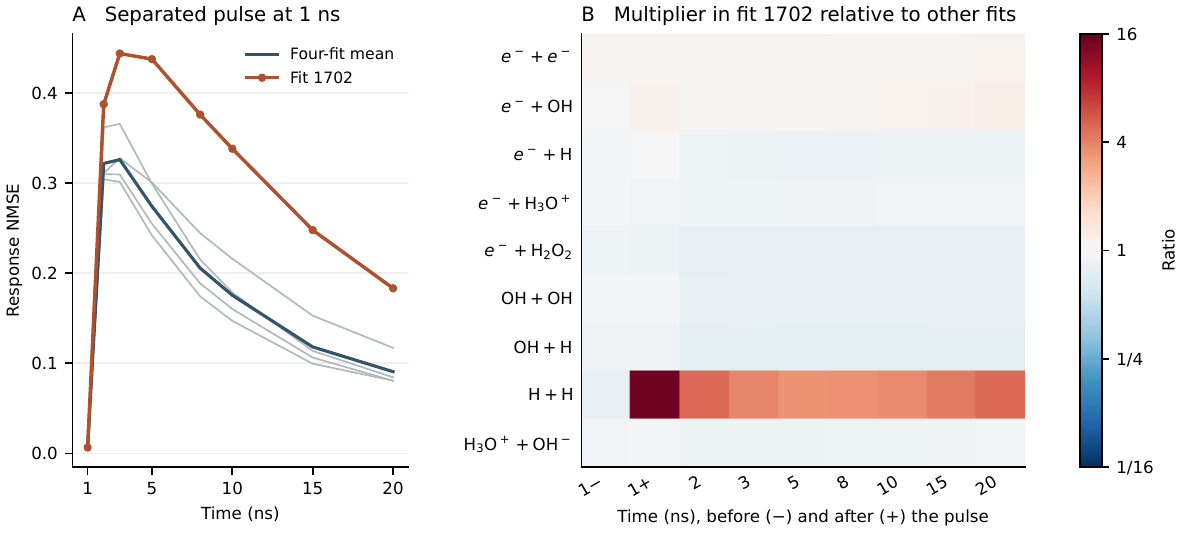}
\caption{Response error and rate corrections for a separated pulse at 400 eV. Panel A shows fit 1702 and the four other moment fits (thin lines) on 64 families. Panel B gives each pulse-branch multiplier in fit 1702 relative to the geometric mean of the other fits on the same family. Colors show the median of these ratios. The two columns marked 1 ns show the states immediately before and after injection.}
\label{fig:seed-jump}
\end{figure}

\subsection{Fit-level results, computation and numerical accuracy}
The following tables give each fit's held-out scores, validation checkpoint and measured training cost. Numerical checks compare the same trained weights under two integration settings. The archived family-level records include all paired estimates across the five cohorts.
\begin{table}[!htbp]\centering
\caption{Change in response NMSE under each exploratory control. Negative values favor the first predictor named. Each interval resamples whole families and conditions on the five fixed fits.}
\label{tab:v13-mechanism-contrasts}
\setlength{\tabcolsep}{3pt}
\begin{tabular}{@{}lrrrr@{}}\toprule
 & \multicolumn{2}{c}{400 eV, familiar pulses} & \multicolumn{2}{c}{200 eV, three pulses}\\
Comparison & Difference & 95\% interval & Difference & 95\% interval\\\midrule
Full-width decay minus C64 & $0.0107$ & $[0.0070,0.0154]$ & $0.0008$ & $[0.0004,0.0012]$\\
Narrow minus full width, both with decay & $-0.0342$ & $[-0.0620,-0.0125]$ & $0.0010$ & $[-0.0036,0.0052]$\\
Shared overlap minus learned moments & $0.0543$ & $[0.0392,0.0696]$ & $0.0401$ & $[0.0302,0.0511]$\\
Passive minus active moments & $0.4018$ & $[0.3751,0.4280]$ & $0.0918$ & $[0.0828,0.1008]$\\
Clock-free minus learned moments & $-0.0019$ & $[-0.0067,0.0033]$ & $-0.0007$ & $[-0.0014,0.0001]$\\
\bottomrule\end{tabular}\end{table}

\begin{table}[htbp]
\centering
\setlength{\tabcolsep}{4pt}
\begin{tabular}{lrrrrrrr}
\toprule
Seed & Selected & Final & 100 eV & 200 eV & 400 eV & New delays & Three pulses \\
\midrule
1701 & 120 & 120 & 0.0180 & 0.0407 & 0.2267 & 0.0282 & 0.0638 \\
1702 & 100 & 120 & 0.0194 & 0.0436 & 0.2059 & 0.0291 & 0.0944 \\
1703 & 115 & 120 & 0.0166 & 0.0357 & 0.1741 & 0.0267 & 0.0667 \\
1704 & 105 & 120 & 0.0171 & 0.0414 & 0.1888 & 0.0302 & 0.0601 \\
1705 & 100 & 120 & 0.0184 & 0.0396 & 0.1871 & 0.0299 & 0.0960 \\
\bottomrule
\end{tabular}
\caption{Fixed-fit terminal response NMSE for full-width C64 with decay. New delays and three successive pulses use separate 200 eV action schedules. Selected and final columns are training epochs.}
\label{tab:c64wd-fit-scores}
\end{table}

\begin{table}[htbp]
\centering
\setlength{\tabcolsep}{4pt}
\begin{tabular}{lrrrr}
\toprule
Seed & Validation score & CPU (h) & Wall (h) & Peak RSS (GiB) \\
\midrule
1701 & 0.0727 & 5.62 & 5.79 & 2.08 \\
1702 & 0.0732 & 7.14 & 7.32 & 2.19 \\
1703 & 0.0706 & 8.06 & 8.23 & 1.83 \\
1704 & 0.0737 & 4.91 & 5.07 & 1.91 \\
1705 & 0.0742 & 8.81 & 8.98 & 1.47 \\
\bottomrule
\end{tabular}
\caption{Observed fitting resources for full-width C64 with decay. Timers exclude setup and include training, validation and completed optimizer-step checkpoints.}
\label{tab:c64wd-resources}
\end{table}

\begin{table}[htbp]
\centering
\setlength{\tabcolsep}{4pt}
\begin{tabular}{lrrrrrr}
\toprule
Bank & Response RMS & Path RMS & Count change & Budget drift & Min. count & Pre-pulse \\
\midrule
100 eV & 3.9e-07 & 2.5e-07 & 1.2e-05 & 4.4e-13 & 0 & 0 \\
200 eV & 6.8e-07 & 5.1e-07 & 2.7e-05 & 9.7e-13 & 0 & 0 \\
400 eV & 1.7e-06 & 1.2e-06 & 5.2e-05 & 1.9e-12 & 0 & 0 \\
Action transfer & 5.9e-07 & 3.3e-07 & 2.6e-05 & 9.9e-13 & 0 & 0 \\
\bottomrule
\end{tabular}
\caption{Numerical qualification for full-width C64 with decay. Values take maxima over seeds and action groups, except for minimum count. RMS changes use training species scales. Count change is the largest difference in molecules between base and refined solves. Budget drift covers atom, charge and reducing equivalents. Pre-pulse compares forced and passive counts.}
\label{tab:c64wd-qualification}
\end{table}

\begin{table}[htbp]
\centering
\setlength{\tabcolsep}{4pt}
\begin{tabular}{lrrrrrrr}
\toprule
Seed & Selected & Final & 100 eV & 200 eV & 400 eV & New delays & Three pulses \\
\midrule
1701 & 50 & 90 & 0.0218 & 0.0416 & 0.1857 & 0.0312 & 0.0731 \\
1702 & 120 & 120 & 0.0184 & 0.0359 & 0.1727 & 0.0284 & 0.1104 \\
1703 & 115 & 120 & 0.0187 & 0.0397 & 0.1581 & 0.0288 & 0.0730 \\
1704 & 85 & 120 & 0.0183 & 0.0401 & 0.1497 & 0.0291 & 0.0586 \\
1705 & 120 & 120 & 0.0198 & 0.0409 & 0.1456 & 0.0292 & 0.0708 \\
\bottomrule
\end{tabular}
\caption{Fixed-fit terminal response NMSE for narrow C64 with decay. New delays and three successive pulses use separate 200 eV action schedules. Selected and final columns are training epochs.}
\label{tab:c64narrowwd-fit-scores}
\end{table}

\begin{table}[htbp]
\centering
\setlength{\tabcolsep}{4pt}
\begin{tabular}{lrrrr}
\toprule
Seed & Validation score & CPU (h) & Wall (h) & Peak RSS (GiB) \\
\midrule
1701 & 0.0769 & 3.93 & 4.17 & 2.40 \\
1702 & 0.0723 & 4.55 & 4.81 & 2.39 \\
1703 & 0.0733 & 6.23 & 6.54 & 1.74 \\
1704 & 0.0771 & 4.59 & 4.84 & 1.77 \\
1705 & 0.0726 & 4.42 & 4.67 & 2.38 \\
\bottomrule
\end{tabular}
\caption{Observed fitting resources for narrow C64 with decay. Timers exclude setup and include training, validation and completed optimizer-step checkpoints.}
\label{tab:c64narrowwd-resources}
\end{table}

\begin{table}[htbp]
\centering
\setlength{\tabcolsep}{4pt}
\begin{tabular}{lrrrrrr}
\toprule
Bank & Response RMS & Path RMS & Count change & Budget drift & Min. count & Pre-pulse \\
\midrule
100 eV & 3.5e-07 & 2.3e-07 & 1.2e-05 & 3.1e-13 & 0 & 0 \\
200 eV & 9.2e-07 & 6.2e-07 & 3.2e-05 & 8.2e-13 & 0 & 0 \\
400 eV & 2.3e-06 & 1.5e-06 & 7.1e-05 & 1.6e-12 & 0 & 0 \\
Action transfer & 5.5e-07 & 3e-07 & 2.7e-05 & 9.4e-13 & 0 & 0 \\
\bottomrule
\end{tabular}
\caption{Numerical qualification for narrow C64 with decay. Values take maxima over seeds and action groups, except for minimum count. RMS changes use training species scales. Count change is the largest difference in molecules between base and refined solves. Budget drift covers atom, charge and reducing equivalents. Pre-pulse compares forced and passive counts.}
\label{tab:c64narrowwd-qualification}
\end{table}

\begin{table}[htbp]
\centering
\setlength{\tabcolsep}{4pt}
\begin{tabular}{lrrrrrrr}
\toprule
Seed & Selected & Final & 100 eV & 200 eV & 400 eV & New delays & Three pulses \\
\midrule
1701 & 120 & 120 & 0.0222 & 0.0358 & 0.1319 & 0.0355 & 0.0741 \\
1702 & 115 & 120 & 0.0218 & 0.0418 & 0.1435 & 0.0372 & 0.0735 \\
1703 & 120 & 120 & 0.0230 & 0.0437 & 0.1468 & 0.0379 & 0.0521 \\
1704 & 120 & 120 & 0.0217 & 0.0410 & 0.1252 & 0.0385 & 0.0656 \\
1705 & 120 & 120 & 0.0239 & 0.0378 & 0.1247 & 0.0360 & 0.0628 \\
\bottomrule
\end{tabular}
\caption{Fixed-fit terminal response NMSE for moments with shared overlap. New delays and three successive pulses use separate 200 eV action schedules. Selected and final columns are training epochs.}
\label{tab:momentsisooverlap-fit-scores}
\end{table}

\begin{table}[htbp]
\centering
\setlength{\tabcolsep}{4pt}
\begin{tabular}{lrrrr}
\toprule
Seed & Validation score & CPU (h) & Wall (h) & Peak RSS (GiB) \\
\midrule
1701 & 0.0796 & 5.90 & 6.19 & 2.18 \\
1702 & 0.0820 & 6.04 & 6.34 & 2.18 \\
1703 & 0.0804 & 5.97 & 6.28 & 1.57 \\
1704 & 0.0825 & 6.14 & 6.44 & 1.77 \\
1705 & 0.0796 & 5.99 & 6.28 & 2.15 \\
\bottomrule
\end{tabular}
\caption{Observed fitting resources for moments with shared overlap. Timers exclude setup and include training, validation and completed optimizer-step checkpoints.}
\label{tab:momentsisooverlap-resources}
\end{table}

\begin{table}[htbp]
\centering
\setlength{\tabcolsep}{4pt}
\begin{tabular}{lrrrrrr}
\toprule
Bank & Response RMS & Path RMS & Count change & Budget drift & Min. count & Pre-pulse \\
\midrule
100 eV & 5.6e-07 & 4.7e-07 & 1.1e-05 & 1.7e-13 & 0 & 0 \\
200 eV & 1.1e-06 & 9.1e-07 & 3.2e-05 & 3.7e-13 & 0 & 0 \\
400 eV & 1.5e-06 & 1.5e-06 & 6.1e-05 & 6.8e-13 & 0 & 0 \\
Action transfer & 6.1e-07 & 6.5e-07 & 3.5e-05 & 3.4e-13 & 0 & 0 \\
\bottomrule
\end{tabular}
\caption{Numerical qualification for moments with shared overlap. Values take maxima over seeds and action groups, except for minimum count. RMS changes use training species scales. Count change is the largest difference in molecules between base and refined solves. Budget drift covers atom, charge and reducing equivalents. Pre-pulse compares forced and passive counts.}
\label{tab:momentsisooverlap-qualification}
\end{table}

\begin{table}[htbp]
\centering
\caption{Solver and conservation checks across the five passive-trained moments fits. Response and trajectory columns give maximum standardized RMS discrepancies between solver settings. Count gives the largest change in molecules. Budget and redox give maximum invariant drift. Minimum gives the smallest species count and pre-pulse gives the largest discrepancy between forced and passive trajectories before the first pulse. Both temporal rows include the shared internal-step extrema.}\label{tab:moments-passive-numerics}
\begin{tabular}{lrrrrrrr}
\toprule
Cohort & Response & Trajectory & Count & Budget & Redox & Minimum & Pre-pulse \\
\midrule
Three source energies & 2.0e-06 & 2.0e-06 & 5.8e-05 & 6.3e-13 & 2.6e-13 & 0 & 0 \\
Unseen delays & 6.8e-07 & 9.5e-07 & 3.0e-05 & 3.7e-13 & 1.5e-13 & 0 & 0 \\
Three pulses & 1.0e-06 & 9.9e-07 & 4.3e-05 & 3.7e-13 & 1.4e-13 & 0 & 0 \\
\bottomrule
\end{tabular}
\end{table}
\begin{table}[htbp]
\centering
\caption{Raw 20 ns response NMSE for passive-trained moments. The last two columns use the held-out 200 eV temporal bank.}\label{tab:moments-passive-per-fit}
\begin{tabular}{lrrrrr}
\toprule
Seed & 100 eV & 200 eV & 400 eV & New delays & Three pulses \\
\midrule
1701 & 0.0429 & 0.1457 & 0.5295 & 0.0800 & 0.1098 \\
1702 & 0.0528 & 0.1426 & 0.4563 & 0.0701 & 0.2065 \\
1703 & 0.0411 & 0.1326 & 0.4711 & 0.0703 & 0.0805 \\
1704 & 0.0413 & 0.1259 & 0.4229 & 0.0673 & 0.0826 \\
1705 & 0.0488 & 0.1497 & 0.5297 & 0.0841 & 0.1071 \\
\midrule
Mean & 0.0454 & 0.1393 & 0.4819 & 0.0743 & 0.1173 \\
\bottomrule
\end{tabular}
\end{table}
\begin{table}[htbp]
\centering
\caption{Paired NMSE differences for passive-trained moments minus the original active moment fits. Intervals resample whole families and condition on the five fixed fits.}\label{tab:moments-passive-paired}
\begin{tabular}{lrr}
\toprule
Cohort & Difference & 95\% interval \\
\midrule
100 eV & 0.0275 & [0.0245, 0.0305] \\
200 eV & 0.1139 & [0.1030, 0.1254] \\
400 eV & 0.4018 & [0.3751, 0.4280] \\
New delays & 0.0510 & [0.0448, 0.0573] \\
Three pulses & 0.0918 & [0.0828, 0.1008] \\
\bottomrule
\end{tabular}
\end{table}
\begin{table}[htbp]
\centering
\caption{Validation selection and training time for passive-trained moments.}\label{tab:moments-passive-resources}
\begin{tabular}{lrrrrr}
\toprule
Seed & Best epoch & Last epoch & Validation & CPU (h) & Wall (h) \\
\midrule
1701 & 20 & 60 & 0.1565 & 1.28 & 1.32 \\
1702 & 75 & 115 & 0.1478 & 2.48 & 2.55 \\
1703 & 25 & 65 & 0.1453 & 1.39 & 1.43 \\
1704 & 15 & 55 & 0.1378 & 1.18 & 1.22 \\
1705 & 15 & 55 & 0.1576 & 1.20 & 1.24 \\
\bottomrule
\end{tabular}
\end{table}
\begin{table}[htbp]
\centering
\caption{Solver and conservation checks across the five moments without clock inputs fits. Response and trajectory columns give maximum standardized RMS discrepancies between solver settings. Count gives the largest change in molecules. Budget and redox give maximum invariant drift. Minimum gives the smallest species count and pre-pulse gives the largest discrepancy between forced and passive trajectories before the first pulse. Both temporal rows include the shared internal-step extrema.}\label{tab:moments-autonomous-numerics}
\begin{tabular}{lrrrrrrr}
\toprule
Cohort & Response & Trajectory & Count & Budget & Redox & Minimum & Pre-pulse \\
\midrule
Three source energies & 1.7e-06 & 1.7e-06 & 4.9e-05 & 6.8e-13 & 4.5e-13 & 0 & 0 \\
Unseen delays & 5.6e-07 & 8.7e-07 & 3.0e-05 & 3.7e-13 & 1.5e-13 & 0 & 0 \\
Three pulses & 6.9e-07 & 8.7e-07 & 3.3e-05 & 3.7e-13 & 1.2e-13 & 0 & 0 \\
1 $\mu$s horizon & 2.7e-07 & 1.2e-06 & 3.3e-05 & 2.4e-12 & 9.7e-13 & 0 & 0 \\
\bottomrule
\end{tabular}
\end{table}
\begin{table}[htbp]
\centering
\caption{Raw 20 ns response NMSE for moments without clock inputs. The last two columns use the held-out 200 eV temporal bank.}\label{tab:moments-autonomous-per-fit}
\begin{tabular}{lrrrrr}
\toprule
Seed & 100 eV & 200 eV & 400 eV & New delays & Three pulses \\
\midrule
1701 & 0.0171 & 0.0295 & 0.0807 & 0.0245 & 0.0258 \\
1702 & 0.0193 & 0.0293 & 0.0827 & 0.0229 & 0.0249 \\
1703 & 0.0193 & 0.0296 & 0.0745 & 0.0236 & 0.0243 \\
1704 & 0.0182 & 0.0288 & 0.0730 & 0.0243 & 0.0253 \\
1705 & 0.0189 & 0.0302 & 0.0803 & 0.0234 & 0.0240 \\
\midrule
Mean & 0.0186 & 0.0295 & 0.0782 & 0.0238 & 0.0249 \\
\bottomrule
\end{tabular}
\end{table}
\begin{table}[htbp]
\centering
\caption{Paired NMSE differences for moments without clock inputs minus the original active moment fits. Intervals resample whole families and condition on the five fixed fits.}\label{tab:moments-autonomous-paired}
\begin{tabular}{lrr}
\toprule
Cohort & Difference & 95\% interval \\
\midrule
100 eV & 0.0006 & [0.0003, 0.0010] \\
200 eV & 0.0041 & [0.0031, 0.0051] \\
400 eV & -0.0019 & [-0.0067, 0.0033] \\
New delays & 0.0004 & [-0.0001, 0.0009] \\
Three pulses & -0.0007 & [-0.0014, 0.0001] \\
\bottomrule
\end{tabular}
\end{table}
\begin{table}[htbp]
\centering
\caption{Validation selection and training time for moments without clock inputs.}\label{tab:moments-autonomous-resources}
\begin{tabular}{lrrrrr}
\toprule
Seed & Best epoch & Last epoch & Validation & CPU (h) & Wall (h) \\
\midrule
1701 & 100 & 120 & 0.0677 & 6.61 & 6.77 \\
1702 & 115 & 120 & 0.0672 & 6.60 & 6.76 \\
1703 & 100 & 120 & 0.0680 & 6.59 & 6.75 \\
1704 & 115 & 120 & 0.0674 & 6.35 & 6.49 \\
1705 & 120 & 120 & 0.0671 & 6.20 & 6.32 \\
\bottomrule
\end{tabular}
\end{table}
\begin{table}[htbp]
\centering
\caption{Response NMSE at $1\,\mu$s in the four previously inspected families per energy. The original long-horizon evaluation used one preselected fit.}\label{tab:autonomous-long}
\begin{tabular}{lrrrr}
\toprule
Energy & Autonomous (5 fits) & Autonomous (1701) & Learned (1701) & Physical \\
\midrule
100 eV & 0.0732 & 0.0896 & 0.0878 & 0.0801 \\
200 eV & 0.1511 & 0.1480 & 0.2632 & 0.1451 \\
400 eV & 0.4857 & 0.4726 & 0.9609 & 0.3882 \\
\bottomrule
\end{tabular}
\end{table}

\begin{table}[htbp]
\centering
\caption{Clock sensitivity on the same four 400 eV source families at $1\ \mu$s. Each cell averages the two archived pulse schedules. The three learned predictors all use seed 1701.}\label{tab:clock-matched-autonomous}
\begin{tabular}{lrrrrr}
\toprule
Predictor & 40096 & 40097 & 40098 & 40099 & Mean \\
\midrule
Original learned clock & 0.9154 & 1.1218 & 1.1018 & 0.7045 & 0.9609 \\
Clock clamped at 20 ns & 0.6010 & 0.5983 & 0.4659 & 0.3768 & 0.5105 \\
Autonomous fit & 0.4105 & 0.6271 & 0.5616 & 0.2912 & 0.4726 \\
Physical moments & 0.3001 & 0.3479 & 0.4418 & 0.4631 & 0.3882 \\
\bottomrule
\end{tabular}
\end{table}
At this horizon, the autonomous fit minus the clamped fit gave a paired difference of -0.0379 (four-family percentile interval $[-0.1381, 0.0622]$).

\endgroup


\clearpage
\begin{thebibliography}{27}
\providecommand{\natexlab}[1]{#1}
\providecommand{\url}[1]{\texttt{#1}}
\expandafter\ifx\csname urlstyle\endcsname\relax
  \providecommand{\doi}[1]{doi: #1}\else
  \providecommand{\doi}{doi: \begingroup \urlstyle{rm}\Url}\fi

\bibitem[Allison et~al.(2016)Allison, Amako, Apostolakis, Arce, Asai,
  et~al.]{allison2016geant4}
J.~Allison, K.~Amako, J.~Apostolakis, P.~Arce, M.~Asai, et~al.
\newblock Recent developments in {Geant4}.
\newblock \emph{Nuclear Instruments and Methods in Physics Research Section A},
  835:\penalty0 186--225, 2016.
\newblock \doi{10.1016/j.nima.2016.06.125}.
\newblock URL \url{https://doi.org/10.1016/j.nima.2016.06.125}.

\bibitem[Bernal et~al.(2015)Bernal, Bordage, Brown, Dav{\'i}dkov{\'a}, Delage,
  et~al.]{bernal2015trackstructure}
M.~A. Bernal, M.~C. Bordage, J.~M.~C. Brown, M.~Dav{\'i}dkov{\'a}, E.~Delage,
  et~al.
\newblock Track structure modeling in liquid water: A review of the
  {Geant4-DNA} very low energy extension of the {Geant4} {Monte Carlo}
  simulation toolkit.
\newblock \emph{Physica Medica}, 31:\penalty0 861--874, 2015.
\newblock \doi{10.1016/j.ejmp.2015.10.087}.
\newblock URL \url{https://doi.org/10.1016/j.ejmp.2015.10.087}.

\bibitem[Chappuis et~al.(2024)Chappuis, Tran, Gon{\c{c}}alves~Jorge, Zein,
  Kyriakou, Emfietzoglou, Bailat, Bochud, Incerti, and
  Desorgher]{chappuis2024pulse}
Flore Chappuis, Hoang~Ngoc Tran, Patrik Gon{\c{c}}alves~Jorge, Sara~A. Zein,
  Ioanna Kyriakou, Dimitris Emfietzoglou, Claude Bailat, Fran{\c{c}}ois Bochud,
  S{\'e}bastien Incerti, and Laurent Desorgher.
\newblock Investigating ultra-high dose rate water radiolysis using the
  {Geant4-DNA} toolkit and a {Geant4} model of the {Oriatron eRT6} electron
  linac.
\newblock \emph{Scientific Reports}, 14\penalty0 (1):\penalty0 26707, 2024.
\newblock \doi{10.1038/s41598-024-76769-0}.
\newblock URL \url{https://www.nature.com/articles/s41598-024-76769-0}.

\bibitem[Ernst et~al.(2018)Ernst, Bartol, Sejnowski, and
  Mjolsness]{ernst2018boltzmann}
Oliver~K. Ernst, Thomas Bartol, Terrence Sejnowski, and Eric Mjolsness.
\newblock Learning dynamic boltzmann distributions as reduced models of spatial
  chemical kinetics.
\newblock \emph{The Journal of Chemical Physics}, 149\penalty0 (3):\penalty0
  034107, 2018.
\newblock \doi{10.1063/1.5026403}.
\newblock URL \url{https://doi.org/10.1063/1.5026403}.

\bibitem[Ernst et~al.(2019)Ernst, Bartol, Sejnowski, and
  Mjolsness]{ernst2019moment}
Oliver~K. Ernst, Thomas~M. Bartol, Terrence~J. Sejnowski, and Eric Mjolsness.
\newblock Learning moment closure in reaction-diffusion systems with spatial
  dynamic {Boltzmann} distributions.
\newblock \emph{Physical Review E}, 99\penalty0 (6):\penalty0 063315, 2019.
\newblock \doi{10.1103/PhysRevE.99.063315}.
\newblock URL
  \url{https://journals.aps.org/pre/abstract/10.1103/PhysRevE.99.063315}.

\bibitem[{Geant4 Collaboration}(2026)]{geant4v1143}
{Geant4 Collaboration}.
\newblock {Geant4} source code, release {v11.4.3}.
\newblock \url{https://github.com/Geant4/geant4/tree/v11.4.3}, 2026.

\bibitem[Gonzalez~Laiz and Schneider(2026)]{gonzalezlaiz2026limits}
Rodrigo Gonzalez~Laiz and Steffen Schneider.
\newblock Understanding the limits of latent world models, 2026.
\newblock URL \url{https://arxiv.org/abs/2610.37588}.
\newblock Preprint.

\bibitem[Gottlieb and Shu(1998)]{gottlieb1998tvd}
Sigal Gottlieb and Chi-Wang Shu.
\newblock Total variation diminishing {Runge--Kutta} schemes.
\newblock \emph{Mathematics of Computation}, 67\penalty0 (221):\penalty0
  73--85, 1998.
\newblock \doi{10.1090/S0025-5718-98-00913-2}.

\bibitem[Incerti et~al.(2010)Incerti, Baldacchino, Bernal, Capra, Champion,
  et~al.]{incerti2010project}
S.~Incerti, G.~Baldacchino, M.~Bernal, R.~Capra, C.~Champion, et~al.
\newblock The {Geant4-DNA} project.
\newblock \emph{International Journal of Modeling, Simulation, and Scientific
  Computing}, 1:\penalty0 157--178, 2010.
\newblock \doi{10.1142/S1793962310000122}.
\newblock URL \url{https://doi.org/10.1142/S1793962310000122}.

\bibitem[Karamitros et~al.(2014)Karamitros, Luan, Bernal, Allison, Baldacchino,
  Davidkova, et~al.]{karamitros2014diffusion}
M.~Karamitros, S.~Luan, M.~A. Bernal, J.~Allison, G.~Baldacchino, M.~Davidkova,
  et~al.
\newblock Diffusion-controlled reactions modeling in {Geant4-DNA}.
\newblock \emph{Journal of Computational Physics}, 274:\penalty0 841--882,
  2014.
\newblock \doi{10.1016/j.jcp.2014.06.011}.
\newblock URL \url{https://doi.org/10.1016/j.jcp.2014.06.011}.

\bibitem[Kim et~al.(2021)Kim, Ricke, and Van~Voorhis]{kim2021correlation}
Changhae~Andrew Kim, Nathan~D. Ricke, and Troy Van~Voorhis.
\newblock Machine learning dynamic correlation in chemical kinetics.
\newblock \emph{The Journal of Chemical Physics}, 155\penalty0 (14):\penalty0
  144107, 2021.
\newblock \doi{10.1063/5.0065874}.
\newblock URL \url{https://doi.org/10.1063/5.0065874}.

\bibitem[Kreipl et~al.(2009)Kreipl, Friedland, and
  Paretzke]{kreipl2009interaction}
Maximilian~Stephan Kreipl, Werner Friedland, and Herwig~G. Paretzke.
\newblock Interaction of ion tracks in spatial and temporal proximity.
\newblock \emph{Radiation and Environmental Biophysics}, 48\penalty0
  (4):\penalty0 349--359, 2009.
\newblock \doi{10.1007/s00411-009-0234-z}.
\newblock URL \url{https://doi.org/10.1007/s00411-009-0234-z}.

\bibitem[Le et~al.(2024)Le, Tran, Fattori, Phan, and Incerti]{le2024pulse}
Tuan~Anh Le, Hoang~Ngoc Tran, Serena Fattori, Viet~Cuong Phan, and Sebastien
  Incerti.
\newblock Modeling water radiolysis with {Geant4-DNA}: Impact of the temporal
  structure of the irradiation pulse under oxygen conditions, 2024.
\newblock URL \url{https://arxiv.org/abs/2409.11993}.
\newblock Preprint.

\bibitem[Peherstorfer et~al.(2016)Peherstorfer, Willcox, and
  Gunzburger]{peherstorfer2016multifidelity}
Benjamin Peherstorfer, Karen Willcox, and Max Gunzburger.
\newblock Optimal model management for multifidelity {Monte Carlo} estimation.
\newblock \emph{SIAM Journal on Scientific Computing}, 38\penalty0
  (5):\penalty0 A3163--A3194, 2016.
\newblock \doi{10.1137/15M1046472}.

\bibitem[Perkins et~al.(2026)Perkins, Webb, and Currell]{perkins2026miracle}
Charlie Perkins, Marcus Webb, and Fred~J. Currell.
\newblock Simulation of radiation chemistry by a one-shot hybrid
  continuum/{Monte Carlo} method.
\newblock \emph{Digital Discovery}, 2026.
\newblock \doi{10.1039/D6DD00003G}.
\newblock URL \url{https://doi.org/10.1039/D6DD00003G}.

\bibitem[Ramos-M{\'e}ndez et~al.(2020)Ramos-M{\'e}ndez, Shin, Karamitros,
  Dom{\'i}nguez-Kondo, Tran, Incerti, Villagrasa, Perrot, {\v S}t{\v e}p{\'a}n,
  Okada, Moreno-Barbosa, and Faddegon]{ramosmendez2020irt}
Jos{\'e} Ramos-M{\'e}ndez, Wook-Geun Shin, Mathieu Karamitros, Jorge
  Dom{\'i}nguez-Kondo, Ngoc~Hoang Tran, Sebastien Incerti, Carmen Villagrasa,
  Yann Perrot, V{\'a}clav {\v S}t{\v e}p{\'a}n, Shogo Okada, Eduardo
  Moreno-Barbosa, and Bruce Faddegon.
\newblock Independent reaction times method in {Geant4-DNA}: Implementation and
  performance.
\newblock \emph{Medical Physics}, 47\penalty0 (11):\penalty0 5919--5930, 2020.
\newblock \doi{10.1002/mp.14490}.
\newblock URL \url{https://pmc.ncbi.nlm.nih.gov/articles/PMC7891885/}.

\bibitem[Regazzoni et~al.(2024)Regazzoni, Pagani, Salvador, Ded{\`e}, and
  Quarteroni]{regazzoni2024ldnets}
Francesco Regazzoni, Stefano Pagani, Matteo Salvador, Luca Ded{\`e}, and Alfio
  Quarteroni.
\newblock Learning the intrinsic dynamics of spatio-temporal processes through
  latent dynamics networks.
\newblock \emph{Nature Communications}, 15:\penalty0 1834, 2024.
\newblock \doi{10.1038/s41467-024-45323-x}.
\newblock URL \url{https://www.nature.com/articles/s41467-024-45323-x}.

\bibitem[Sanchez-Gonzalez et~al.(2020)Sanchez-Gonzalez, Godwin, Pfaff, Ying,
  Leskovec, and Battaglia]{sanchez2020gns}
Alvaro Sanchez-Gonzalez, Jonathan Godwin, Tobias Pfaff, Rex Ying, Jure
  Leskovec, and Peter Battaglia.
\newblock Learning to simulate complex physics with graph networks.
\newblock In \emph{Proceedings of the 37th International Conference on Machine
  Learning}, volume 119 of \emph{Proceedings of Machine Learning Research},
  pages 8459--8468. PMLR, 2020.
\newblock URL
  \url{https://proceedings.mlr.press/v119/sanchez-gonzalez20a.html}.

\bibitem[Schwarz(1969)]{schwarz1969spur}
Harold~A. Schwarz.
\newblock Applications of the spur diffusion model to the radiation chemistry
  of aqueous solutions.
\newblock \emph{The Journal of Physical Chemistry}, 73\penalty0 (6):\penalty0
  1928--1937, 1969.
\newblock \doi{10.1021/j100726a047}.
\newblock URL \url{https://doi.org/10.1021/j100726a047}.

\bibitem[Si and Yan(2026)]{si2026rdjepa}
Chenhao Si and Ming Yan.
\newblock {RD-JEPA}: Predictive latent pretraining for few-trajectory transfer
  across reaction--diffusion equations, 2026.
\newblock URL \url{https://arxiv.org/abs/2609.29403}.
\newblock Preprint.

\bibitem[Thorpe et~al.(2026)Thorpe, Tretiakov, Hsiao, Low, Li, Iqbal, Bhatt,
  Topcu, and Kumar]{thorpe2026viable}
Adam~J. Thorpe, Stepan Tretiakov, Cheng-Hsi Hsiao, Su~Ann Low, Xingjian Li,
  Hassan Iqbal, Neel~P. Bhatt, Ufuk Topcu, and Krishna Kumar.
\newblock Physically viable world models: A case for query-conditioned embodied
  {AI}, 2026.
\newblock URL \url{https://arxiv.org/abs/2605.30542}.

\bibitem[Tran et~al.(2021)Tran, Ramos-M{\'e}ndez, Shin, Perrot, Faddegon,
  Okada, Karamitros, Dav{\'i}dkov{\'a}, {\v S}t{\v e}p{\'a}n, Incerti, and
  Villagrasa]{tran2021synchronousirt}
Hoang~Ngoc Tran, Jos{\'e} Ramos-M{\'e}ndez, Wook-Geun Shin, Yann Perrot, Bruce
  Faddegon, Shogo Okada, Mathieu Karamitros, Marie Dav{\'i}dkov{\'a},
  V{\'a}clav {\v S}t{\v e}p{\'a}n, S{\'e}bastien Incerti, and Carmen
  Villagrasa.
\newblock Assessment of {DNA} damage with an adapted independent reaction time
  approach implemented in {Geant4-DNA} for the simulation of
  diffusion-controlled reactions between radio-induced reactive species and a
  chromatin fiber.
\newblock \emph{Medical Physics}, 48\penalty0 (2):\penalty0 890--901, 2021.
\newblock \doi{10.1002/mp.14612}.
\newblock URL \url{https://pmc.ncbi.nlm.nih.gov/articles/PMC7986154/}.

\bibitem[Tran et~al.(2024)Tran, Archer, Baldacchino, Brown, Chappuis,
  et~al.]{tran2024chemistryreview}
Hoang~Ngoc Tran, Jay Archer, G{\'e}rard Baldacchino, Jeremy M.~C. Brown, Flore
  Chappuis, et~al.
\newblock Review of chemical models and applications in {Geant4-DNA}: Report
  from the {ESA BioRad III} project.
\newblock \emph{Medical Physics}, 51\penalty0 (9):\penalty0 5873--5889, 2024.
\newblock \doi{10.1002/mp.17256}.
\newblock URL \url{https://aapm.onlinelibrary.wiley.com/doi/10.1002/mp.17256}.

\bibitem[Villar et~al.(2021)Villar, Hogg, Storey-Fisher, Yao, and
  Blum-Smith]{villar2021scalars}
Soledad Villar, David~W. Hogg, Kate Storey-Fisher, Weichi Yao, and Ben
  Blum-Smith.
\newblock Scalars are universal: Equivariant machine learning, structured like
  classical physics.
\newblock In \emph{Advances in Neural Information Processing Systems},
  volume~34, 2021.
\newblock URL
  \url{https://proceedings.neurips.cc/paper/2021/file/f1b0775946bc0329b35b823b86eeb5f5-Paper.pdf}.

\bibitem[Wei et~al.(2026)Wei, Hsu, Niu, Li, and Li]{wei2026distinguish}
Rongzhe Wei, Hans Hao-Hsun Hsu, Peizhi Niu, Yifan Li, and Pan Li.
\newblock What must a world model distinguish for planning?, 2026.
\newblock URL \url{https://arxiv.org/abs/2609.33030}.

\bibitem[Yin et~al.(2021)Yin, Le~Guen, Dona, de~B{\'e}zenac, Ayed, Thome, and
  Gallinari]{yin2021aphynity}
Yuan Yin, Vincent Le~Guen, J{\'e}r{\'e}mie Dona, Emmanuel de~B{\'e}zenac,
  Ibrahim Ayed, Nicolas Thome, and Patrick Gallinari.
\newblock Augmenting physical models with deep networks for complex dynamics
  forecasting.
\newblock In \emph{International Conference on Learning Representations}, 2021.
\newblock URL \url{https://openreview.net/forum?id=kmG8vRXTFv}.

\bibitem[Zaheer et~al.(2017)Zaheer, Kottur, Ravanbakhsh, P{\'o}czos,
  Salakhutdinov, and Smola]{zaheer2017deepsets}
Manzil Zaheer, Satwik Kottur, Siamak Ravanbakhsh, Barnab{\'a}s P{\'o}czos,
  Ruslan Salakhutdinov, and Alexander~J. Smola.
\newblock Deep sets.
\newblock In \emph{Advances in Neural Information Processing Systems},
  volume~30, 2017.
\newblock URL
  \url{https://proceedings.neurips.cc/paper_files/paper/2017/hash/f22e4747da1aa27e363d86d40ff442fe-Abstract.html}.

\end{thebibliography}
\end{document}